\documentclass[usenames,lettersize,journal,romanappendices, dvipsnames]{IEEEtran}

\usepackage{microtype}            
\PassOptionsToPackage{warn}{textcomp}  
\usepackage{textcomp}                  
\usepackage[dvipsnames]{xcolor}

\usepackage{mathptmx}                  
\usepackage{times}                     
\usepackage{cite}                      
\usepackage{tabu}                      
\usepackage{booktabs}                  
\usepackage{wrapfig}
\usepackage[T1]{fontenc}

\usepackage{graphicx}
\usepackage{comment}
\usepackage{multirow} 
\usepackage{caption}
\usepackage{subcaption}
\usepackage{epstopdf}
\usepackage{amsmath}
\usepackage{array}
\usepackage{tikz}
\usepackage[export]{adjustbox}
\usepackage{hyperref}
\hypersetup{
   pdfpagemode=UseNone,
   pdftitle={\@title},
   pdfauthor={\@author},
   pdfsubject={},
   pdfkeywords={},
   pageanchor=true,
   plainpages=false,       
   hypertexnames=false,    
   bookmarksnumbered=false,
   bookmarksopen=true,     
   bookmarksopenlevel=3,   
   pdfpagemode=UseNone,    
   pdfstartview=Fit,       
   pdfborder={0 0 0},      
   breaklinks=true,        
   colorlinks=true,       
   linkcolor={rgb:red,8;green,111;blue,200}, %
   citecolor={rgb:red,8;green,111;blue,200}, %
   urlcolor={rgb:red,8;green,111;blue,200},  %
   nesting=true
}
\usepackage{xspace}

\usepackage{tabu}                      
\usepackage{booktabs}                  
\usepackage{lipsum}                    
\usepackage{mwe}                       

\usepackage{mathptmx}                  
\usepackage{wrapfig}
\usepackage{xcolor}
\usepackage{setspace} 

\usepackage{blindtext}
\usepackage{multicol}

\usepackage{tcolorbox}
\usepackage{microtype}                 
\PassOptionsToPackage{warn}{textcomp}  
\usepackage{textcomp}                  
\usepackage{times}                     
\usepackage{wrapfig}
\usepackage{calc}
\usepackage{todonotes}
\usepackage{graphicx}
\usepackage{comment}
\usepackage{multirow} 
\usepackage{caption}
\usepackage{subcaption}
\usepackage{epstopdf}
\usepackage{amsmath}
\usepackage{array}
\usepackage{tikz}
\usepackage{xspace}
\usepackage[percent]{overpic}
\usepackage{enumitem}
\usepackage{ccicons}
\usepackage{wrapfig}

\usepackage[export]{adjustbox}

\usepackage{fancybox, graphicx}

\usepackage{nicematrix}
\usetikzlibrary{patterns,shadings}

\definecolor{Black}{RGB}{0,0,0}

\newcommand{\tochange}[1]{\protect\textcolor{blue}{#1}}
\renewcommand{\tochange}[1]{#1}

\newcommand{\et}{essential stimuli\xspace}

\newcommand{\typename}{{VisTypes}\xspace}

\newcommand{\ipc}{``\textit{image purpose}''\xspace}

\newcommand{\expertUs}{ExpertResearchers\xspace}
\newcommand{\expertExternal}{ExpertViewers\xspace}

\usepackage[utf8]{inputenc}
\newcommand{\vistypesgooglelink}{\href{https://docs.google.com/spreadsheets/d/1YCfS9xgjPzEP8-oNy9OIToSk8KqydGHTi0Qrl12qhpA/edit?usp=sharing}{\texttt{go.osu.edu/vistypes}}\xspace}

\newcommand{\vistypeImagegooglelink}{\href{http://go.osu.edu/vis30k}{\texttt{go.osu.edu/vis30k}}\xspace}

\newcommand{\vistypegooglelink}{\href{https://go.osu.edu/vistypesdatapaperfigures}{\texttt{go.osu.edu/vistypesdatapaperfigures}}\xspace}

\newcommand{\osflink}{\href{https://osf.io/dxjwt}{\texttt{osf\discretionary{}{.}{.}io\discretionary{/}{}{/}dxjwt}}\xspace}

\newcommand{\massvislink}{\href{go.osu.edu/vistypescodedmassvis}
{\texttt{go.osu.edu/vistypescodedmassvis}}\xspace}

\newcommand{\imagelink}{\href{https://visimagenavigator.github.io/}{\texttt{visimagenavigator.github.io}}\xspace}

\definecolor{VividBurgundy}{RGB}{159,29,53}
\definecolor{dev}{RGB}{29,159,53}
\definecolor{picolor}{HTML}{2a9d8f}
\definecolor{jccolor}{RGB}{159,89,53}
\definecolor{wecolorc}{RGB}{255, 153, 51}
\definecolor{theycolorc}{RGB}{0, 191, 255}
\definecolor{dengcolorc}{RGB}{24, 214, 14}
\newcommand{\wecolor}[1]{\protect\textcolor{wecolorc}{#1}}
\newcommand{\theycolor}[1]{\protect\textcolor{theycolorc}{#1}}
\newcommand{\dengcolor}[1]{\protect\textcolor{dengcolorc}{#1}}

\newcommand{\codingdecisions}[1]{\textit{#1}}
\newcommand\drpi[1]{{\color{Black}#1}}

\newcommand\jcfr[1]{\protect\textcolor{blue}{#1}}
\renewcommand{\jcfr}[1]{#1}
\newcommand\jctg[1]{\protect\textcolor{black}{#1}}

\newcommand{\eg}{e.\,g.}
\newcommand{\ie}{i.\,e.}

\newcommand{\obj}{object\xspace}

\newcommand{\vislength}{``generalized bar representations''\xspace}
\newcommand{\visbar}{\vislength\xspace}
\newcommand{\vispoint}{``generalized point representations''\xspace}
\newcommand{\visline}{``generalized line representations''\xspace}
\newcommand{\visnodelink}{``generalized node-link trees\discretionary{/}{}{/}graphs, networks, meshes''\xspace}
\newcommand{\visarea}{``generalized area representations''\xspace}
\newcommand{\vissurface}{``generalized sur\-face and volume representations''\xspace}
\newcommand{\visgrid}{``generalized matrix and grid''\xspace}
\newcommand{\visglyph}{``generalized glyph representations''\xspace}
\newcommand{\vistext}{``generalized text representations''\xspace}
\newcommand{\viscolor}{``generalized continuous color/grey-scale/texture representations''\xspace}

\newcommand{\visschematic}{``sche\-ma\-tic representations and concept illustrations''\xspace}
\newcommand{\visgui}{``GUI (screen\-shots)\discretionary{/}{}{/}user interface depiction''\xspace}

\newcommand{\barabbr}{\textit{Bar}\xspace}
\newcommand{\lineabbr}{\textit{Line}\xspace}
\newcommand{\pointabbr}{\textit{Point}\xspace}
\newcommand{\areaabbr}{\textit{Area}\xspace}
\newcommand{\nodelinkabbr}{\textit{Node-link}\xspace}
\newcommand{\gridabbr}{\textit{Grid}\xspace}
\newcommand{\glyphabbr}{\textit{Glyph}\xspace}
\newcommand{\textabbr}{\textit{Text}\xspace}
\newcommand{\colorabbr}{\textit{Cont.-ColorPattern}\xspace}
\newcommand{\surfacevolumeabbr}{\textit{Surface/Volume}\xspace}
\newcommand{\schematicabbr}{\textit{Schematic}\xspace}

\newcommand{\scholar}{scholars\xspace}

\newlength{\pictureheight}

\newcommand{\papersSeven}{695\xspace} 
\newcommand{\imagesfigSeven}{6,833\xspace} 

\newcommand{\imagestype}{4,070\xspace} 

\newcommand{\totaltypelabels}{6,252\xspace} 
\newcommand{\totalhardnesslabels}{13,647\xspace} 
\newcommand{\totaldimlabels}{7,174\xspace} 
\newcommand{\totalfunctiontypelabels}{2,754\xspace} 

\newcommand{\lineprop}{$17.7\%$}

\newcommand{\surfacecon}{$74\%$\xspace}
\newcommand{\barcon}{$67\%$\xspace}  
\newcommand{\linecon}{$65\%$\xspace}
\newcommand{\pointcon}{$57\%$\xspace}

\newcommand{\patterncon}{$42\%$\xspace}

\newcommand{\textcon}{$39\%$\xspace}
\newcommand{\glyphcon}{$35\%$\xspace}

\newcommand{\guicon}{$70\%$\xspace}

\newcommand{\numschematicOrg}{1,929\xspace} 
\newcommand{\numguiOrg}{825\xspace}  

\renewcommand*{\thesection}{\arabic{section}}
\newlength\myheight
\newlength\mydepth
\newcommand*\wordimg[1]{%
  \settototalheight\myheight{Xygp}%
  \settodepth\mydepth{Xygp}%
  \raisebox{-\mydepth}{\includegraphics[height=\myheight]{#1}}%
}

\newcommand{\appref}[1]{\hyperref[#1]{Appx.}~\ref{#1}}

\begin{document}

\newcommand{\mytitle}{
An Image-based Typology for Visualization}
\title{\tochange{\mytitle}}

\author{Jian Chen\hspace{1pt}\href{https://orcid.org/0000-0002-1599-0831}{\raisebox{3pt}[0pt][0pt]{\includegraphics[width=2.7mm]{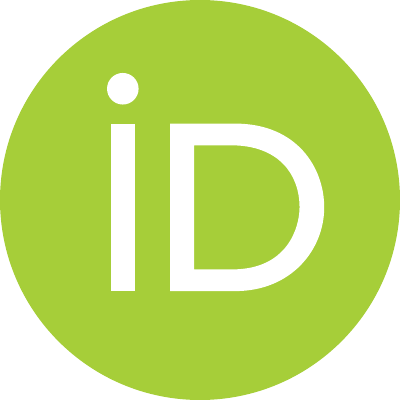}}}, 
Petra Isenberg\hspace{1pt}\href{https://orcid.org/0000-0002-2948-6417}{\raisebox{3pt}[0pt][0pt]{\includegraphics[width=2.7mm]{orcid}}}, 
Robert S. Laramee\hspace{1pt}\href{https://orcid.org/0000-0002-3874-6145}{\raisebox{3pt}[0pt][0pt]{\includegraphics[width=2.7mm]{orcid}}}, 
Tobias Isenberg\hspace{1pt}\href{https://orcid.org/0000-0001-7953-8644}{\raisebox{3pt}[0pt][0pt]{\includegraphics[width=2.7mm]{orcid}}},\\ 
Michael Sedlmair\hspace{1pt}\href{https://orcid.org/0000-0001-7048-9292}{\raisebox{3pt}[0pt][0pt]{\includegraphics[width=2.7mm]{orcid}}}, 
Torsten M{\"o}ller\hspace{1pt}\hspace{1pt}\href{https://orcid.org/0000-0003-1192-0710}{\raisebox{3pt}[0pt][0pt]{\includegraphics[width=2.7mm]{orcid}}}, 
Rui Li\hspace{1pt}\href{https://orcid.org/0000-0002-8166-9667}{\raisebox{3pt}[0pt][0pt]{\includegraphics[width=2.7mm]{orcid}}}
\thanks{This paper was produced by The Ohio State University, USA; Université Paris-Saclay, CNRS, Inria, LISN, France; University of Nottingham, UK; University of Stuttgart, Germany; University of Vienna, Austria}
\thanks{Manuscript received March --, 2024; revised --, --.}}

\markboth{submitted to IEEE Transactions on Visualization and Computer Graphics}%
{Chen \MakeLowercase{\textit{et al.}}: \typename: An Image-Based Typology for Visualization: A Survey}


\maketitle

\begin{abstract}%
\jctg{We present and discuss the results of a qualitative analysis of visualization images to derive an image-based typology of visualizations.} For each image, we seek to \drpi{identify its main focus or the essential stimuli.
}
\jctg{As a result,
we derived 
10 image-based
visualization types.
}
\jctg{We describe 
coding decisions we made in the 
derivation process.
}
\jctg{The resulting image typology 
can serve a number of purposes: 
enabling researchers and practitioners to identify visual design styles,
facilitating the categorization of visualization images for the purpose of research and teaching, enabling researchers to study the evolution of the community and its research output over time, 
and facilitating a discussion of standardization in visualization. 
In addition, the tool and dataset enable \scholar to closely examine the images and how they are published and communicated in our community.}
\osflink presents a pre-registration
and all supplemental materials.

\end{abstract}

\begin{IEEEkeywords}
Typology, image-based, visual representation, evaluation, image types. 
\end{IEEEkeywords}

\begin{figure*}[!tp]
  \centering  
  \includegraphics[width=\linewidth]{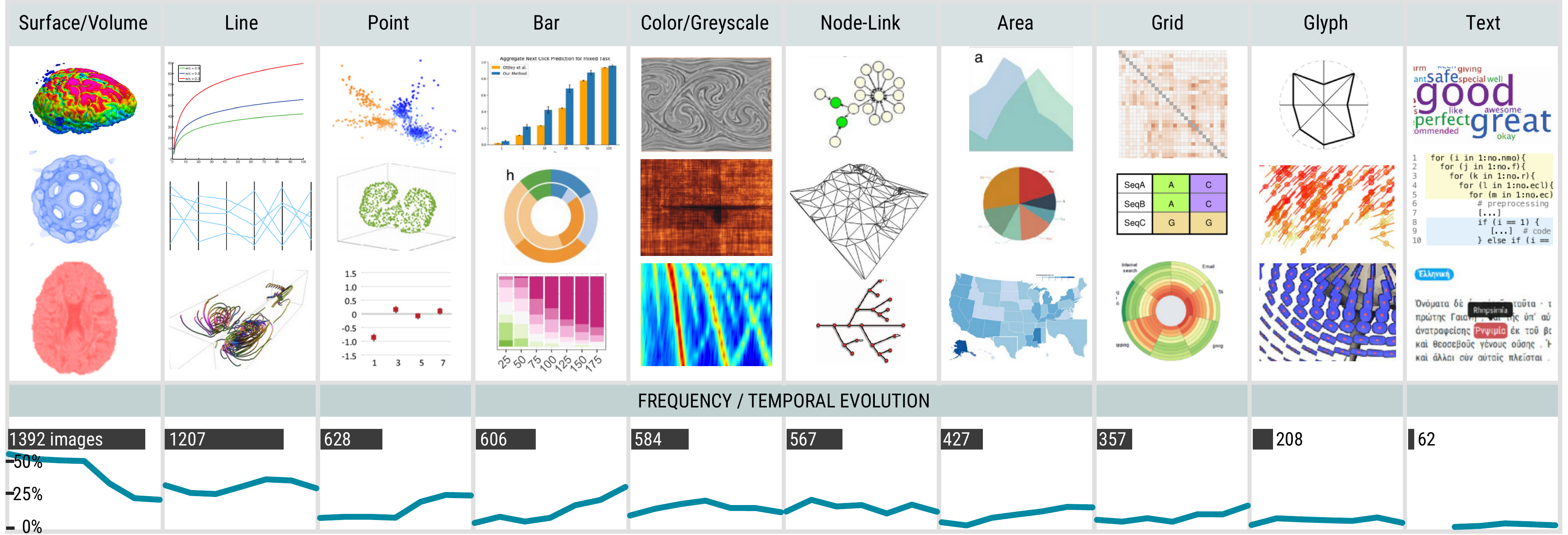}
 \caption{\tochange{\textbf{Our 10 visualization types (\typename).} 
 They characterize the associations of what we see from 
 images. 
Here we show
the overall image counts and relative frequency trends in \% for 1990--2020. Surface-based techniques \& volumes are most common, followed by line-based techniques and points.
Together, these three main types make up about half of the images we coded
\tochange{
(See our online tool}
 \imagelink for additional examples).
Each type represents a family of visual appearances of \et, the main focus of a visualization image.
} 
} 
\label{fig:teaser}
\end{figure*}

\section{Introduction}
\label{sec:intro}

\IEEEPARstart
{W}{hile} characterizations of visualization representations have been proposed in the literature, they focus on
\jctg{what designers construct rather than what viewers see from images.}
We attempt to fill this gap by describing an image-based typology.
Our goal is to provide a framework for categorizing visualization images with a focus on the \textit{visual appearance} of the \textit{\et}---\jctg{\textit{the main visual focus of an
image. 
}}

\textbf{Why is it important to categorize visualization images?
} 
\tochange{Categorization is 
``\textit{an abstraction of things and ideas into groups and most if not all categories have verbal labels}''~\cite[p.\ 318]{ware2020information}.
Categorization allows us to reason about a phenomenon, discover order, and simplify discussion and experience \cite{Jacob:2004:Classification}.}
Visualization research has taken advantage of these benefits throughout its history. 
Existing categorizations enable us to understand the field, its practices, and progress better and inspire 
future work; for example, to study representations for specific data types \cite{aigner2011visualization, Schulz:2022:TreeVis}, research topics and keywords \cite{isenberg2016visualization}, evaluation goals and practices \cite{isenberg2013systematic,lam2011empirical},  
interaction techniques \cite{yi2007toward}, or tasks and activities \cite{dimara2021critical,amar2005low}. 
Finally, in the era of artificial intelligence, one can also use our 
categories to index knowledge and to benchmark human or artificial information processors, to eventually assist humans~\cite{sucholutsky2023getting, rensink2021visualization}.

\textbf{Why a categorization based on visual appearance?}
\IEEEpubidadjcol
We consider visualization images to be standalone entities, and we posit that categorizing them enables us to derive new insights. For example, when developing a new research method for various applications, testing the method on a representative set of images is important; for an example see the development of the BeauVis scale \cite{He:2023:BeauVis}.  
In addition, a categorization focused on visual appearance also allows us to conduct a dedicated historical analysis. We can observe and compare the evolution of rendering styles and contextual representations within a category~\cite{koesten2023subjective}, compare design styles across categories, and attempt to reason about possible influences such as individual papers, designers, or rendering hardware.  Finally, for writing overview articles, textbooks, or lecture series, categorizations based on the main visual focus of attention can highlight the variety of approaches to specific techniques or identify aspects that need further investigation.

\textbf{Why a system of main focus or essential stimuli by visual similarity?}
Approaches for categorizing data visualizations often rely on evaluating them with respect to the amount of \textit{data} they represent (\eg, a few or millions) \cite{keim2000designing}, with respect to the difference in \textit{coordinate systems} (a part of the visual implantation by Bertin~\cite{bertin1983semiology}, also see Claessen and Van Wijk's work \cite{claessen2011flexible}), with respect to what aspect of the \textit{represented data is being visually encoded} (\eg, multivariate \& multifaceted \cite{Kehrer2013a}), or with respect to \textit{continuity} \cite{collins2009bubble}. These considerations belong to the fundamental aspects of a representational system \cite{palmer1978fundamental}. 
Here, we take a different approach that is neither a \textit{top-down} categorization (focusing on techniques and/or rendering methods) nor a categorization based on designer's basic drawing elements (grouping ``marks and channels'' that may not be able to distinguish image similarity). Instead, we use \et as \textit{prototype representations}, where a \textit{representation} is ``\textit{something static, something we can see and contemplate from images}''; and \textit{prototype} \cite{rosch1973natural} means ``\textit{essential stimuli that take a salient position in the formation of the category}''~\cite{tversky2019mind}.
\drpi{This categorization is human-centric, based on the observations and discussions of six senior visualization researchers. In later parts of the paper we show, however, that the categorizations can also be successfully applied by others.}

\textbf{Our method:}
The artifacts we categorized 
are images from papers published in IEEE VIS (or known as VisWeek between 2008–2012). 
\jctg{Specifically, we coded \imagesfigSeven~figures from \papersSeven~papers published in 1990, 1995, 2000, 2005, 2010, 2015, and 2020, spanning a 30-year history of the conference
\cite{chen2021vis30k}.}
To prioritize the main focus or essential stimuli in these images, we did not consult the captions or original papers. 
We follow 
an 
interpretation of categorization \cite{Jacob:2004:Classification} that 
\jctg{supports graded 
structures and allows categories 
to vary depending on contextual use.}
Thus, 
images
can be members of a category to varying degrees.
That is, some members of a category may be seen as more 
or less 
representative of a category than others,  a view that 
a taxonomy might prohibit. 
Also, fuzzy boundaries between categories mean that 
images can belong to more than one category.
Finally, we also used our background knowledge to influence the discussion of images that were not clearly in or out of certain categories. 
As such, our work is considered a \emph{categorization} rather than a \emph{classification} \cite{Jacob:2004:Classification}. 

\textbf{Results and validation:} Our final 
typology, \typename,
describes 10 visualization types (\autoref{fig:teaser}), two 
\jctg{purposes}
(schematic and GUI), and their perceived 
spatial dimensionality
(2D and 3D).
We also report on the results of a user study to validate the usability of our typology (\autoref{sec:validationstudy}). Finally, we
discuss how to apply \typename in the form of 
\textit{three 
scenarios}:  
to explore 
new empirical study methods,
to analyze visualization evolutions,
and to promote standardization (\autoref{sec:cases}).

\begin{table*}[!tb]
\caption{{\textbf{Definitions of our 10 labels.
}
They define the visual appearance of the \et
independent of data, tasks, or rendering techniques.
We color-coded the features in our codes in \wecolor{orange}.
The \obj
informs a viewer what to see to process the information in an image properly in order to build meaning from data represented in an image. Our code types are independent of data, tasks, or rendering techniques. 
}
}
  \label{tab:12schema}
  \scriptsize%
	\centering%
\begin{tabular}{p{0.004\textwidth}p{0.185\textwidth}p{.44\textwidth}p{.295\textwidth}}
  \toprule
  &\textbf{Visualization Type Codes} &  \textit{Description} & \textit{Examples}\\
  \midrule
(1) &Generalized Bar Representations (\barabbr) & Graphs that represent data with straight bars that can be arranged on a straight or curved baseline and whose \wecolor{heights or lengths are proportional to the values} they represent. & bar charts, stacked bar charts, 
box plots, or sunburst diagrams.
 \\
 (2) &{Point-based
 Representations
 (\pointabbr)} & Representations that 
 \wecolor{use point locations}. These locations are often shown using dots or circles, but also \wecolor{other shapes} such as 3D spheres, triangles, stars, etc. & scatterplots, point clouds, dot plots, or bubble charts.\\
 (3)&{
 Line-based 
 Representations (\lineabbr)} & Representations where information is emphasized through \wecolor{(straight or curved) lines}.
 & line charts, parallel coordinates, contour lines, radar/spider charts, streamlines, or tensor field lines.\\
(4)&
Node-link Trees/Graphs, Networks, Meshes (\nodelinkabbr) & 
 Representations using points for and \wecolor{explicit connections 
 between these points} to convey relationships between data values. & node-link diagrams, node-link trees, node-link graphs, meshes, arc diagrams, or Sankey diagrams.\\
(5)& 
Area-based Representations (\areaabbr) & Representations with a focus on areas of 2D space or 2D surfaces including 
\wecolor{sub-sets of these surfaces}. Areas can be geographical regions or polygons whose size or shape represents abstract data. 
 & (stacked) area chart, 
    streamgraph,
    ThemeRiver,
    violin plot, 
    cartograms,
    \jcfr{histograms},
    ridgeline chart,
    Voronoi diagram,
    treemaps,
    pie chart.\\
 (6) & {
 Surface-based
   and Volume Representations (\surfacevolumeabbr)} & 
  Representations of the \wecolor{inner and/or outer features} and/or \wecolor{boundaries of a continuous spatial phenomenon or object} in
  3D physical space or 4D space-time, or slices thereof.  
 & terrains,
 isosurfaces, 
 stream surfaces,
 volume rendering using transfer functions, 
 slices through a volume (\eg, X-ray, CT slice).\\
 (7) & Generalized Matrix / Grid (\gridabbr) & 
 Representations that \wecolor{separate data into a \textit{discrete} spatial} grid structure.  
 The grid often has rectangular cells but may also use other shapes such as hexagons or cubes. Elements such as glyphs or a color encoding can appear in the grid cells. 
  &
  network matrices, 
  discrete heatmaps, 
  scarf or strip plots, space-time cubes, or matrix-based network visualizations.\\
 (8) &
 Continuous Color and Grey-scale, {and Textures (\colorabbr)} 
& 
  Representations of structured \wecolor{patterns across an image or atop a geometric 3D object}. These patterns can be evoked by changes in intensity, changes in hue, brightness, and/or saturation. The changes are typically \wecolor{smooth (continuous)} but could show \wecolor{sharp transitions} as well.  
  & Directional patterns such as Line Integral Convolution (LIC), Spot Noise, and Image-Space Advection (ISA) to show flow fields, continuous heatmaps, intensity fields, or even a binary image. \\
 (9) &
 Glyph-based 
 Representations (\glyphabbr) & 
    \wecolor{Multiple small independent visual representations} (often encoded by position and additional dimensions using color, shape, or other geometric primitives) that depict multiple attributes (dimensions) of a data record. Placement is usually meaningful and typically multiple glyphs are displayed for comparison.  
   & 
   Star glyphs, 3D glyphs, Chernoff faces, vector field glyphs\\
 (10) & {
 Text-based 
 Representations (\textabbr)} &  \wecolor{Representations of data} (often text itself) that use \wecolor{varying properties of letters\discretionary{/}{}{/}words} such as font size, color, width, style, or type to encode data.
  & 
 Tag clouds, word trees, parallel tag clouds, typomaps.\\
  \bottomrule
  \end{tabular}\vspace{-1ex}%
\end{table*}

\tochange{
In summary, we make three contributions:
}
\begin{itemize}[nosep,left=0pt .. \parindent]
\item 
\tochange{
\textit{Method.} 
We have, for the first time, considered visualizations from the perspective of 
the \et in images. }
\item 
\tochange{
\textit{Types and the \typename Dataset.} 
We develop the currently largest carefully annotated dataset about visualization types. 
}
\item
\textit{User study.} 
We validate the usability of \typename. 
\end{itemize}

\section{Terms and Related Work}
\label{sec:relatedWork}

\noindent 
Related work includes visualization categorization studies of construction guidelines, data types, tasks, and work that analyzes research figures. We will review these areas next.

\subsection{\tochange{Visualization Typology: Definitions}}

\noindent 
The \textit{visualization images} we discuss refer to figures in papers, which we visually assessed. 
We 
differentiate \emph{visual element} from the term \emph{visual stimulus}.
By \emph{visual elements}, we refer to the individual components or building blocks used by visualization designers or practitioners to create an effective composition in an image. Visual elements often include marks and channels such as lines, shapes, colors, textures, icons, and typography (see Munzner's textbook~\cite{munzner2014visualization}). Furthermore, we take a psychophysics perspective and use the term \emph{visual stimulus} to mean any visual input or signal that stimulates the visual system
~\cite[pp.\ 394--395]{ware2012information}.
It is a broad term that can include objects, patterns, colors, and any other low-level visual input (\eg, lines)~\cite{wolfe2021guided}. As a result, while both \emph{visual element} and \emph{visual stimulus} are related to visual representations, we use \emph{visual element} as input to create visual design compositions and \emph{visual stimulus} to mean visual input perceived by visual information processors.
Given this distinction, \jctg{we use the primary visual focus to represent a stimulus that takes a salient position in a visualization image.}

\subsection{\tochange{Categorization as an analogy of ``What is it \textit{like}?''}} 

\noindent
There is a long history of 
categorization methods
in philosophy and 
psychology \cite{givon1986prototypes}.
Wittgenstein \cite{wittgenstein2010philosophical} argues that users are fluid about categories: different people may provide different answers, and the same person may respond with different answers at different times. 
As a result, a top-down categorization that draws clear boundaries of shared properties to those of other categories is 
often
too strict to represent how people understand categories.
Instead, a ``this-looks-like-that'' bottom-up association can be used. 
Rosch \cite{rosch1973natural} later proposed that natural categories exist (of colors or forms) which are abstracted to \textit{natural prototypes} \tochange{which take a salient position in the formation of a category} from a bottom-up cluster of similar instances. 
Psychologists (\eg, Medin and Schaffer \cite{medin1978context}, Nosofsky~\cite{nosofsky1986attention}, and Krushki \cite{kruschke2020alcove}) have gone further
and argued for what they call the \textit{sample-based theory} of categorization, where humans store 
\textit{instances} and \textit{examples}; 
similarities and relatedness between entities allow us to learn \textit{associations} between these examples.
\jcfr{For \textit{objects} that are closely associated, 
they \textit{look like} a cluster of learned examples and belong to the same category.} 
Inspired by these works, we summarize and provide examples of visualization images and 
what
we see in them. We make these associations the basis of our typology.

\setlength{\pictureheight}{2.2cm}

\begin{figure*}[!t]
\centering
\subfloat[ \theycolor{Map} vs. \wecolor{\textabbr}.]{ 
\includegraphics[height=1.1\pictureheight]{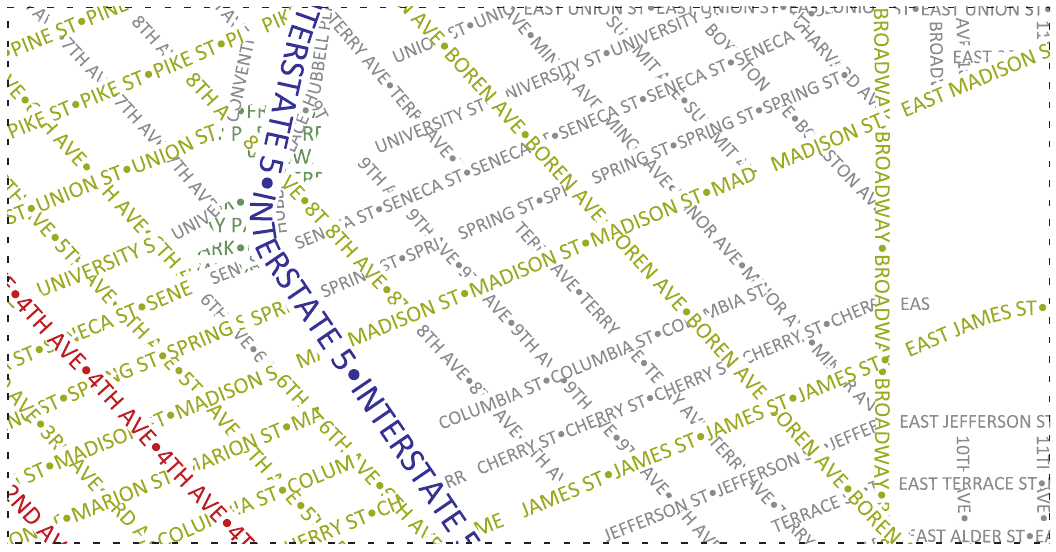}
} 
\hfill
\subfloat[\theycolor{Lines} vs. \wecolor{Schematic}.]{
\includegraphics[height=0.8\pictureheight]{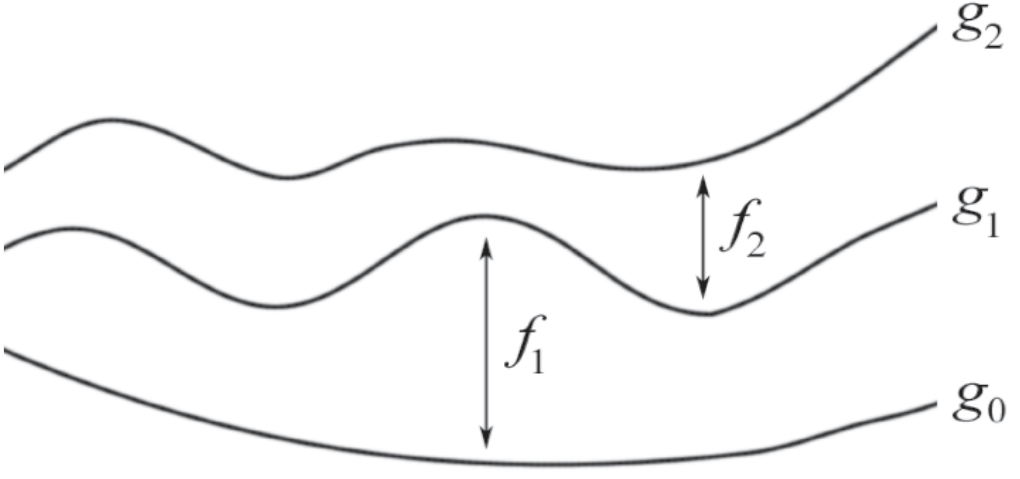}
} %
\hfill
\subfloat[
\theycolor{\barabbr} 
vs.\ \wecolor{\barabbr\,$\&$\,\pointabbr}.]
{ 
~\includegraphics[height=1.2\pictureheight]{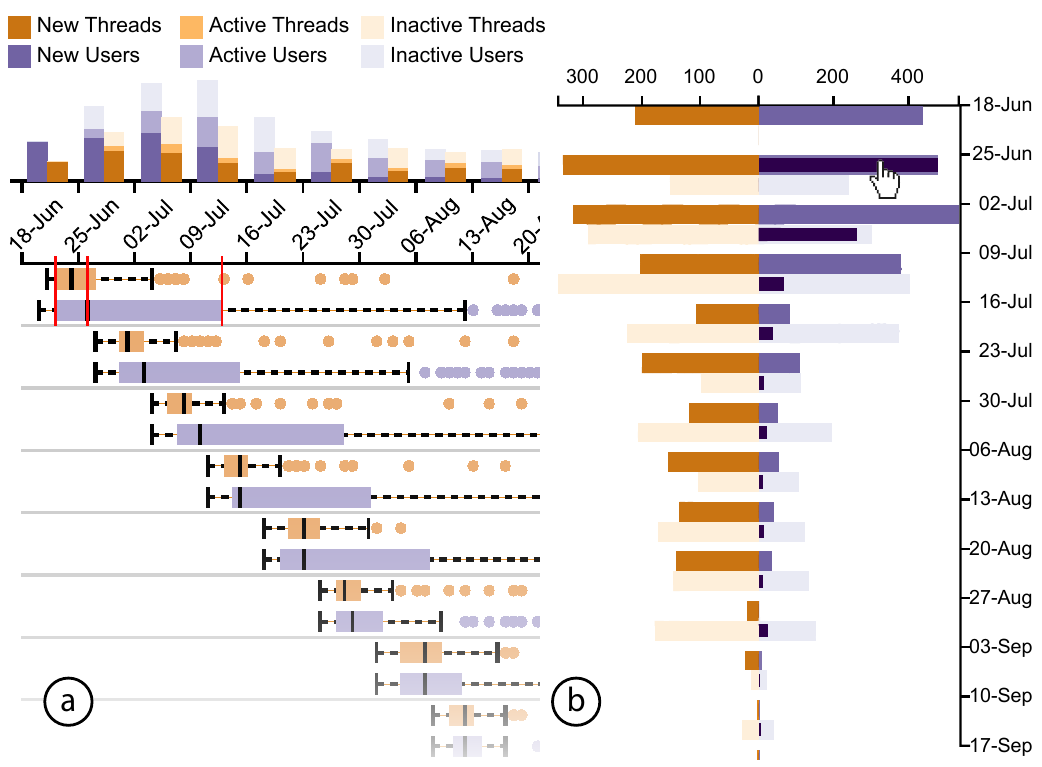}~
} %
\hfill
\subfloat[\dengcolor{Donut chart} vs. \wecolor{\barabbr $\&$ \nodelinkabbr}.]
{
\includegraphics[trim={600 0 0 0},clip, height=1.2\pictureheight]{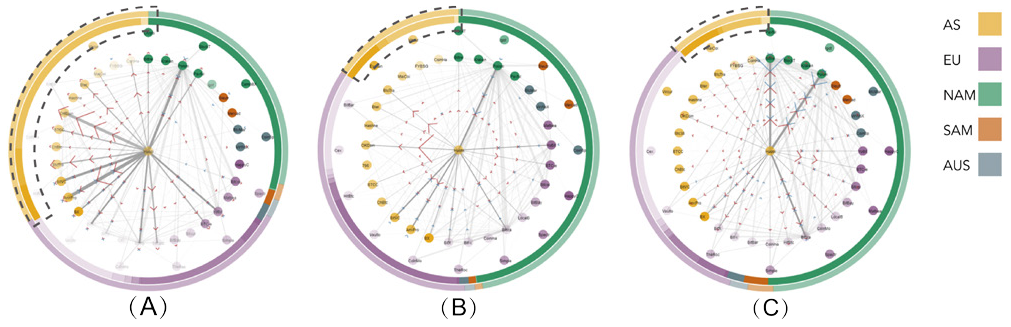}
} 
\caption{\tochange{\textbf{Code type comparisons.} We compare codes between \theycolor{Borkin et al.~\cite{borkin2013makes}'s \textit{types}}
color-coded in \theycolor{cyan}  
and \wecolor{ours} (color-coded in \wecolor{orange}) in (a)-(c) and between \dengcolor{Deng et al.~\cite{deng2020visimages}'s types} (color-coded in \dengcolor{green}) and ours in (d). 
We label image (a)~\cite{afzal2012spatial} \wecolor {text}---because the \et that represent the geospatial data and metadata here are text. 
 Our typology does not include ``map'' because maps are a representation technique that may have very different visual appearances (\eg, consist of only points, areas, or text as seen here).
We label image (b)~\cite{byron2008stacked} as a \wecolor{schematic representation}, elucidating its conceptual illustration;
actual data are not the main focus. 
Image (c) has two labels \wecolor{\barabbr and \pointabbr}.
{We have more complete essential stimuli to signify the importance of seeing both  \wecolor{\barabbr and \pointabbr}.} 
Image (d)~\cite{yue2018bitextract} emphasizes data types (networks), and we use \wecolor{\barabbr \& \nodelinkabbr}, beyond the technique label of \dengcolor{Donut-chart}.
}
}
\label{fig:typecomparisons}
\end{figure*}

\subsection{Visualization categorizations}

\noindent 
Textbooks, in particular, often rely on categorizations to 
structure content \cite{rees:survey}. While some (older) books \cite{Brinton:1939:GP} are collections of 
graphical representations, many of today's textbooks regularly use one of a few structures:

\textbf{Focus on construction rules in design or techniques.}
A seminal approach to characterizing visual designs is Bertin's visual semiotics \cite{bertin1983semiology}. 
He discussed the fundamental building blocks of a visualization that are modified by 
visual elements (channels) that encode data. 
Similarly, researchers have proposed to describe visual designs through the lens of a visual language with a set of syntactic rules.  Examples include Wilkinson's Grammar of Graphics (GoG) \cite{Wilkinson:2005:GrammarOfGraphics}, Engelhardt's Language of Graphics \cite{Engelhardt:2002:LanguageOfGraphics}, and Mackinlay's automatic design \cite{mackinlay1986automating}.
Applying rules formulated in a visual language can yield a broad range of visual
designs \cite{munzner2014visualization} and 
tools and libraries are based on them, \eg, Tableau \cite{mackinlay2007show}, D3 \cite{bostock2011d3}, 
Vega-Lite \cite{satyanarayan2016vega}, and Draco \cite{moritz2018formalizing}. Others, \eg, Tufte's Envisioning Information \cite{Tufte1990Envisioning}, differentiate techniques by higher-level \textit{construction rules}, 
\eg, small multiples, or \textit{principles}. 
\eg,
layering and separation.
Again others use concrete technique names such as Lohse et al.'s \cite{lohse1994classification}
11 categories: graphs, tables, time charts, network charts, diagrams (process and structure diagrams), maps, cartograms, icons, and photorealistic images. 
\drpi{This type of categorization sometimes mixes data-type specific representations (maps, network charts, time charts) and data-type agnostic representation (icons, diagrams, tables).} 
In the case of Lohse et al.~\cite{lohse1994classification},  there are 60 images inspired by Bertin~\cite{bertin1983semiology} and Tufte~\cite{tufte1998visual}, viewers were 16 students and staff at the University of Michigan with no special subject training.  In our case, we selected out input images from the VIS proceedings, and our coders were visualization experts with extensive experience in visualization research. Experts are generally more suitable to define  categories~\cite{tanaka1991object}.

\textbf{Focus on data types.}
Other researchers have categorized prototypical  
{visual designs} 
based on the data type. 
This approach makes sense as, in a typical iterative 
design process, data are systematically mapped, winnowed, and refined to visual encoding \cite{Sedlmair:2012:DSM}.
Ward et al. \cite{Ward:2015:IDV}, \eg, 
classify
visualization techniques for spatial data, geospatial data, time-oriented data, multivariate data, trees, graphs, and networks, and text and document visualizations. Heer et al.'s visualization zoo \cite{Heer:2010:TVZ} classifies time series, statistical data, maps, hierarchies, and networks.  
Brodlie \cite[p. 40]{Brodlie2012scientific_chap3} classifies techniques into those for point, scalar, vector, and tensor data. 
\jctg{Franke and Haehn~\cite{franke2020modern} present implementations on the web.}
Compared to these characterizations, where data are input and visual images are output, we attempt to characterize images, \jctg{where an image is input, and the primary foci or essential stimuli are the output,} without necessarily knowing the characteristics of the data that led to the final image.  
\tochange{For example, we make no distinction between a line chart that shows temporal data and one that shows, \eg, point relationships in parallel coordinates.} 
\jctg{Finally, our categorization only concerns the visual stimuli that  
represent data, and the visual representations taking a salient position in the image; we also excluded the non-essential stimuli that do not directly map to data (\eg, axis and labels). 
}

\textbf{Focus on task and analysis question types.}
Another set of textbooks introduces visualizations by linking representation and analysis tasks/questions. Fisher and Meyer \cite{Fisher:2018:MDV}, for instance, group techniques such as histograms and boxplots under the analysis question of ``showing how data is distributed.'' Maciejewski \cite{Maciejewski:2011:DRT} also takes this approach in his grouping of techniques.
Again, a focus on analysis questions considers a-priori criteria to choose and categorize visualization techniques in the same vein as data and construction rules do. Visually similar techniques are thus considered in separate categories; for example, Fisher and Meyer \cite{Fisher:2018:MDV} categorize bar charts under ``visualizations that show how groups differ'' and histograms under ``visualizations that show how data is distributed.'' 

\textbf{A cross-disciplinary perspective for categorization.}
We 
are certainly not the first to realize the 
differences between \textit{what we design} and \textit{what we see}. 
For example, 
the 
vision science community has realized that composition solutions from low-level features (\eg, Canny edge detector \cite{canny1986computational}, orientation map \cite{malik1997computing}, and HOG algorithms \cite{felzenszwalb2010object}) to composite high-level objects do not align with \textit{how} people see~\cite{wolfe2021guided, potter2004pictorial}.
Recent deep convolutional neural networks (DNNs) can assign categories to items because the categories are treated as a continuous space of related 
concepts
\cite{chen2019looks, malisiewicz2008recognition, ye2022visatlas}.
\tochange{For example, BioCLIP~\cite{stevens2024bioclip} combined a 
large-
language model and images to produce biological concepts. 
However, the authors worked on natural scenes rather than on abstract
images.
\jctg{As we show in this paper, however, the diversity of
representations from images is large: a single category can be broad and contain various spatial arrangements, compositions, and viewpoints.}}

\subsection{The role of graphs in scientific communication} 

\noindent{Even though our approach to categorizing images differs from that of many textbooks, 
our method of studying them to evaluate scientific advances has been used before. 
Latour~\cite{latour2012visualisation} presented graph features that make them essentially a pervasive form of visualization and a specialized vocabulary for transforming and analyzing data to represent scientific findings. The pervasiveness and centrality of scientific figures led Latour to conclude that scientists exhibit a ``graphical obsession'' and suggest that graphs distinguish scientific domains. 
}

Other image datasets~\cite{savva2011revision, battle2017position} are available, but they focus on what designers draw rather than
what viewers should see.

Instead, our work is 
closely related to
Borkin et al.~\cite{borkin2013makes} and Li and Chen~\cite{li2018toward} who studied high-level visualization categories from large image datasets. 
Borkin et al.~\cite{borkin2013makes}, \eg, suggested a taxonomy of techniques. Theirs is a mix of encoding (area, bar, \dots), data (text),
layout (circle), and analysis-focused (distribution) categories. Similarly, Deng et al.~\cite{deng2020visimages} and Lu et al.~\cite{lu2024VAID} added data and technique labels.

In our categorization we instead attempted to separate data, techniques, and \et. 
\tochange{For example,
we considered whether \et are more similar to each other within a category than they are compared to other categories. For example, lines in \autoref{fig:typecomparisons}(a) are a schematic view as they do not encode real data and bars and points are both \et in \autoref{fig:typecomparisons}(b). %
These examples show that our \typename typology provides
an abstract,
and a general way to organize and store visual designs, independent of the underlying techniques and data (as in \autoref{fig:typecomparisons}(b) and (c)).
}

\section{The Image Coding Process}
\label{sec:process}
\begin{figure*}[!thp]
    \centering

    
\includegraphics[width=0.9\textwidth]{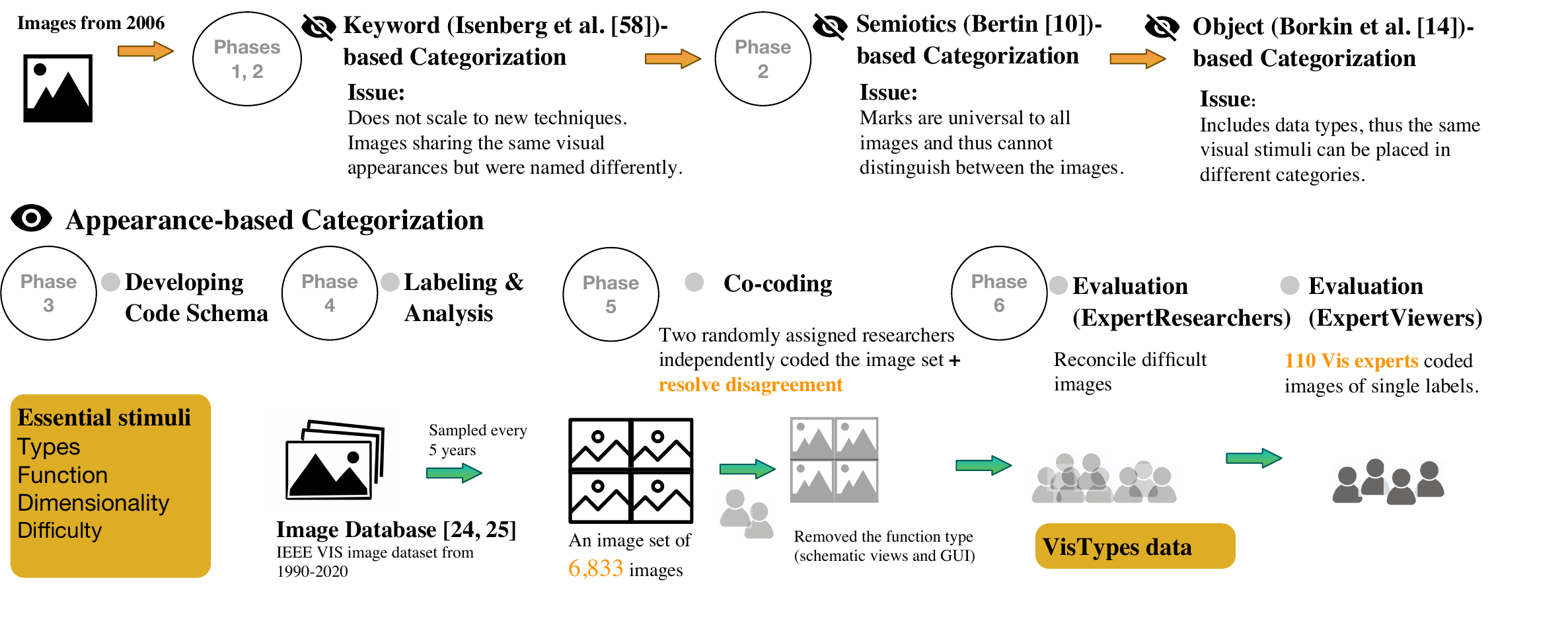}

\caption{\tochange{\textbf{The image coding process.}
We developed the method, identified the type codes, and performed the image labeling and curation in a multi-year team effort. 
\jctg{
\jcfr{This lengthy process includes three failed attempts
(Phases 1 $\&$ 2).
We finally used, as shown in Phases 3--6, a similarity-based appearance categorization to derive the \et. 
\textbf{Observation}: visualization techniques and design elements are distinct from what we see for making conscious judgments of  representations.
    }}}}
    \label{fig:codingprocess}
\end{figure*}

\noindent 
\jctg{We begin with an overview of our process that led to \typename; graphically represented in \autoref{fig:codingprocess}.
We did not originally start out with a focus on essential stimuli. Instead, we applied other categorizations that used specific techniques (\eg, Isenberg et al.~\cite{isenberg2016vispubdata}), design features (\eg, ``marks and channels''~\cite{bertin1983semiology}), or a mix of them (\eg, Borkin et al.~\cite{borkin2013makes}) in Phases 1-2. 
A consequence of these failed attempts was that we rejected the idea that ``marks and channels'' or ``techniques'' could describe what we perceived in these images. Visual designs are often composed of many marks, channels, and techniques combined together. It became very difficult to label each element and describe which ones formed the visual focus of each image. 
In Phases 3--6 we then developed our focus on essential stimuli to support categorize what we see.
} \jctg{In this section we describe many difficult decisions that were part of this development process. Readers who are not interested in the decision process can skip this section and continue with \autoref{sec:visTypesResults} (\textit{Results}) directly.}

\subsection{Goals of our image coding}
\label{sec:goals}

\noindent These primary goals guided our coding process in Phases 3--6:

\textbf{Provide a small set of broad categories.}
We purposefully wanted to create only a few categories that would remain manageable despite the detailed and often complex images produced in the community. 
The type abstraction should enable analyses between and within categories to support researchers while 
analyzing the diversity of representations
and inferring the prevalence of rendering methods, 
algorithms, or dimensions.

\textbf{Focus on visual \jctg{similarity}}. 
As described earlier, our typology focuses on describing visual similarity via \et. 
\jctg{
Our types will appear visually similar regardless of the usage of data or tasks.
Low-level marks and channels are not always enough although some of our types share similarities with some marks and channels. Yet, the layout and their morphology are irrelevant for our types.}

\textbf{Collect experiences on the difficulties of categorizing visualization images.} We documented our multi-stage process to derive a relatively high-level categorization of images and describe inherent uncertainty, failed attempts, and current limitations. We also recorded how difficult it was to understand images taken out of the context of the text and captions since many examples did not necessarily contain a descriptive title and clear attribute annotations. 


\subsection{Visualization image data source}

\noindent
{\textit{Image Sampling.}}
We developed our typology 
using
the VIS30K \cite{chen2021vis30k} collection of images and its associated VisPubData \cite{isenberg2016vispubdata} meta data.
This dataset largely 
represents visualization as a field because it contains every visualization image published at 
IEEE VIS 
(including Visual Analytics, Information Visualization, and Scientific Visualization) since 1990. 
\tochange{Our domain's research progress and the scientific solutions reveal themselves visually in the images in these papers and thus provide a focused perspective without managing the diverse publishing goals of journals and conferences.} 
Also, based on our code development process (described below), it became clear that we would not be able to 
code all 30,000 images.
\tochange{We thus prioritized covering images across time to understand historic changes.
We 
labeled images in five-year intervals for our full coding process, starting with 1990 and up until 2020 (inclusive), 
yielding \imagesfigSeven
images 
from \papersSeven IEEE VIS 
full papers (incl.\ case studies).
}

\tochange{\textit{Removal of GUI and Schematic Representations.}}
\tochange{We also suppressed two 
\jctg{\ipc}
categories that we 
initially used: \textit{\visgui} and \textit{\visschematic}. GUIs are screenshots of visualization tools 
showing multiple categories and schematics 
do not encode any data.  This choice allowed us to focus on 
data representations instead of compositions.}

\tochange{\textit{Exclusion of Symbolic Representations.} Algorithms, mathematical equations, and tables are symbolic representations often used to structure and interpret unstructured text data. These symbolic representations do not visually represent the texts, and we have thus excluded them. 
Our \textit{text-based representation} in our categorization~(\autoref{tab:12schema}) is only applicable when the text itself encodes data. \jcfr{Thus, it excludes cases where text is used solely for labeling and annotation (\eg, Stokes et al.~\cite{stokes2022striking}) or cases in which text serves as the underlying data source but its properties are not encoded as text (\eg, word or document embedding coded as \vispoint~\cite{ji2019visual} or \visarea~\cite{dai2016unlocking}).}
}

\subsection{Image classification process}
\label{sec:classificationProcess}

\noindent{\jctg{Included in our process description are 
failed attempts, key decisions, and iterative coding refinement (\autoref{fig:codingprocess}) that precede the final codes of primary foci or essential stimuli (\autoref{tab:12schema}).}
}

\tochange{%
\textbf{Phase 1---Initial image classification based on keywords}
(from circa Aug.\ 2020, approx.\ 2 months):
We began our work focusing on \emph{visualization techniques}, where we considered that technique names (such as treemaps, parallel coordinates, etc.) could describe the content of the images we analyzed well. To improve objectivity and reduce bias, we 
tagged images with the most common technique names 
extracted from author keywords used for VIS papers. We ranked the author keywords extracted in prior work \cite{isenberg2016visualization} to derive
the initial top-21 keywords for specific techniques
\jcfr{(\appref{sm:keywords})}.
In addition to the encoding techniques used in each image, we added two code categories that 
described
additional image characteristics: the rendering dimensionality (\ie, 2D or 3D) and the functional purposes of creating the image
(\ie, the reason why the authors created each image, for example, illustration of a visualization technique, experiment results, or screenshot of GUI. (\autoref{tab:funcDim} in \appref{sm:web-interfaces}). 
\drpi{The initial label set thus included 25 codes}.}

\tochange{%
\textbf{Phase 2---Initial coding}
(from circa Oct.\ 2020, approx.\ 1 month): 
To test our initial code set, each coder categorized visualization images from 2006.
We used these images only in this phase to refine our code set. 
We subsequently introduced new technique codes by merging techniques that had been tagged ``other,'' 
and added ``schematic diagram'' to the list of image purposes, resulting in 22 technique codes.  
We discussed the definition of these terms and gave all coders (we call ourselves ``\expertUs'') written instructions and example images from each category
(\autoref{fig:web-interface-phase2} in \appref{sm:web-interfaces}).
One coder initially labeled each image in this stage, and a second coder validated the image. We based the validation assignment of the second coder on their respective expertise to verify all images included or excluded in a specific category 
(\eg, volume depictions were coded and verified by someone with a background in volume graphics). 
We removed false positives, avoided false negatives, and ensured image categorization consistency with these steps.}

\tochange{%
\textbf{Phase 3---Consolidation, seeing by association and analogies}
(from circa Nov.\ 2020, approx.\ 1 month): 
We then discussed what worked well, what did not, and why. 
\drpi{The codes that focused on visualization techniques quickly became difficult to apply as the number of techniques grew increasingly large. 
We had difficulties defining when a technique should receive its own label or be covered under ``other.'' Also, some seemingly similar images where coded under different techniques. 
Point clouds and 3D scatterplots, \eg, both render points according to underlying coordinates in 3D space, with the main difference that scatterplots typically include reference structures such as axes and gridlines.}}
These conflicts led us to re-frame our code set using higher-level, more \textit{general} visualization type codes.

We
decided to focus on describing the \textit{main focus of a given image} or \textit{\et} 
and to create codes that enable the viewer to distinguish graphical 
similarities and differences. 
We thus re-grouped and merged the codes, which shared similar visual characteristics, into a more general code category.
For example, we put \emph{isosurface} into a more general \emph{surface-based techniques} category and grouped \emph{point clouds} and \emph{scatterplots} into a more abstract \emph{point-based techniques} category
(in \appref{sm:consolidation} we list more decisions). 

This consolidation process resulted in 10 high-level visualization-type codes that 
became part of our final set shown in \autoref{tab:12schema}. \drpi{We reduced our 
\ipc
category to just two codes: GUI (Screenshots) and Schematics, as these were visually identifiable without requiring knowledge of the underlying data.} Both categories represent the codes from ``other representations'' from Phase~1, which did not appear on our technique-focused labels. 
Given the challenges of labeling the visualization images we encountered, we also
decided to collect subjective ratings of difficulty (easy, neutral, and hard).
see \autoref{fig:web-interface-phase3} in \appref{sm:web-interfaces}.

\tochange{%
\textbf{Phase 4---Code calibration}
(from circa Dec.\ 2020, approx.\ 3 months): 
In the first two months after having arrived at the new labels, we discussed, debated, and coded two sets of 50 (\ie, 100 in total) randomly chosen images from our seven-year target image data set to calibrate our collective understanding.
During this exercise, we clarified code definitions and discussed ambiguities. 
We assigned these 100 images to each expert researcher for quality control, aligning our decisions, and discussing potential pitfalls with our new code set.
We 
implemented a dedicated web-based labeling tool 
(\autoref{fig:web-interface-phase4-5} in \appref{sm:web-interfaces}) 
for the coding in this phase. 
At the end of this phase, our project was 
one year old. We had almost weekly meetings and discussions throughout this time and
reached a consensus on the 
typology.}

\tochange{%
\textbf{Phase 5---Result coding and validation}
(from circa Apr.\ 2021, approx.\ 6 months): 
In this phase, we coded all \imagesfigSeven~images in our chosen dataset based on the refined definitions and characterizations, using largely the same coding tool as before
(\appref{sm:web-interfaces}). 
\drpi{We used the 10 visualization type labels, two 
\ipc
labels (GUI, Schematics), two labels for the dimensionality (2D vs.\ 3D), and three difficulty labels (easy, neutral, hard) that capture how difficult the categorization was for the coder.
We first looked at the function of an image. We assigned the respective label if the image showed either a schematic or a GUI. If not, the 
\ipc
was implicitly considered a ``visualization example,'' and we proceeded to assign a visualization type label. 
Both categories shared an ``I cannot tell'' code assigned when neither explicit 
\jctg{purpose}
nor a visualization type could be assigned. In addition, coders could freely add new labels for visualization or image types when they found something new that was not covered by existing categories. For dimensionality, we also added a code called ``I cannot tell,'' which could be checked when coders were unsure whether the visualization was a 2D or 3D rendering. Coders could assign multiple types and dimensionalities to one image because many images show more than one visual design.}}

\tochange{%
In the coding process, we also recoded the 100 images we had previously labeled during Phase 4. We randomly assigned two coders to each image to ensure high-quality results and capture potential difficulties in applying the labels. This phase was 
laborious due to the number of images we coded. \drpi{We met regularly to resolve further questions that arose during coding and to discuss or clarify code descriptions further.}}

\tochange{%
\textbf{Phase 6---Verification}
(from circa Oct.\ 2021, approx.\ 6 months): 
In this final phase, the two coders assigned to each image worked to reach an agreement when their labels did not match. 
For this purpose, we developed 
more web-based visual interfaces 
(see \autoref{fig:web-interface-consistency} in \appref{sm:web-interfaces})
that focused on conflict resolution and on providing an overview of applied labels.
We then filtered the results so that only the inconsistently coded images were shown, and we used this process to resolve all disagreements.
This verification was also a lengthy and laborious process that required constant discussion. 
We analyzed difficult and ambiguous cases as a team every week until we could agree on a solution (we describe some of the most difficult decisions in more detail in \autoref{sec:ambiguous}).  
As part of these discussions, we added 
our last visualization type
\textit{\viscolor}, listed as point (8) in \autoref{tab:12schema}. 
We also further refined our definitions. 
Consequently, we went through all previously labeled images again to check if they had to be re-coded for consistency and resolved potential resulting disagreements as part of our discussion process.}

\section{Results}
\label{sec:visTypesResults}

\begin{figure}[!bt]
    \centering
    \includegraphics[clip, trim={0 0 12 20}, height=0.48\columnwidth]{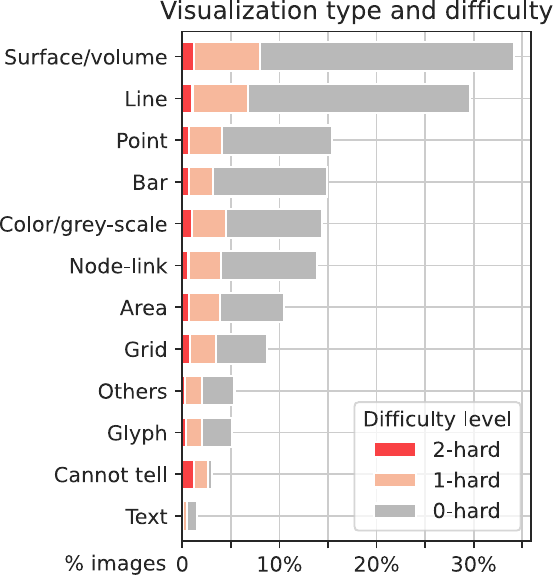}
    \hfill
    \includegraphics[clip, trim={0 0 12 20}, height=0.48\columnwidth]{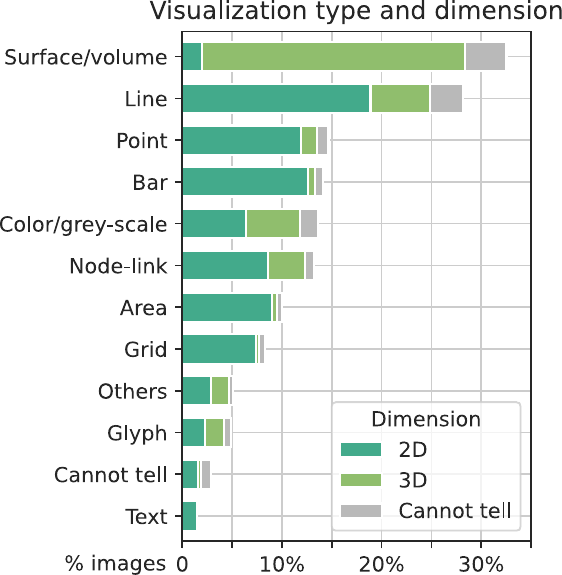}
    \caption{The proportion of applied image codes for each category, relative to the total \imagestype
    visualization images (after excluding the pure schematic and GUI images).
    We can see that the most common visualization types were \vissurface and ``line-based representations.''}
    \label{fig:typeDistribution}
\end{figure}

\noindent%
Our coding process resulted in \totaltypelabels image type labels, 
{\totalfunctiontypelabels function labels (GUI and Schematics),
} \totaldimlabels 2D/3D dimensionality labels, and \totalhardnesslabels difficulty labels for the \imagesfigSeven images. 
\autoref{fig:teaser} and 
\autoref{fig:typeDistribution} 
show the type distribution. We now describe each type of \et in more detail and show canonical or typical examples for each.
\autoref{fig:consistency}, \autoref{fig:surfaceline3D}, and \autoref{fig:coocc} in \appref{sm:resultsExpertUs} show additional results.

\newlength{\figverticaloffset}
\newlength{\figureheight}
\newlength{\figheightminusverticaloffset}

\subsection{Visualization types}
\newlength{\spaceaftercanonicalexamples}
\setlength{\spaceaftercanonicalexamples}{2ex}

\noindent\textbf{Surface-based Representations and Volumes} 
\label{sec:vc.surfacevolume}

\emph{represent the inner and/or outer features and/or boundaries of a continuous spatial phenomenon or object in 3D physical space (\eg, isosurfaces~\cite{gregorski:adaptive}), 4D space-time, or slices thereof.} Frequent characteristics 
in this category include surface or volume-based rendering 
with lighting, shading, camera perspectives, shadows, and/or transparency; or 2D medical imaging slices. Common applications include medical images, 
terrain surfaces, and images of physical simulations (\eg, ~\cite{edmunds:surface}).

Volume rendering has been one of the most important areas of visualization in the early years
\cite{isenberg2016visualization}. As such, it is perhaps not surprising that surface and volume representations were the most common techniques in our images. 
\jcfr{They are also the easiest to identify. This category has the highest coder consistency, at \surfacecon, among all types.}
It is also the only representation technique with primarily 3D renderings. The few 2D renderings we found included volume slices (\eg, MRIs or CTs) or X-rays. Similar to what was shown in prior work on keywords \cite{isenberg2016visualization}. \jcfr{There has been a significant decrease in the proportion of surface and volume images, dropping from approximately 47\% in 2005 to around 20\% recently. 
}
Canonical examples, 
from \cite{hummel2010iris}, \cite{tikhonova2010visualization},
\cite{nguyen2021modeling}, are:

\setlength{\pictureheight}{2.1cm}
\noindent\begin{minipage}{\columnwidth}\vspace{1ex}%
\centering
\includegraphics[clip, trim={0 40 560 0},height=\pictureheight]{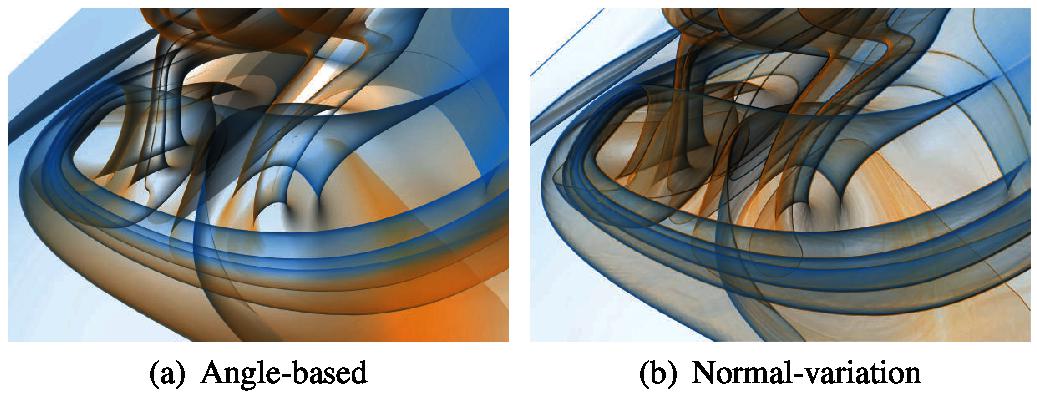}\hspace{7mm}%
\includegraphics[clip, trim={400 240 1200 0},height=\pictureheight]{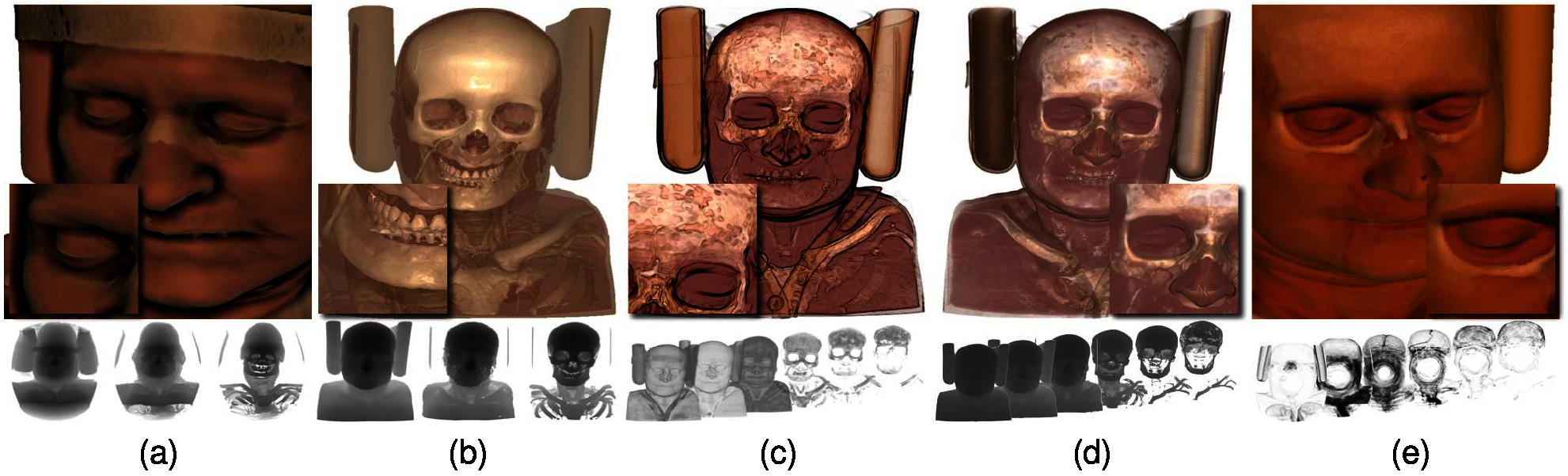}\hspace{5mm}%
\includegraphics[clip, trim={600 0 0 0}, height=\pictureheight]{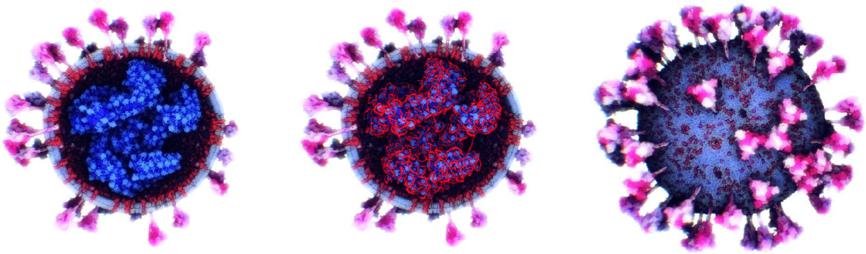}%
\vspace{\spaceaftercanonicalexamples}\end{minipage}

\textbf{Line-based Representations} 
\textls[-1]{\emph{are graphs where information is emphasized through straight or curved lines.} Canonical vi\-su\-a\-li\-za\-tion techniques are line charts, parallel coordinates, ra\-dar\discretionary{/}{}{/}spi\-der charts, con\-tour lines, streamlines, or tensor field lines.}

Lines, edges, and curves are the second most common representations indicated by our image labels.
We did not code lines that simply delineate screen-space areas as edges. 
About a sixth (\lineprop) of line-based representations are rendered in 3D. \jcfr{
For a sizable proportion of $12\%$ of the line chart images, the coders also could not tell whether a line chart is 3D or 2D due to a lack of clear depth cues.} 
\jcfr{Coder consistency ranks among the top three, on par with \vissurface and ``generalized bar representations,'' at a consistency level of \linecon.}
Most line charts are the typical 2D line charts that most often represent temporal data. 
Canonical line-based representations, taken from \cite{athawale2020direct}, \cite{stoll2005visualization},
\cite{verma2000flow}, are:

\setlength{\pictureheight}{2.1cm}
\noindent\begin{minipage}{\columnwidth}\vspace{1ex}%
\centering
\includegraphics[clip, trim={0 50 500 0}, height=\pictureheight]{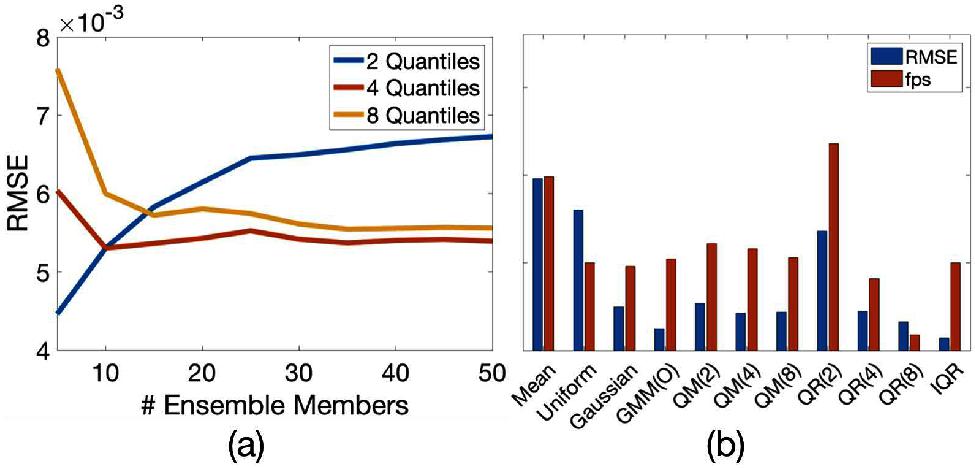}\hspace{7mm}%
\includegraphics[clip, trim={0 0 340 0}, height=\pictureheight]{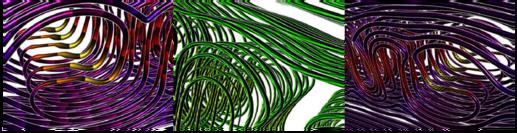}\hspace{7mm}%
\includegraphics[clip, trim={0 0 460 0}, height=\pictureheight]{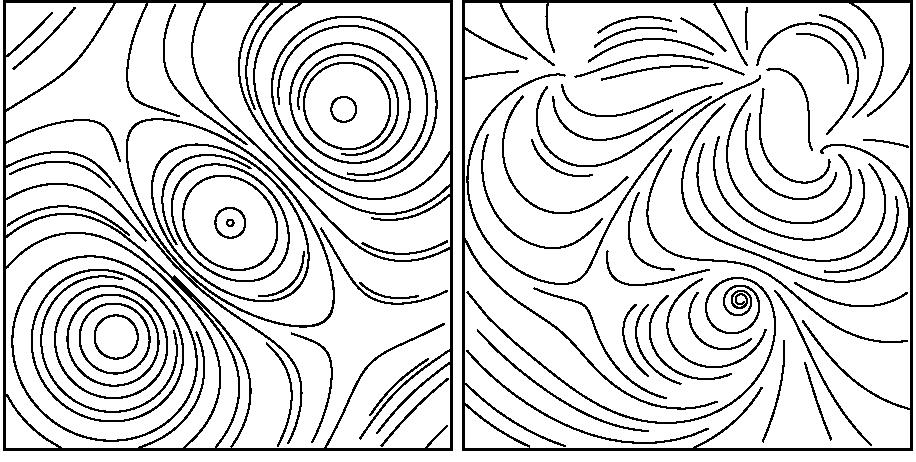}%
\vspace{\spaceaftercanonicalexamples}\end{minipage}

\textbf{Point-based Representations}  \emph{encode point locations. These locations are often shown using dots or circles, but also other shapes such as 3D spheres, triangles, stars, etc.}

\textls[-1]{Similar to Bertin~\cite{bertin1983semiology}, we considered point-based representation to encode point locations in a 2D or 3D space. Point marks could be small circles but also 3D spheres and sometimes other shapes like triangles, stars, etc. Canonical examples of visualization techniques of this type are scatterplots, (volumetric) point clouds, or dot plots. Point-based representations were the third most common visualization type according to our coding. 
We found that identifying them was slightly more difficult than \visbar we discuss next, and the overall consistency of the coders was \pointcon. Surprisingly, we saw only a small percentage of 3D point-based representations, perhaps due to a large amount of work on scatterplots or using scatterplot-like representations of, \eg, dimensionality reduction or clustering results. 
Canonical examples, taken from \cite{Pu:2000:AlgorithmVis}, \cite{xia2020smap}, \cite{ushizima2012augmented}, are:}
%
\setlength{\pictureheight}{2.1cm}
\noindent\begin{minipage}{\columnwidth}\vspace{1ex}%
\centering
\includegraphics[height=\pictureheight]{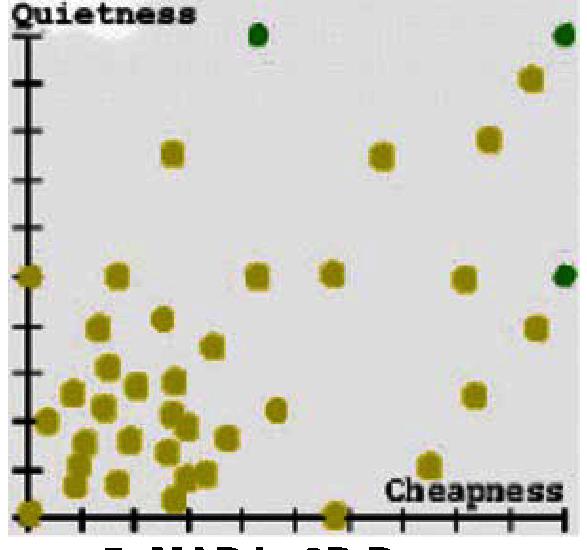}
\hspace{7mm}%
\includegraphics[clip, trim={280 45 30 550}, height=\pictureheight]{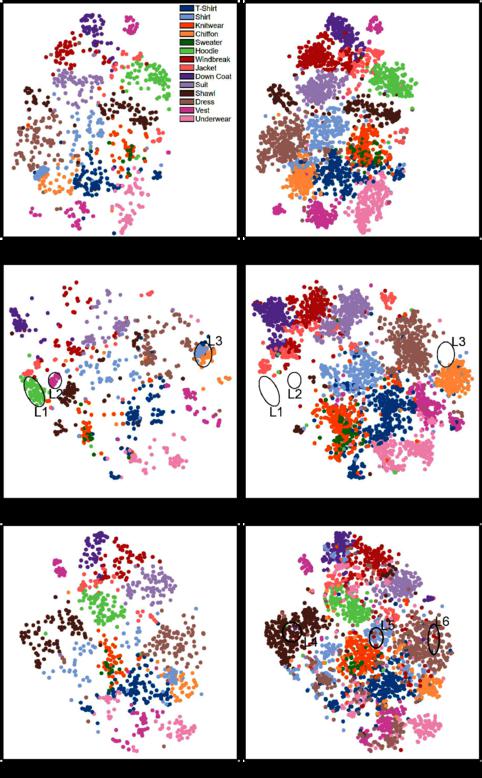}
\hspace{7mm}%
\includegraphics[clip, trim={380 15 0 290}, height=\pictureheight]{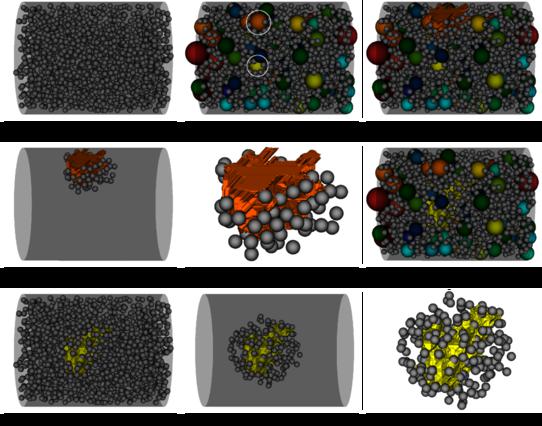}%
\vspace{\spaceaftercanonicalexamples}\end{minipage}

\textbf{Generalized Bar Representations} 
%
\emph{are images that represent data with straight visual encodings (bars) that can be arranged on a straight or curved baseline and whose heights or lengths are proportional to the values that they represent. Bars can be stacked, have error bars, or appear as histograms.}

Generalized bar charts are the third most common visualization type and have increased in number in recent years. We find that they are among the easiest to identify, and the consistency between coders is at \barcon.
3D generalized bar charts are extremely rare, we found them either in the early years or, more recently, to depict data on 3D surfaces such as on a globe. \tochange{In contrast to pie charts, we include doughnut charts here as they use the segment length to encode data~\cite{skau2016arcs}.}  Canonical examples, taken from \cite{Weiss:2021:Revisited}, \cite{Ye:2021:ShuttleSpace}, \cite{Kim:2021:Githru}, are: 

\setlength{\pictureheight}{1.5cm}
\noindent\begin{minipage}{\columnwidth}\vspace{1ex}%
\centering
\includegraphics[height=\pictureheight]{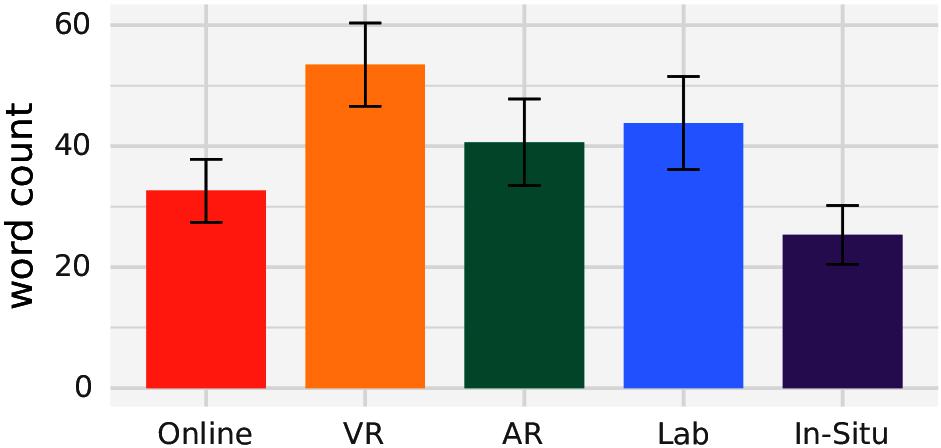}\hfill%
\includegraphics[height=\pictureheight]{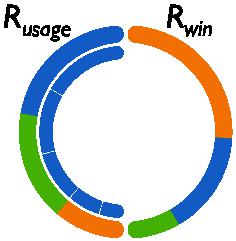}\hfill%
\includegraphics[height=\pictureheight]{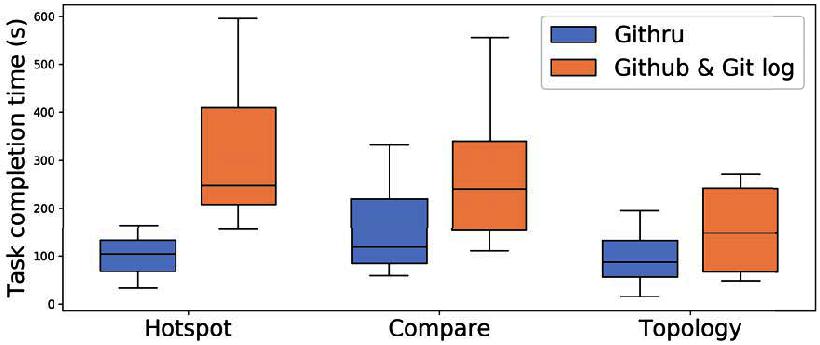}%
\vspace{\spaceaftercanonicalexamples}\end{minipage}

\textbf{Continuous Color and Grey-scale, \jctg{and Texture} Representations}
\label{sec:vc.pattern}
\textls[-5]{\emph{show structured color patterns across an image or atop a geometric 3D object. These patterns \textls[-10]{can be evoked by changes in intensity, hue, brightness, and\discretionary{/}{}{/}or saturation}. The changes are typically smooth (continuous) but can also show sharp transitions.} Typical examples include directional patterns (\eg, flowfields), continuous heatmaps, intensity fields, or, in extreme cases, even a binary image. 
\jcfr{Coders' consistency was among the lowest across all categories, marked at \patterncon.} 
Canonical examples, taken from
\cite{garcke2000continuous,van2003image,feng2020topology}, are:}
\setlength{\pictureheight}{2.1cm}
\noindent\begin{minipage}
{\columnwidth}\vspace{1ex}%
\centering
\includegraphics[height=\pictureheight]{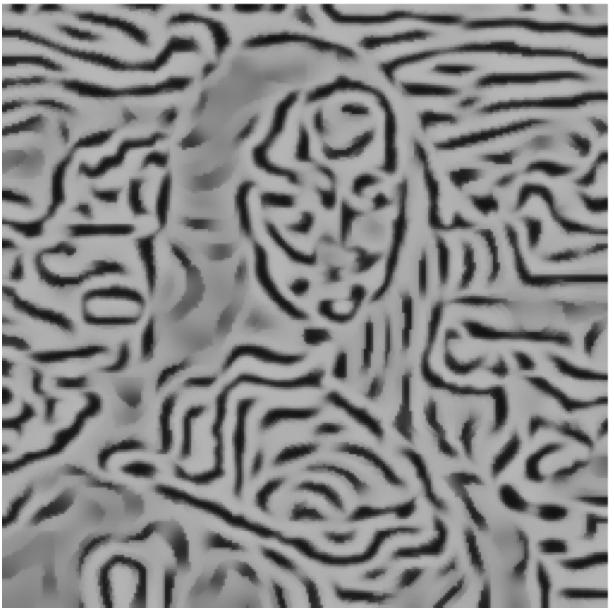}\hspace{7mm}%
\includegraphics[clip, trim={0 0 500 0}, height=\pictureheight]{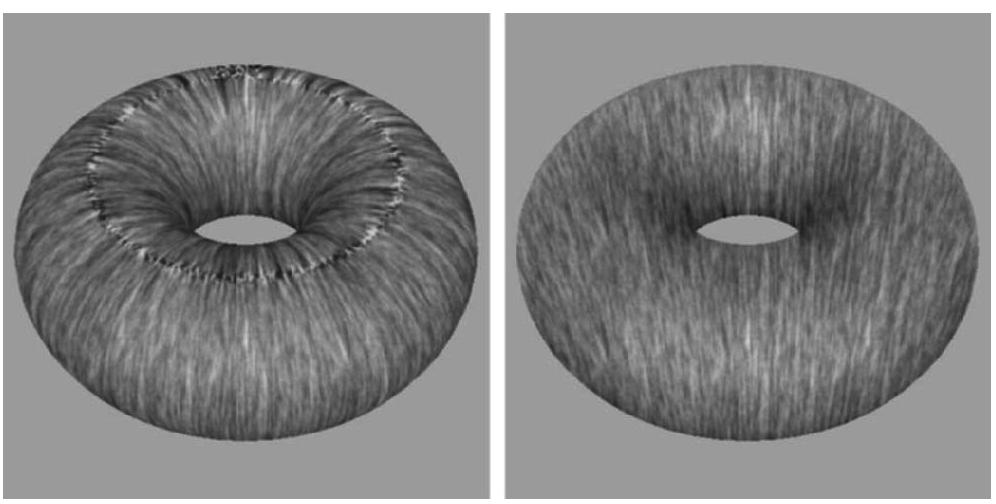}\hspace{6mm}%
\includegraphics[clip, trim={50 20 780 100},height=\pictureheight]{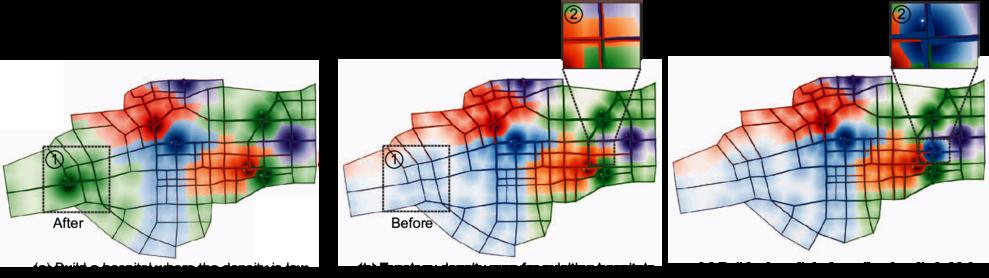}
\vspace{\spaceaftercanonicalexamples}\end{minipage}

\textbf{Node-link Trees/Graphs, Networks, Meshes} 
\label{sec:vc.nodelink}
\textls[-5]{\emph{depict points and explicit connections between these points to convey relationships between data items. Node-link relationships can be found in trees, graphs, networks, and meshes.} 
\jctg{Both nodes (in a network graph) or vertices (in a wireframe mesh) represent topological points in the structure: nodes represent abstract entities in network graphs, while in geometry, they correspond to vertices),  and connections represent the relationships between these nodes. Thus, they are visually similar for the creation of a network-based structure.} Examples include node-link networks or trees, topological graphs, or wireframe meshes.}

Node positions can be given, \eg, as geospatial locations or be derived from the data (\eg, projections). Connections can be continuous (\eg, in a Reeb graph, as the topological structure is given by showing continuous functions in space) or discrete (\eg, edges in a tree). Representations of this type were the 6\textsuperscript{th} most common representation type and their representation has stayed relatively stable at $\approx$\,10--20\% of images per year. 
Most images were 2D in nature, but \drpi{$27\%$} were 3D images. 
Canonical representations, taken from \cite{phan2005flow}, \cite{yoghourdjian2020scalability}, \cite{bhaniramka2000isosurfacing}, are:

\setlength{\pictureheight}{2.1cm}
\noindent\begin{minipage}{\columnwidth}\vspace{1ex}%
\centering
\includegraphics[height=\pictureheight]{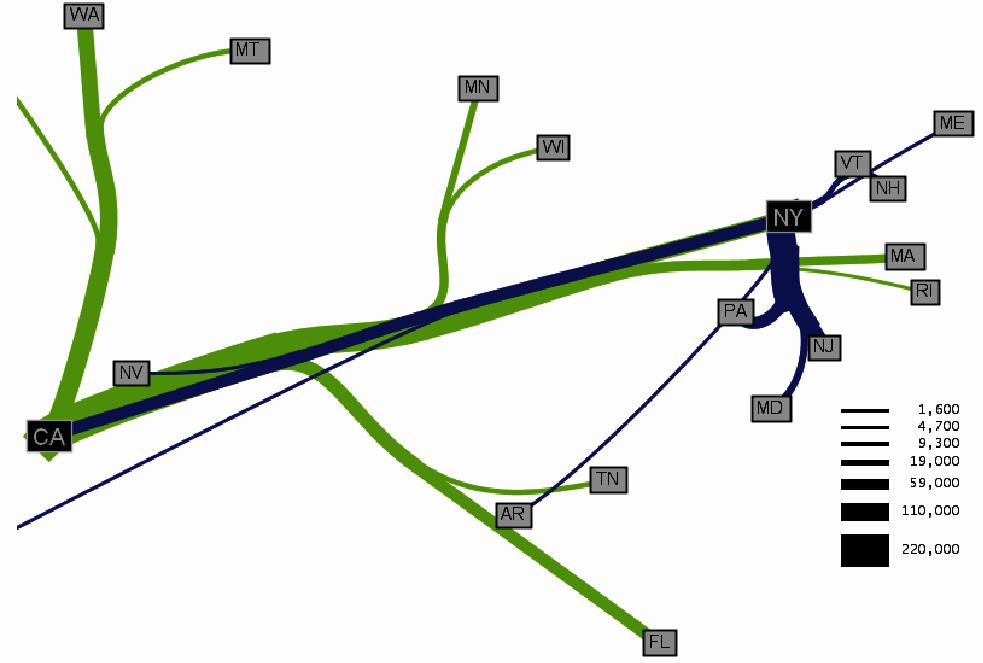}\hspace{7mm}%
\includegraphics[clip, trim={900 1306 300 650}, height=\pictureheight]{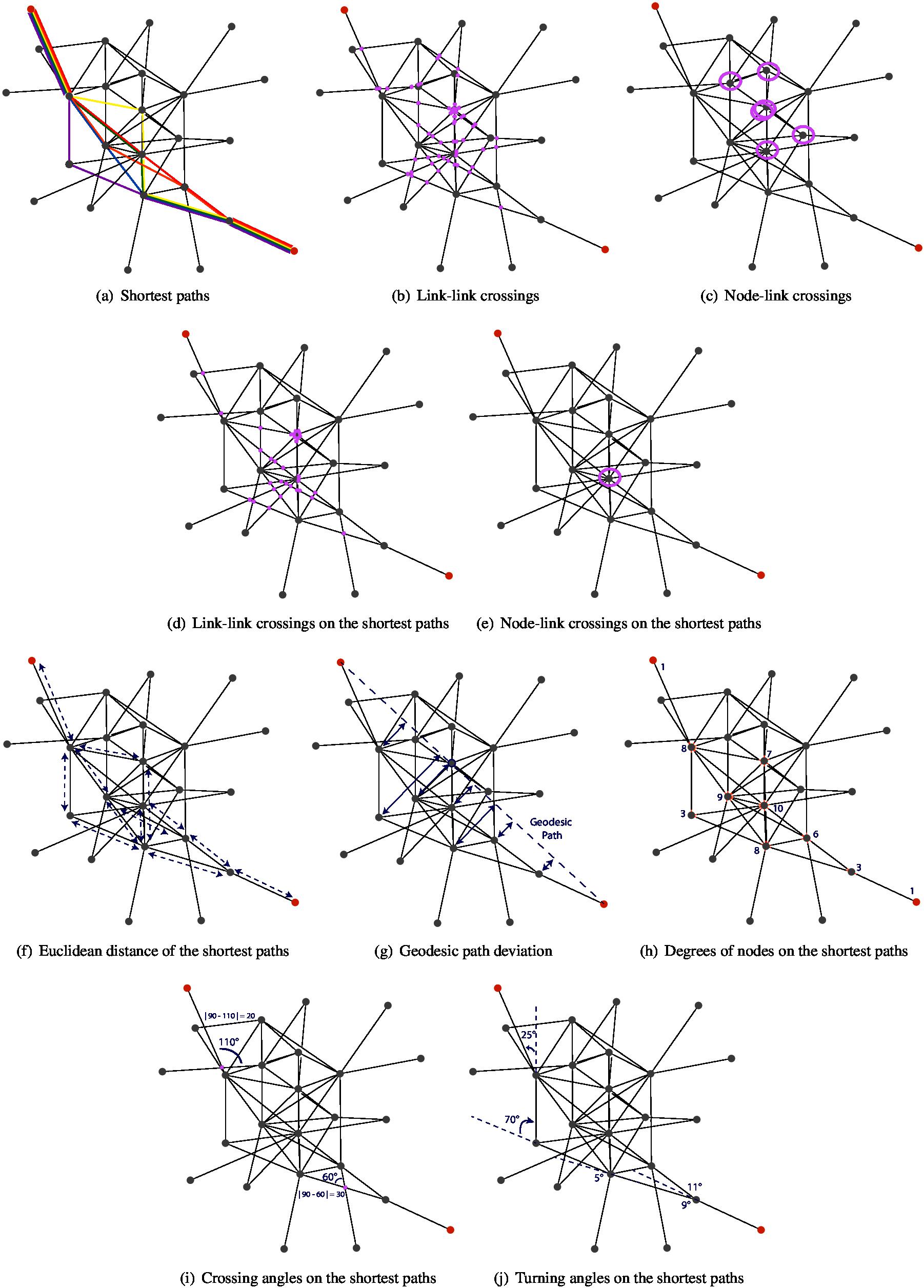}\hspace{7mm}
\includegraphics[clip, trim={500 30 0 0}, height=\pictureheight]{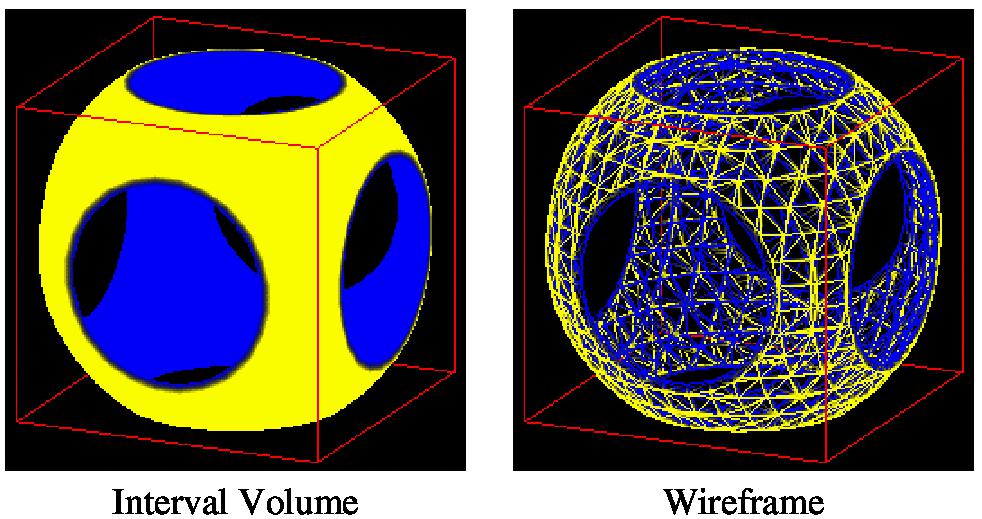}%
\vspace{\spaceaftercanonicalexamples}\end{minipage}

\textbf{Generalized Area Representations} 
\label{sec:vc.area}
\textls[-1]{\emph{are representations with a focus on areas of 2D space or 2D surfaces, including sub-sets of these surfaces. Areas can be geographical regions or polygons whose size or shape represents data. Areas often feature explicit boundaries and, within, are filled with categorical colors or use contrast in luminance and shading to encode attributes of the areas.} Common examples of generalized area charts are pie charts, streamgraphs, (stacked) area charts, treemaps, cartograms, choropleth maps, or violin plots.}


Generalized area charts were the 7\textsuperscript{th} most common type of representation type in the images we coded. Most areas were part of surfaces rendered in 2D. Over the years, the proportion of area charts increased by 1--2\% every 5 years to just under $10\%$ by 2005 and reached about $15\%$ in 2020. 
Canonical examples, taken from \cite{bu2020sinestream},
\cite{shneiderman2001ordered},
\cite{rheingans1995interactive}, include:

\setlength{\pictureheight}{1.5cm}
\noindent\begin{minipage}{\columnwidth}\vspace{1ex}%
\includegraphics[clip, trim={60 700 1120 70}, height=\pictureheight]{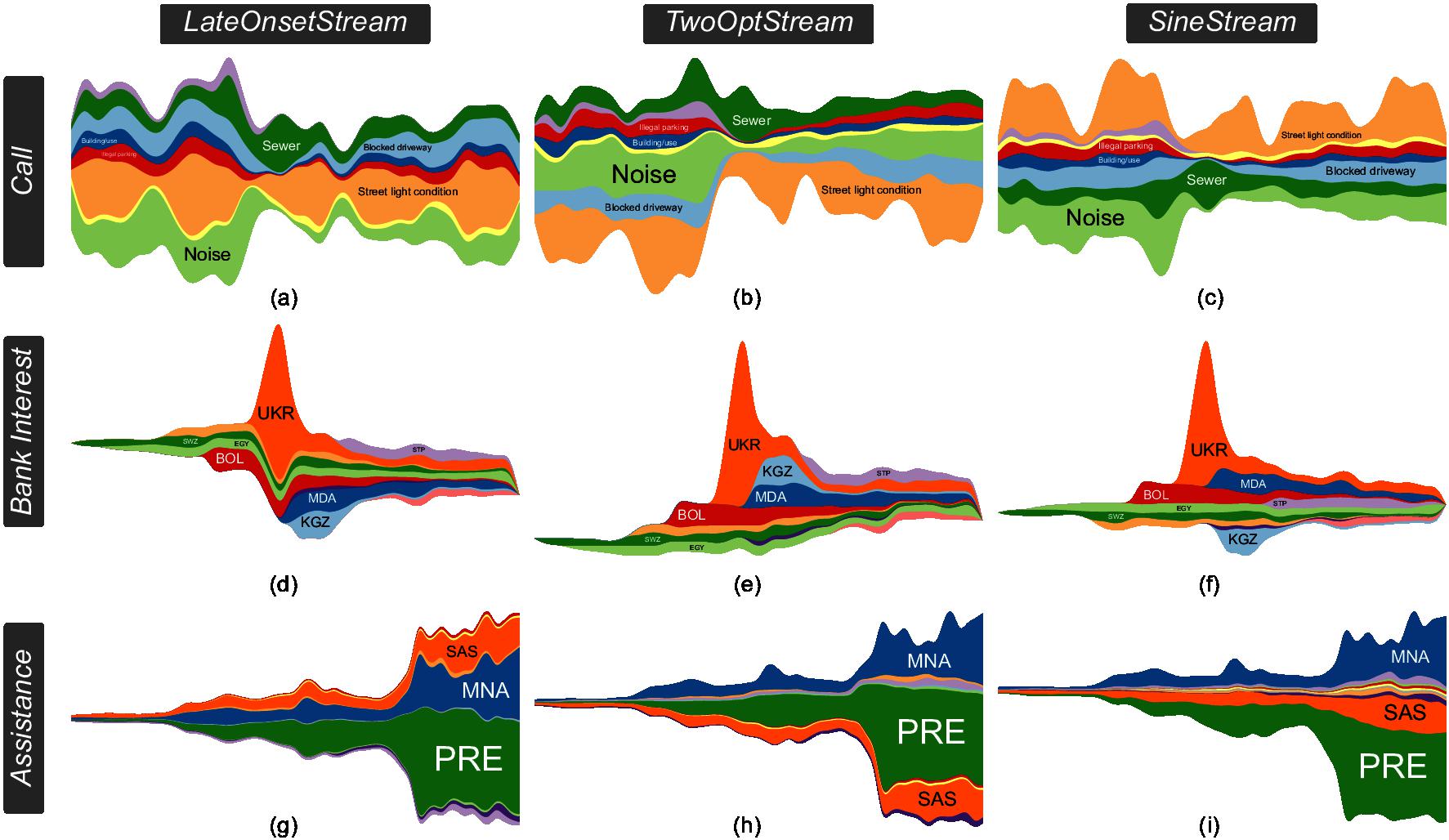}\hfill%
\includegraphics[height=\pictureheight]{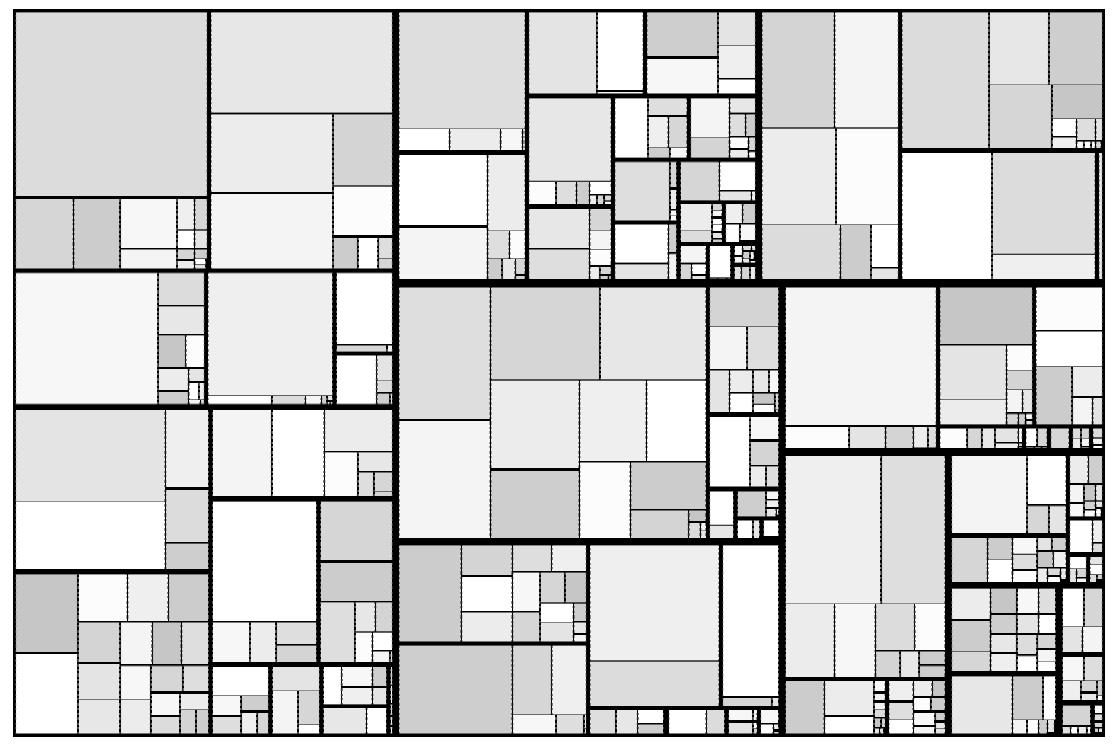}\hfill%
\includegraphics[height=\pictureheight]{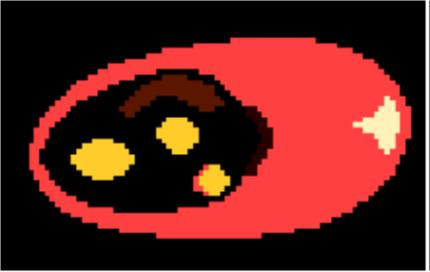}%
\vspace{\spaceaftercanonicalexamples}\end{minipage}

\textbf{Generalized Matrix or Grid} 
\label{sec:vc.grid}
\emph{separate data into a discrete spatial grid structure. The grid often has rectangular cells but may also use shapes like hexagons or cubes. Elements such as glyphs or a color encoding can appear in the grid cells.} Common visualization techniques in this representation type are discrete heatmaps, scarf/strip-plots, space-time cubes, or matrix-based network visualizations.

The grid can vary in resolution.
This type includes figures whose underlying grid is part of the data structure but does not include figures where the underlying grid is merely used as a convenient arrangement of sub-sets of the data (as in small multiples and scatterplot matrices). 
\tochange{It also does not include visualizations such as a treemap~\cite{shneiderman2001ordered}, a representation of hierarchical data using nested rectangles. While the treemap has rectangular cells, each cell \textit{area} corresponds to a data value, and the visualization belongs to the \textit{area} category.} 

For all generalized matrix or grid visualizations, other elements such as color or glyphs can appear at their discrete grid positions (\eg, grid-based vector field visualization). \jcfr{Under 10\% of all images contained generalized matrices/grids, consistently over the years by 2015 and then increased to $15\%$ in 2020.} Of these, \drpi{89\%} were 2D images and \drpi{4.2\%} were 3D (the remaining are unclear). 
Canonical examples, taken from
\cite{ingram2010dimstiller},
\cite{Nielson:1991:TAD},
\cite{garcke2000continuous},
\cite{meulemans2020simple}, include:

\noindent\begin{minipage}{\columnwidth}\vspace{1ex}%
\centering
\includegraphics[clip, trim={500 0 30 30}, height=0.2\columnwidth]{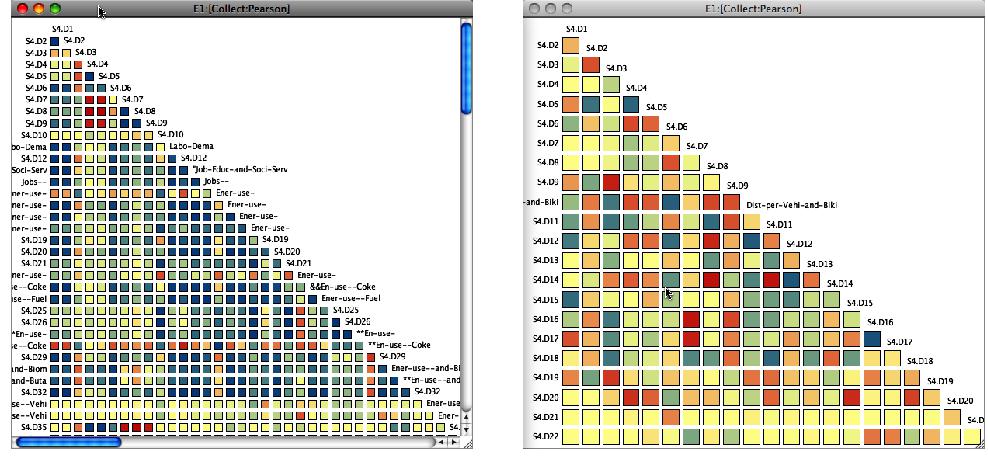}\hfill%
\includegraphics[clip, trim={230 160 240 80},height=0.18\columnwidth]{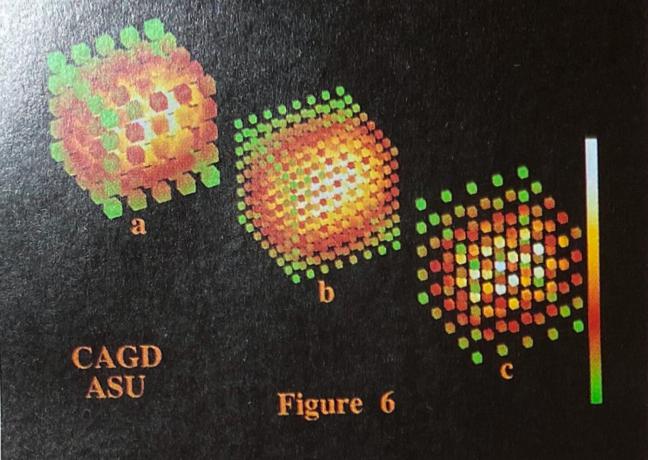}\hfill%
\includegraphics[width=0.2\columnwidth]{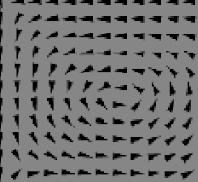}\hspace{1mm}%
\includegraphics[clip, trim={500 150 1000 150}, height=0.2\columnwidth]{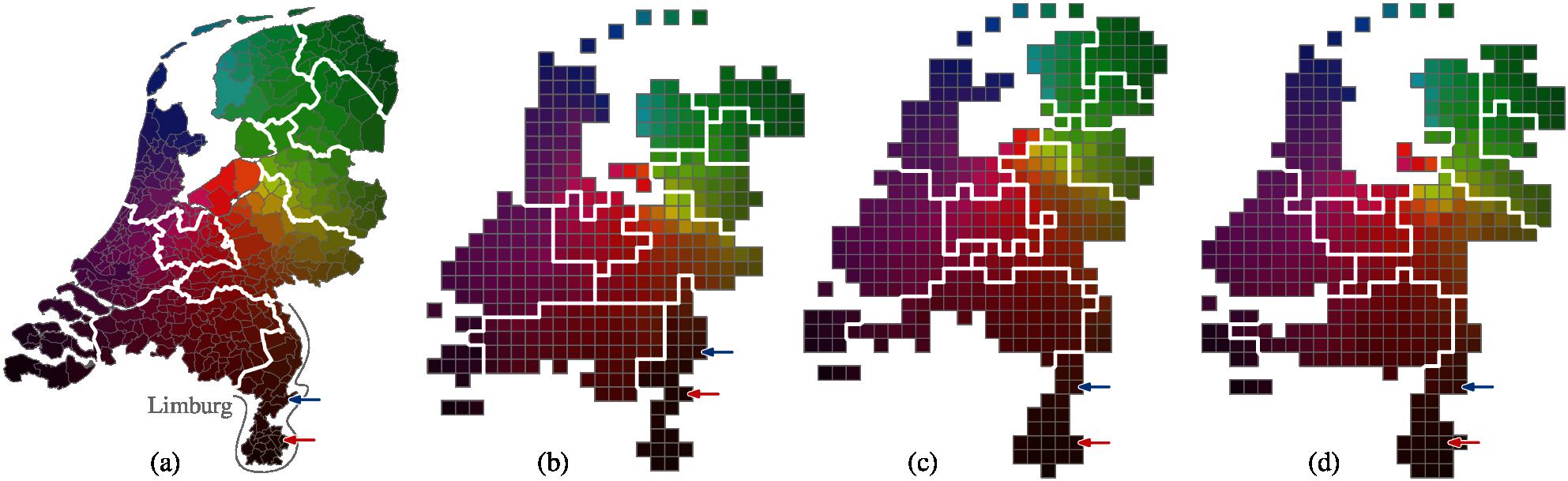}%
\vspace{\spaceaftercanonicalexamples}\end{minipage}

\textbf{Glyph-based Representations} 
\label{sec:vc.glyph}
\emph{commonly include multiple small independent visual representations arranged in space. These representations depict multiple attributes (dimensions) of a data record and often use multiple geometric primitives to encode data.} 
For example, glyphs may consist of a small 3D cuboid where height, width, and depth encode different data dimensions. Other common examples of glyphs are star glyphs, Chernoff faces, whisker glyphs, and so on.

Glyph-based encodings were not particularly frequent ($\approx$5\% of all images contained glyphs), and we saw only slightly more glyphs rendered in 2D than in 3D. Glyphs, however, were difficult to identify and our consistency in coding this visualization type was initially only \glyphcon. 
Canonical examples, taken from
\cite{fanea2005interactive},
\cite{hlawatsch2011flow},
\cite{hlawitschka2005hot},
\cite{garcke2000continuous}, are:

\noindent\begin{minipage}{\columnwidth}\vspace{1ex}%
\centering
\includegraphics[clip, trim={0 0 800 0}, height=\pictureheight]{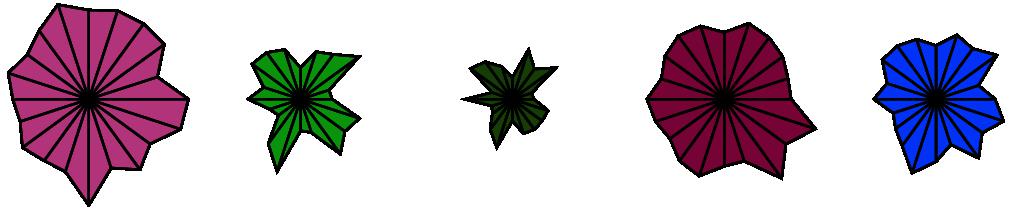}\hfill%
\includegraphics[clip, trim={473 40 0 0}, height=\pictureheight]{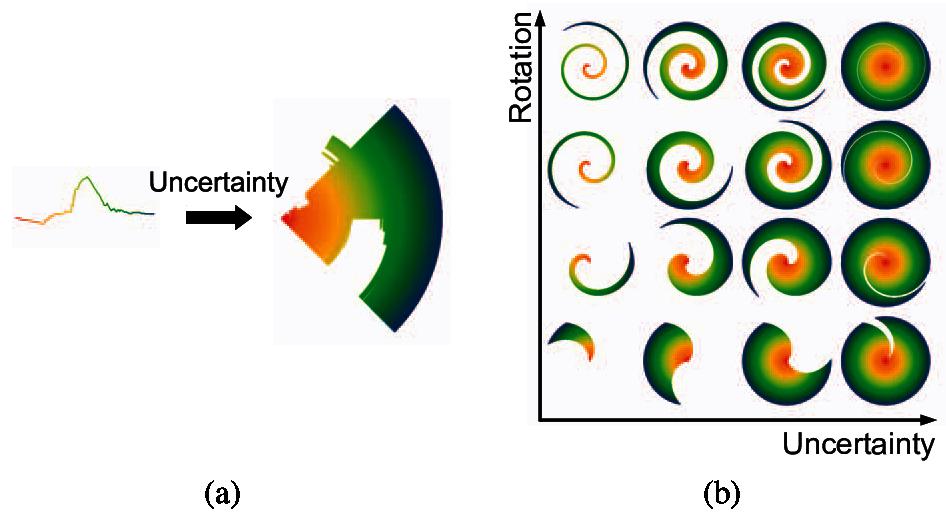}\hfill%
\includegraphics[height=0.5\pictureheight]{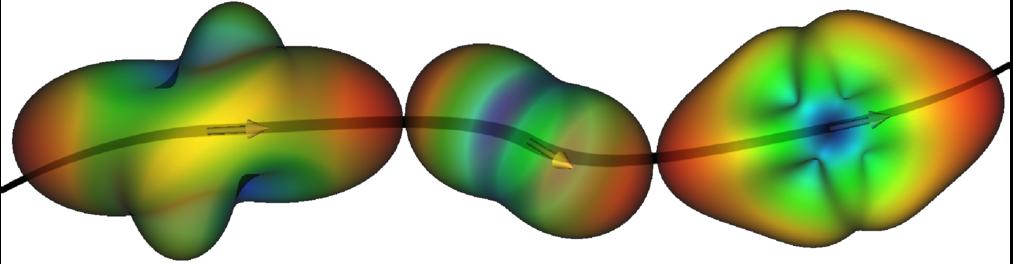}\hfill
\includegraphics[height=\pictureheight]{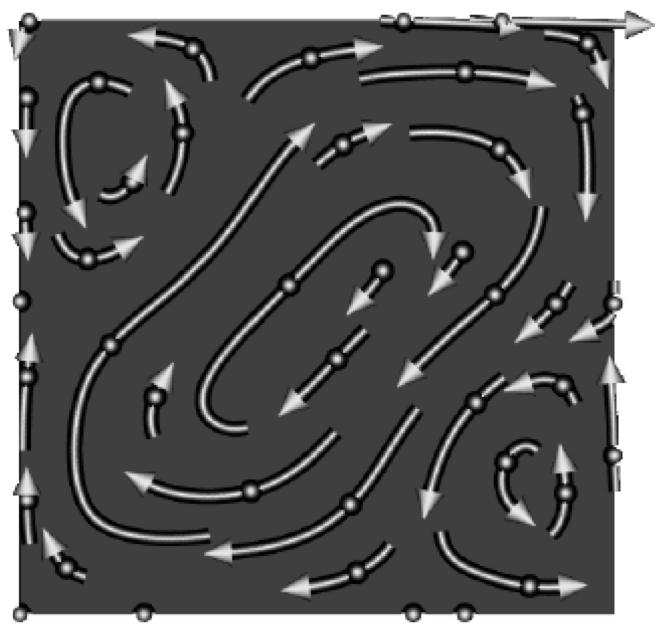}%
\vspace{\spaceaftercanonicalexamples}\end{minipage}

\textbf{Text-based Representations} 
\label{sec:vc.text}
\emph{show data (usually text itself) using varying properties of letters or words such as font size, color, width, style, or type.} Common visualization techniques for this type are tag clouds, word trees, or typomaps.

\textls[-13]{
Text-based representations were the rarest in our coding. All text-based representations were rendered in 2D. The initial coding consistency was low at \textcon, primarily because some coders initially also coded representation where text was used as a data source. 
Example images, taken from \cite{lee2010sparkclouds},}
\cite{wang2017edwordle},
\cite{afzal2012spatial}, include:

\noindent\begin{minipage}{\columnwidth}\vspace{1ex}%
\centering
\includegraphics[clip, trim={100 200 200 200},height=1.4\pictureheight]{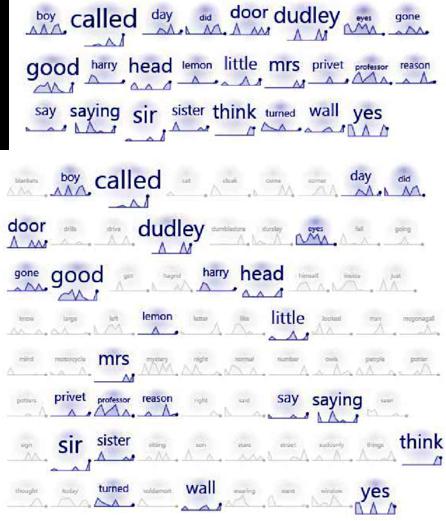}\hspace{7mm}%
\includegraphics[clip, trim={0 50 1500 0}, height=1.4\pictureheight]{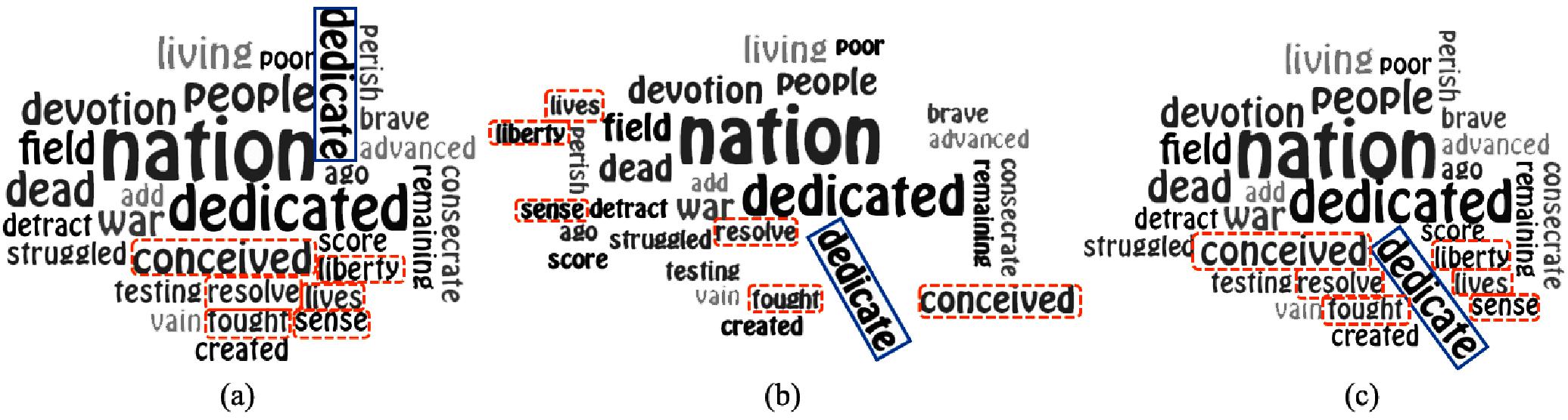}\hspace{7mm}%
\includegraphics[height=1.4\pictureheight]{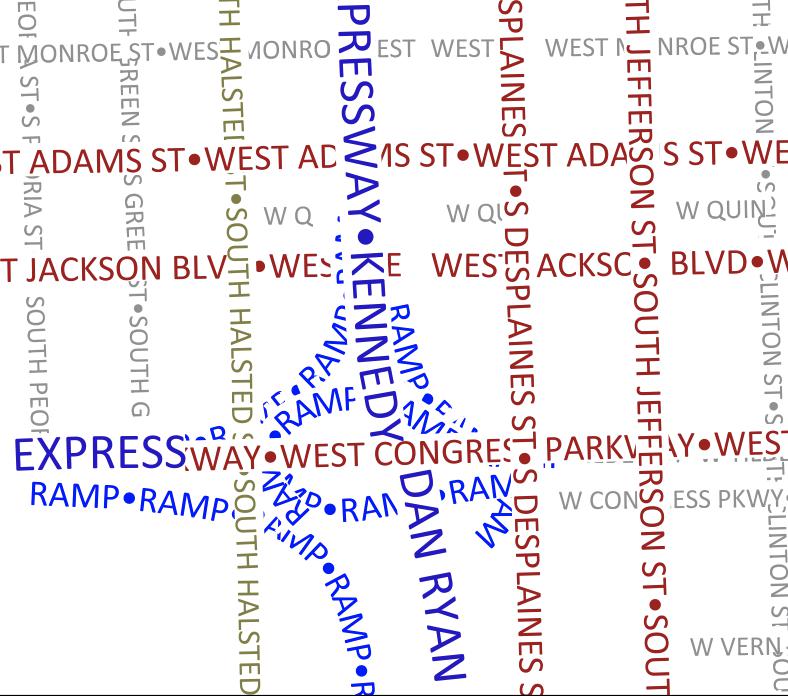}%
\vspace{\spaceaftercanonicalexamples}\end{minipage}

\subsection{Visualization functions}

\noindent To all images showing one of the visualization types above, we implicitly assigned the function to ``showcase a visualization technique.'' This function was by far the most common. In addition to this implicit function, we coded screenshots or images of graphical user interfaces (GUIs) and schematic representations. We did not assign additional individual visualization types for images with both functions. 

\textbf{GUI (Screenshots)/User Interface Depiction.}
\label{sec:vc.gui}
\emph{These images are generally screenshots or photos of a system interface.  GUIs require the presence of window components or other UI widgets such as buttons, sliders, boxes, scroll bars, pointers (\eg, the hand cursor showing interaction), etc.}

Non-WIMP interfaces (\eg, for VR or touch-based applications) are indicated by, \eg, a hand\discretionary{/}{}{/}finger touching a surface or clearly visible interface hardware such as a tablet, a tabletop display, or other types.
We found \numguiOrg GUI images in total, representing about $12\%$ of all images. 
The proportion of GUI images over time is relatively stable. 
Only about $13\%$ are 3D images.
Coders find that identifying GUI images is relatively easy and the overall consistency of coders was \guicon.
Canonical examples, taken from \cite{bergman1995rule}, 
\cite{dasu2020sea},
\cite{yoghourdjian2020scalability}, are:

\noindent\begin{minipage}{\columnwidth}\vspace{1ex}%
\centering
\includegraphics[height=\pictureheight]{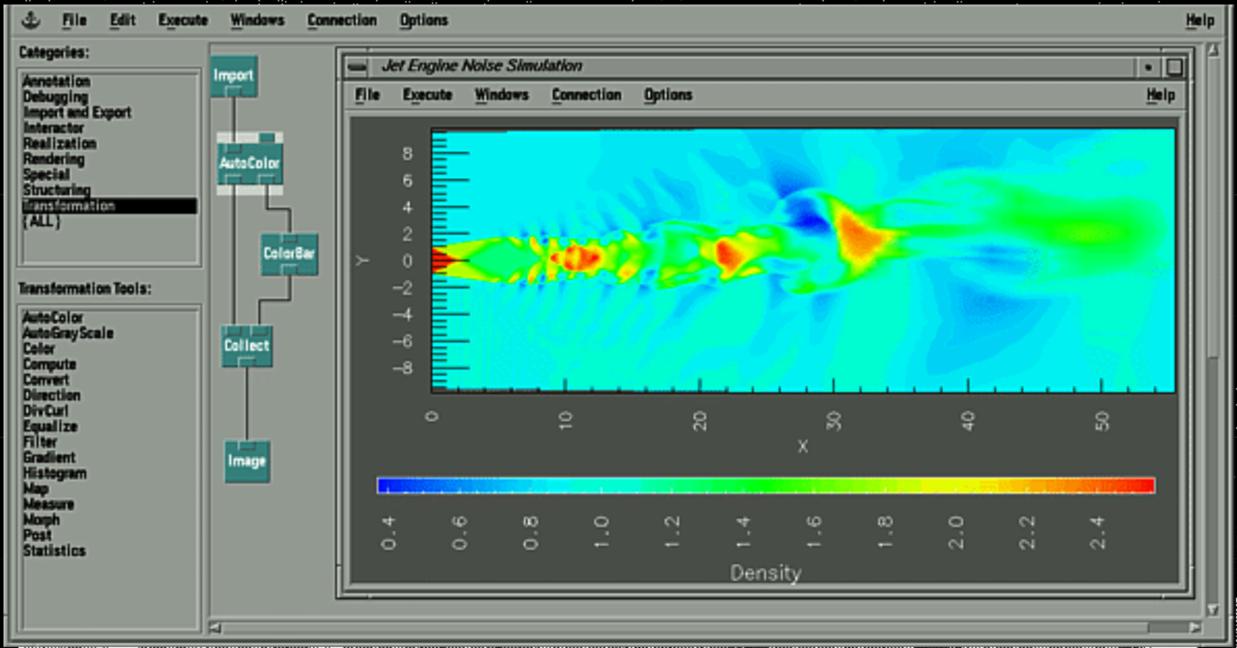}\hfill
\includegraphics[height=\pictureheight]{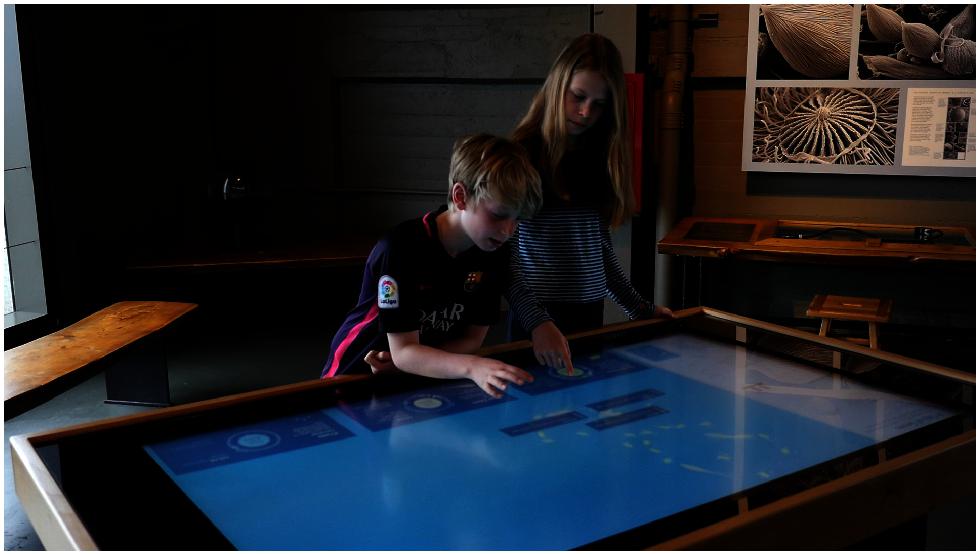}\hfill%
\includegraphics[height=0.17\columnwidth]{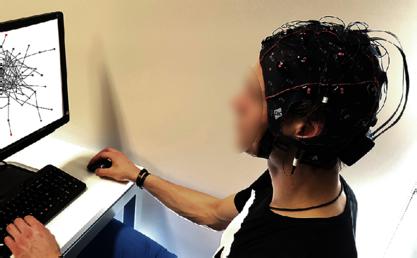}%
\vspace{\spaceaftercanonicalexamples}\end{minipage}

\textbf{Schematic Representation and Concept Illustrations.}
\label{sec:vc.schematic}
\emph{These were often simplified depictions that show the appearance, structure, or logic of a process or concept.}
Typical examples include flowcharts to illustrate algorithms, process diagrams, or sketches.
Schematics and illustrations are common in research papers, not just in visualization papers. 
We coded \numschematicOrg images in this category, 
$79\%$ of which were 2D.
Canonical examples taken from
\cite{lee2020data},
\cite{jeong2010interactive},
\cite{isenberg2010exploratory} are:


\noindent\begin{minipage}{\columnwidth}\vspace{1ex}%
\centering
\includegraphics[height=\pictureheight]{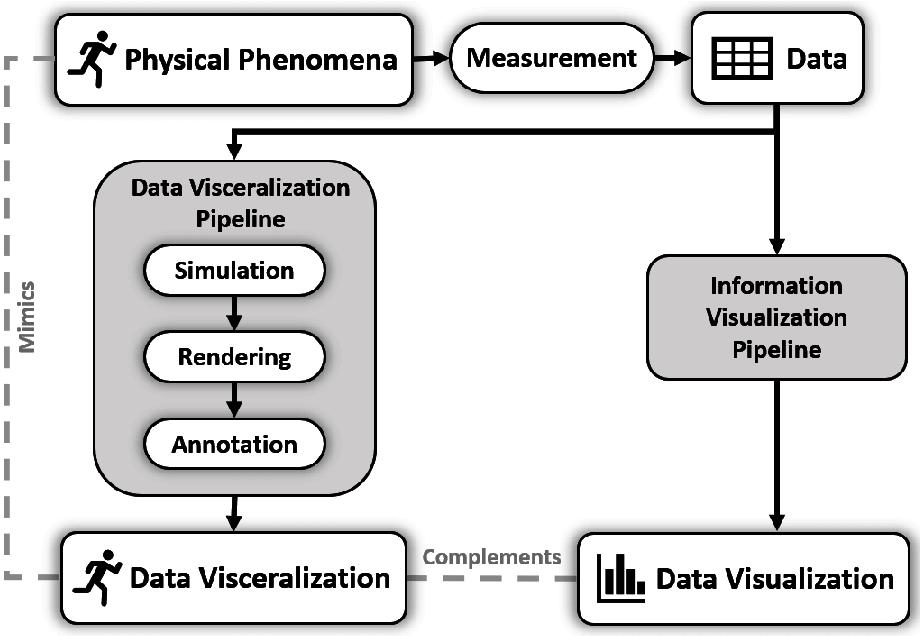}\hfill%
\includegraphics[clip, trim={0 0 0 0},height=\pictureheight]{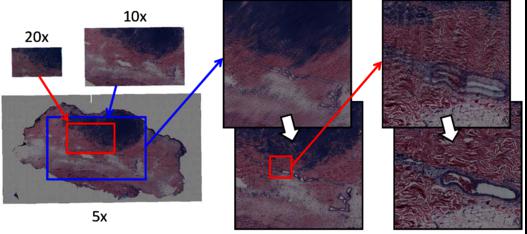}\hfill%
\includegraphics[clip, trim={0 0 1550 0},height=\pictureheight]{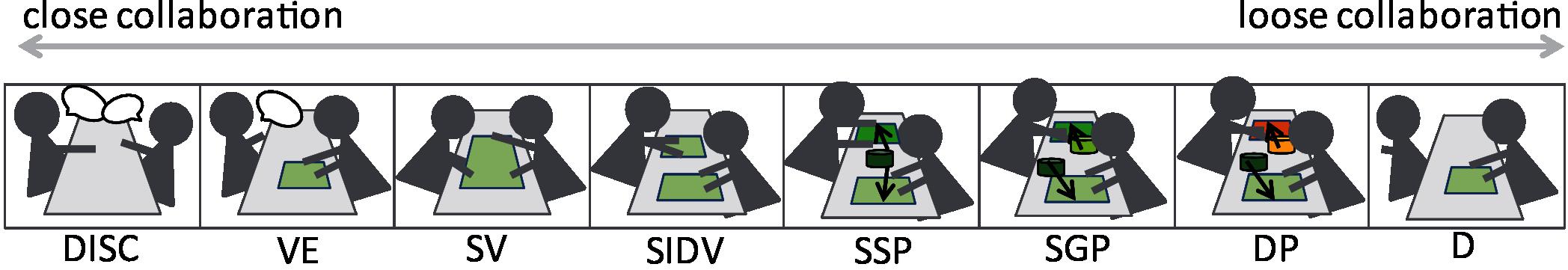}%
\end{minipage}

\subsection{Dimensionality: 2D and 3D}
\label{sec:dim}
\noindent We also coded the spatial dimensionality of the 10 visualization types (Bar, Point, etc.). 
We labeled a flat representation on a 2D plane without perceived depth as 2D, while we classified images that appeared to be in 3D (or volumetric) space as 3D.
To code an image as 3D we looked for depth cues such as occlusion, lighting and shading, parallel and perspective projection, rotation, or any other depth cues.
\begin{wrapfigure}{l}{0.25\textwidth}
\centering
\includegraphics[width=0.16\textwidth]{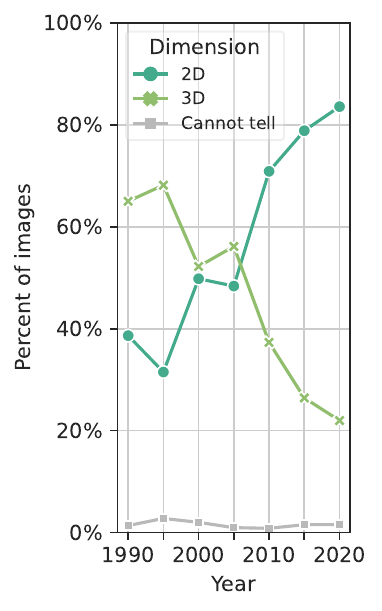}
\vspace{-1ex}
\caption{\tochange{\label{fig:dimDistribution}Percentage of 2D\discretionary{/}{}{/}3D by total images.
The sum in each year is larger than $100\%$ because some images contain both types.}}
\vspace{-1ex}
\end{wrapfigure}
We also required continuous depth with smooth transitions between depth values.  In other words, we did not generally code images as 3D when the content resided in just one or two 2D planes or layers, \eg, an artificial discrete depth. 
Even this 
simple exercise turned out to be non-trivial for many images due to the presence or absence of a mixture of depth cues.
The use of 2D representations became more common than 3D after 2005 (\autoref{fig:dimDistribution})
and had been steadily increasing. 
It is also not surprising that 3D representations were used more broadly for ``surface-based representations and
volumes.''

%
%
\section{\tochange{Discussion of Expert Research Group Findings}}

\label{sec:discussion}
\label{sec:ambiguous}

\noindent
{Reflecting on our results, we (\expertUs) examine the challenges and limitations of our categorization process. 
}

\subsection{\tochange{Choosing the level of abstraction for categories}}
\noindent One of our biggest challenges was to find a categorization we could apply without requiring the details of the data each image represents or what construction rules were used.

\subsubsection{\textbf{Bertin's Marks and Channels as Inspiration}}
\label{sec:bertinMarks}
One of our first attempts was to use Bertin's semiology of graphics \cite{bertin1983semiology} and, particularly, his marks and visual variables for describing visual designs.  This approach resulted in numerous (low-level) codes per image that did not enable us to derive meaningful categories. 
A bar chart, for example, could be coded as line marks with a length encoding for quantity and a position encoding for category
(\autoref{fig:sm:barSurface} in \appref{sm:Bertin}). 
Instead, we needed higher-level categories that could apply to \emph{entire} visualization images
%
Still, Bertin's definition of visual variables and marks
inspired the identification of several of our categories: 
\pointabbr
primarily reference a position, 
\barabbr
a length, and 
\areaabbr
a two-dimensional size. 
\jctg{This result indicates that those design elements (line or position in bars) are 
different from what we see (length in bars). Therefore,
we chose 
not to label according to the individual design elements, but more collective visual foci such as \surfacevolumeabbr, \textabbr, and \colorabbr.
}

\subsubsection{\textbf{Visualization Techniques as a Typology}}
\label{sec:visMarksAsTypology}
\jctg{Attempting to abstract from visualization techniques failed as well. Different techniques could result in similar visuals.}
In addition, identifying techniques sometimes requires knowledge of the data represented. 
Timeline visualizations, for instance, could only be identified when temporal 
attribute
was indicated on the axis labels
(\autoref{fig:sm:timeline} in \appref{sm:Bertin}).
These additional (often insurmountable) difficulties made the visualization techniques approach impractical.
\tochange{
What we retained from this failed attempt, however, was a focus on the \et.
We intentionally chose not to code legends, coordinates, time, labels, or embellishments, and instead focused on the most visually important part of an image \jctg{that show actual data} (as opposed to simpler illustrations).
}

\subsubsection{\textbf{Classification vs.\ Categorization}}
\label{sec:classCat}
We began our work with the goal of deriving a classification of images, where each image would be assigned to a single category. 
In a classification system, each image represents the class as any other.  
By the time we had experimented (and failed) with the first two approaches {(Bertin and visualization techniques)},
we noticed several problems with {trying to come up with a classification}: we saw many visual representations that would form a distinct category (\eg, all that used bars to encode data) but also many that would not. 
While we show representative images in this paper, 
our
\href{https://visimagenavigator.github.io/}{Vis\-Ima\-ge\-Na\-vi\-ga\-tor tool} shows many more that are less clearly members of their assigned category. Therefore, we shifted our focus and turned to a modern view of categorization \cite{Jacob:2004:Classification}, which enables us to assign images to multiple categories, but also identify images as more or less representative of their category.

\subsubsection{\textbf{Is \viscolor a Separate Type?}}
\label{sec:continousColor}
Despite the observation that properties of color, such as hue, saturation, and luminance, are visually dominant in data representations, our first two attempts did not include a dedicated visualization type related to color. Instead, we initially coded most continuous and discrete heatmap type encodings under ``generalized ma\-trix\discretionary{/}{}{/}grid'' based on the logic that these encodings are applied on pixel-level grids. 
In our iterative coding refinement phase, many discussions centered on when a continuous color-based encoding should be coded as a grid, especially when it did not look like a grid. When continuous color scales were applied to 3D geometries, such as streamlines, \eg, the ma\-trix\discretionary{/}{}{/}grid encoding no longer seemed appropriate. 
Therefore, after many discussions, we created a distinct category to recognize continuous colormaps.  To distinguish it from \visgrid, we required the color mapping to be systematic and not a result of illumination or an author-chosen categorical representation. 
At this point, we already had a separate category for texture-based representations that we eventually merged with the new color category as both focus on revealing continuous patterns. 

\subsubsection{\textbf{The Influence of Domain Knowledge}}
\label{sec:influenceDomainKnowledge}
Throughout the creation of our categories, we struggled with identifying and suppressing the influence of our individual 
knowledge. The team members (\expertUs) with a background in volume/surface-based techniques, \eg, often mentioned specific techniques, tasks, or goals for images while creating our categories, to which the other team members had little or no relation. 
Similarly, our definitions of glyphs were not aligned between experts with a background in flow visualization and those in abstract visualization. 
In practice, we constantly reminded ourselves to suppress our knowledge of non-visual aspects of the 
images to resolve ambiguities. 
We resolved many issues through discussions of individual images and revising our type codes iteratively with examples and counter-examples.

\subsection{General coding ambiguity}
\label{sec:generalCodingAmbiguity}
\noindent While we adopted a view of categorization with fuzzy boundaries, we 
engaged in discussions to resolve ambiguities for frequent cases. 
The most important challenges were:

\begin{figure}[!t]
\centering
\setlength{\figverticaloffset}{11pt}
\setlength{\figureheight}{0.25\columnwidth}
\setlength{\figheightminusverticaloffset}{\figureheight}
\addtolength{\figheightminusverticaloffset}{-\figverticaloffset}
\subfloat[~\hspace{\columnwidth}~%
]
{\raisebox{-\figverticaloffset}[\figheightminusverticaloffset][-\figverticaloffset]{~~~~\includegraphics[clip, height=\figureheight]{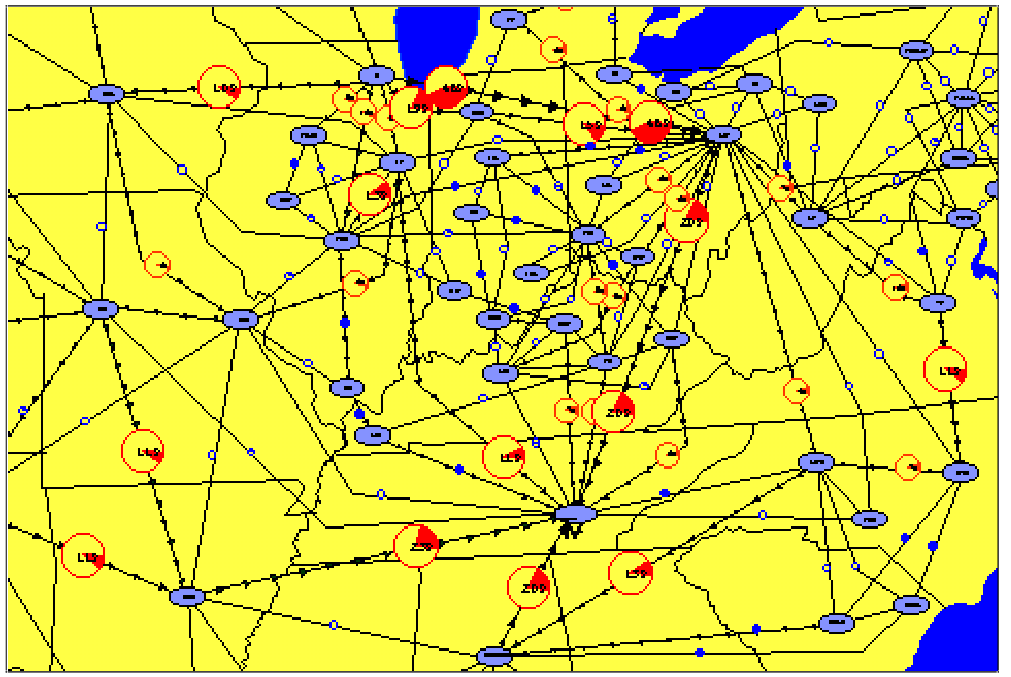}}
\label{fig:areanodelink}}
\hfill%
\subfloat[~\hspace{\columnwidth}~%
]{\raisebox{-\figverticaloffset}[\figheightminusverticaloffset][-\figverticaloffset]{~~~~~\includegraphics[height=\figureheight]{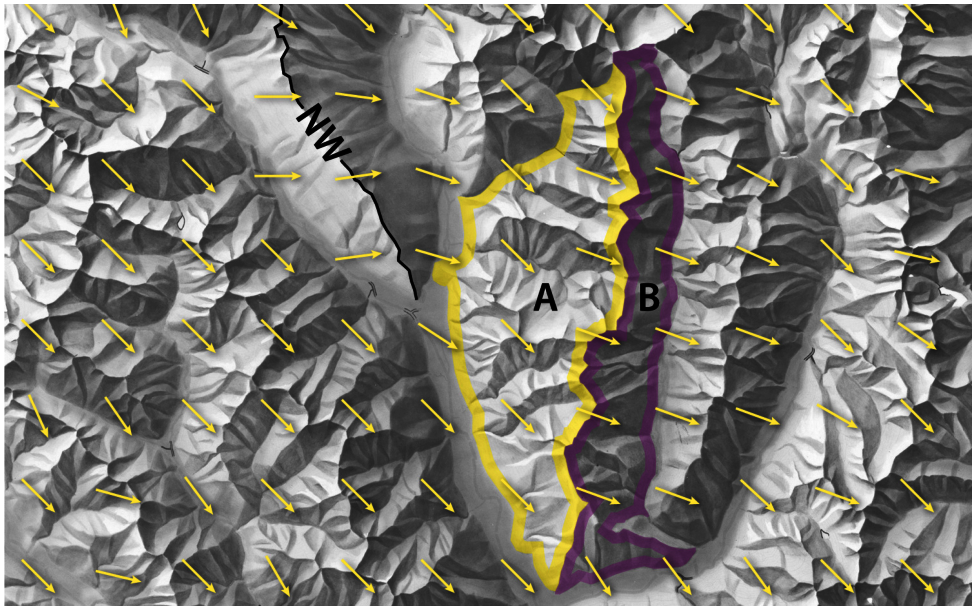}}%
\label{fig:terrain}}%
\vspace{-\figverticaloffset}
\caption{\jctg{Challenging cases.
(a) We choose to code geospatial data as \areaabbr and route connections \nodelinkabbr. (b) We choose  to code the terrain  \surfacevolumeabbr and the arrows on top \glyphabbr. Images from (a) Overbye et al.~\cite{overbye2000new} and (b) Jenny et al.~\cite{jenny2020cartographic}.}}
\label{fig:challengingArea}
\end{figure}

\begin{figure}[!t]
  \centering
\setlength{\figverticaloffset}{11pt}
\setlength{\figureheight}{0.4\columnwidth}
\setlength{\figheightminusverticaloffset}{\figureheight}
\addtolength{\figheightminusverticaloffset}{-\figverticaloffset}
\subfloat[~\hspace{\columnwidth}~]{
\raisebox{-\figverticaloffset}[\figheightminusverticaloffset][-\figverticaloffset]{~~~~\includegraphics[height=\figureheight]{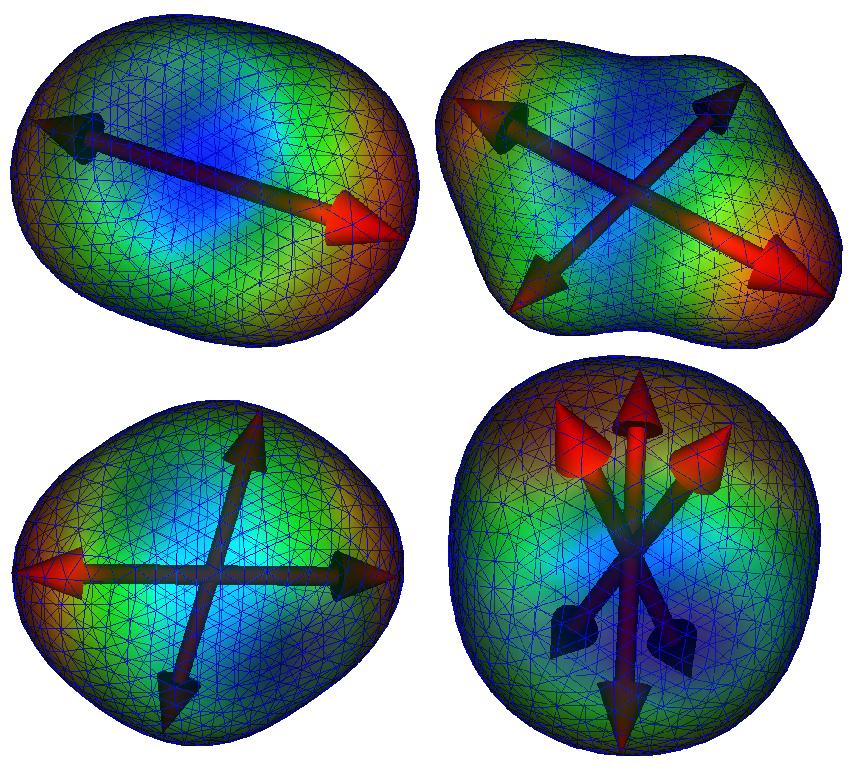}}
\label{fig:sa}
}\hfill%
\subfloat[~\hspace{0.9\columnwidth}~]{%
\raisebox{-\figverticaloffset}[\figheightminusverticaloffset][-\figverticaloffset]{~~~~\includegraphics[height=0.9\figureheight]{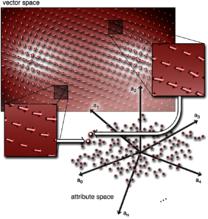}}
\label{fig:sb}
}%
\vspace{-\figverticaloffset}
\caption{Challenging cases of coding \schematicabbr. (a) Is this a glyph-based or schematic representation? (b) We choose not to code the visualization type inside \schematicabbr. Images from (a) Hlawitschka \& Scheuermann~\cite{hlawitschka2005hot} and 
(b) Daniels et al.~\cite{daniels2010interactive}.}
\label{fig:challenging:1}
\end{figure}

\subsubsection{\textbf{Surfaces and Volume Rendering}}
\label{sec:surVol}
Early on, we separated surface and volume renderings into two separate codes. Several of us, however, found it challenging to perceive differences between the two. Volumes, for example, can be rendered semi-transparently (looking cloud-like) or opaquely, the latter having the appearance of a surface. We decided that ``recognizing'' an image type at this level of detail (\eg, whether surfaces are rendered from voxel or mesh data) was unreliable and may not even be important for describing a figure's visual content as the underlying algorithmic techniques
transparent to viewers.  
Eventually, high-level concepts can directly contribute to reasoning~\cite{chen2019looks, wang2020toward}.
Applying this principle and removing the specific details of a technique enabled us to focus on what all instances of a given type have in common, and we thus combined surface and volume techniques.

\subsubsection{\textbf{Ambiguous Area-based Images}}
\label{sec:ambiguousArea}
We tried to avoid using data types in our categorization because we wanted to focus consistently on visual impressions. 
We originally had a ``cartographic map'' category but removed it because it focused on a specific data type and technique. Instead, we decided that depicting areas and their relationships was the underlying principle for many cartographic maps and related techniques such as area charts, stream graphs, etc. (\autoref{fig:challengingArea}). There were, however, exceptions. Route maps, \eg, where lines indicate a direct route, were coded separately as 
networks because the routes encoded topological relationships (\autoref{fig:areanodelink}). 
One difficulty we encountered was the distinction between a ``map'' (cartographical map) and a ``terrain'' (wireframe or surface) (\autoref{fig:terrain}). 
Conceptually, these are very similar. Using visual appearance as our guide, however, map images generally appeared to delineate distinct areas, while terrains showed continuous surfaces with elevations.
Another map-re\-la\-ted difficulty arose when maps were used as a reference structure for data representations layered on top, akin to how gridlines are used on scatterplots. 
In these cases, we had to 
derive elaborate procedures 
for consistently treating reference structures, 
\eg, axis lines. 
If, in addition to maps, other visual encodings are present, we decided to code these
if removing them changes the meaning or information conveyed in the image.
These decisions, however, can be difficult and resulted in some coding inconsistencies.

\subsubsection{\textbf{Ambiguous Schematic Images}}
\label{sec:ambiguousSch}
A large number of figures are schematic representations or concept illustrations. 
It is often challenging to differentiate between schematics and a demonstration of a visual encoding technique. We had to abandon our initial goal of ignoring what data was encoded to be able to 
judge whether the representation showed a ``toy'' dataset.\footnote{Here, a ``toy'' dataset is a synthetic, exemplar dataset to demonstrate a principle for ease of understanding.} 
While toy datasets are common in schematic representations, the frequent absence of context, such as coordinate axes, labels, or scales, made their identification difficult. Many figures did not depict scales, which aligns with observations by Cleveland and McGill~\cite{cleveland1985graphical} in their review of the use of graphics in other scientific journals.

We also struggled with using annotations in figures as identification criteria for schematics. \autoref{fig:sa}, for example, can be coded either as glyph-based (it shows a mathematical tensor) or as a schematic (it illustrates the authors' design idea or a mathematical function)---a majority of us chose to code
\visschematic. 
Schematic images are often meant to be particularly pedagogical and, thus, include several labels and arrows or other annotations. We experimented with specific coding guidelines in which we considered whether or not, after the removal of annotations, we could still see an example of a visualization type---in the former case, the image would have been a schematic. Later, we agreed that the appearance of labels and annotations to explain an image did not automatically mean the image was a schematic. Our general heuristic for schematics involved establishing if we saw (1) a well-known (or toy) dataset, (2) a pedagogical purpose, and (3) an illustration of a concept.

\begin{figure}[!t]
\centering
\setlength{\figverticaloffset}{11pt}
\setlength{\figureheight}{0.33\columnwidth}
\setlength{\figheightminusverticaloffset}{\figureheight}
\addtolength{\figheightminusverticaloffset}{-\figverticaloffset}
\subfloat[~\hspace{\columnwidth}~%
]
{\raisebox{-\figverticaloffset}[\figheightminusverticaloffset][-\figverticaloffset]{~~~~\includegraphics[clip, height=\figureheight]{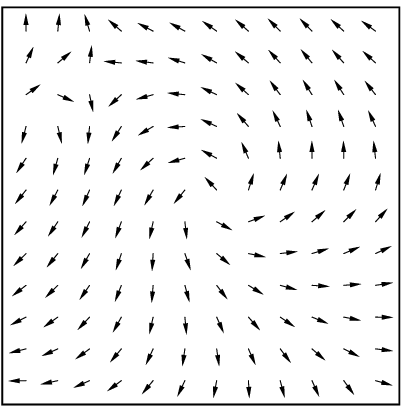}}
\label{fig:glyphOnly}}
\hfill%
\subfloat[~\hspace{\columnwidth}~%
]{\raisebox{-\figverticaloffset}[\figheightminusverticaloffset][-\figverticaloffset]{~~~~~\includegraphics[height=\figureheight]{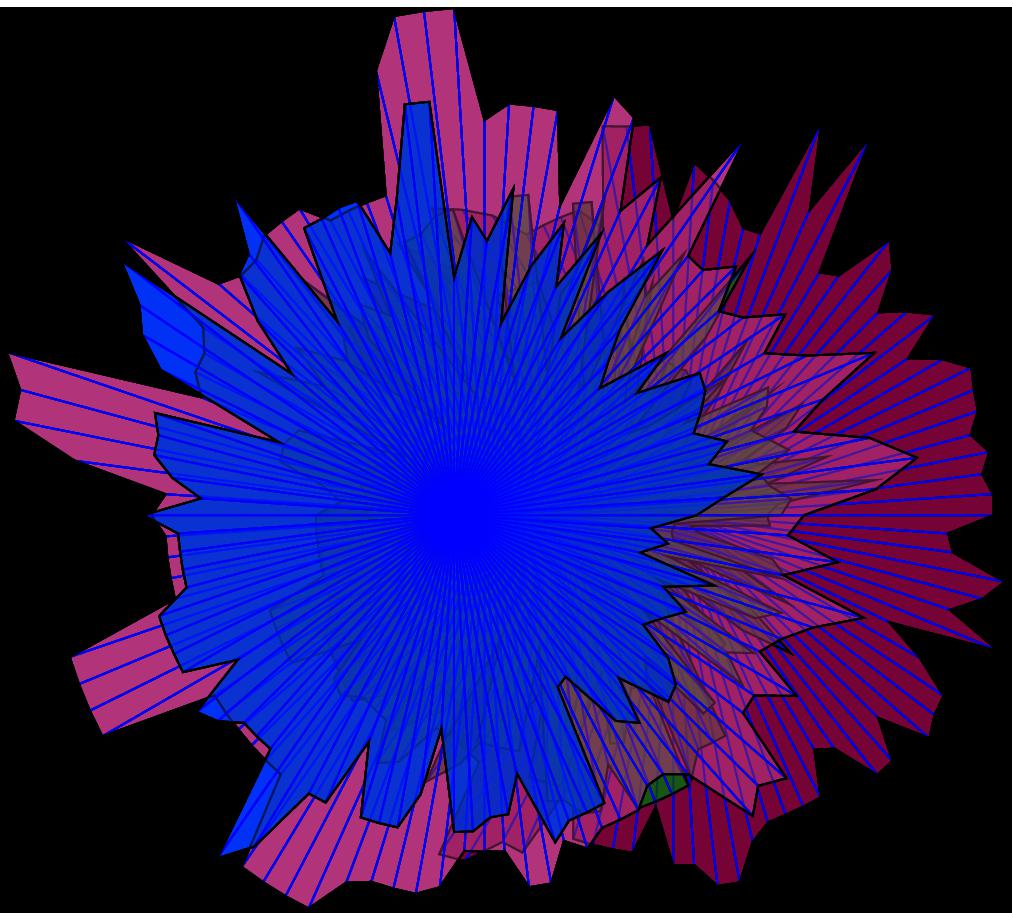}}%
\label{fig:areaGlyphglyph}}%
\vspace{-\figverticaloffset}
\caption{Challenging cases of coding \glyphabbr. (a) The arrows show the 2D position as well as orientation. 
(b) StarGlyph-like dimensional comparisons and thus a type of \glyphabbr.
Images from (a) Bian et al.~\cite{bian2020implicit} and (b) Fanea et al.~\cite{fanea2005interactive}.}
\label{fig:glyphHard}
\end{figure}

\subsubsection{\textbf{Ambiguous Glyph Cases}}
\label{sec:ambiguousGlyph}
Glyphs are notoriously difficult to define. Recent attempts have emphasized different aspects of delineating a glyph from other encodings. Fuchs et al.~\cite{fuchs2013evaluation} defined data glyphs as ``data-driven visual entities, which use different visual channels to encode multiple attribute dimensions.''
Borgo et al.~\cite{borgo2013glyph} followed Ward \cite{ward2008multivariate} to define glyphs as ``a visual representation of a piece of data where the attributes of a graphical entity are depicted by one or more attributes of a data record.'' 
Munzner's \cite{munzner2014visualization} definition is broad and requires a data encoding to be assembled out of multiple marks that encode data.
Every bar in a stacked bar chart~\wordimg{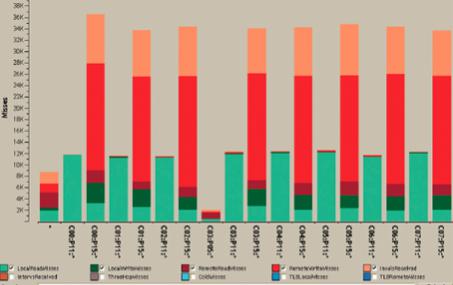}, \eg, would be a ``microglyph'' according to Munzner, because it is a composite object of multiple length-encoding marks. 
In our coding, we used a mix of the given definitions because, for all coders, it was difficult to dissociate their research background on glyphs from the coding.

\textls[-12]{While many existing encodings focus on describing a (single) glyph, glyph-based visualization is often used in practice when multivariate data needs to be presented simultaneously. 

A challenge is thus to determine when a primitive becomes a glyph. 
There is no agreed-upon threshold of how many data dimensions are encoded in a glyph and when a glyph becomes a chart. 
There seems to be, however, a general consensus that a glyph requires a certain level of complexity to be categorized as such.}
We labeled an image as a glyph-based representation if 
multiple representations of data points 
represent both position and additional data dimensions using color, shape, or other geometric primitives. \autoref{fig:glyphHard} illustrates two difficult cases.

\tochange{Finally,
we first had a separate ``tensor glyph'' category. This category, however, required knowledge of affine transforms applied to a basic shape, which could not be concluded from pure perceptual observations. So, we removed this category.}

\subsection{Multiple encoding ambiguity}
\label{sec:challenges:multiple-codes}

\begin{figure}[!t]
\centering
\subfloat[%
Cont. color.
]{\includegraphics[height=0.159\textwidth]{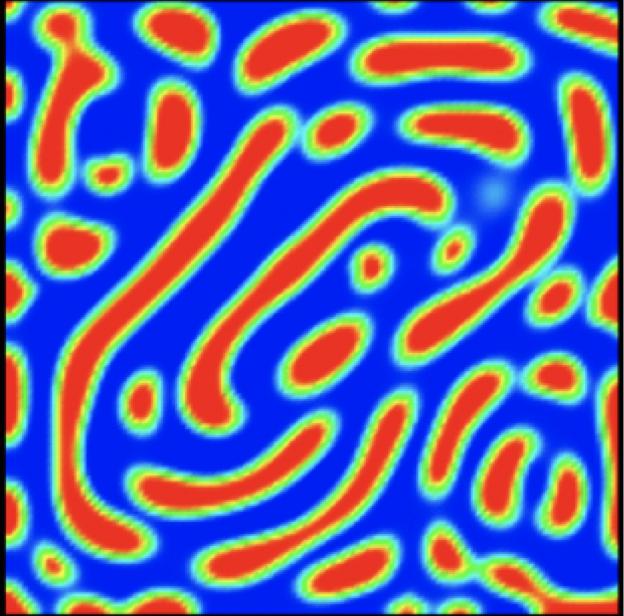}%
\label{fig:c}}\hfill%
\subfloat[%
Cont.\,color\,\&\,\glyphabbr.%
]{\includegraphics[height=0.159\textwidth]{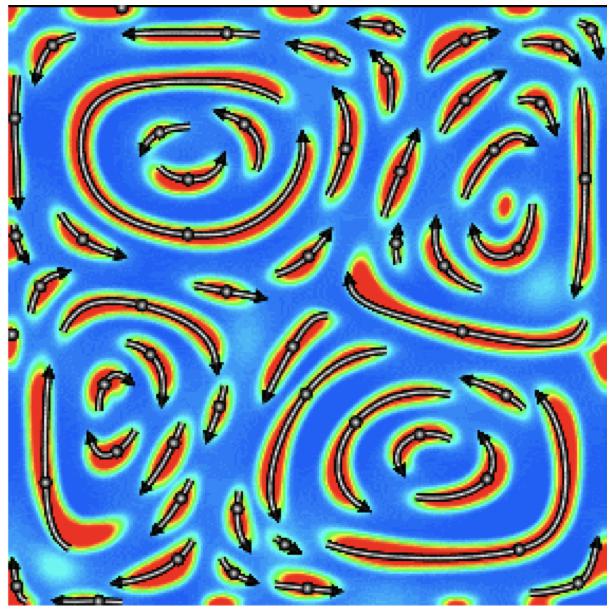}%
\label{fig:gc}}\hfill%
\subfloat[%
Grey-scale\,\&\,Glyph.]{\includegraphics[height=0.159\textwidth]{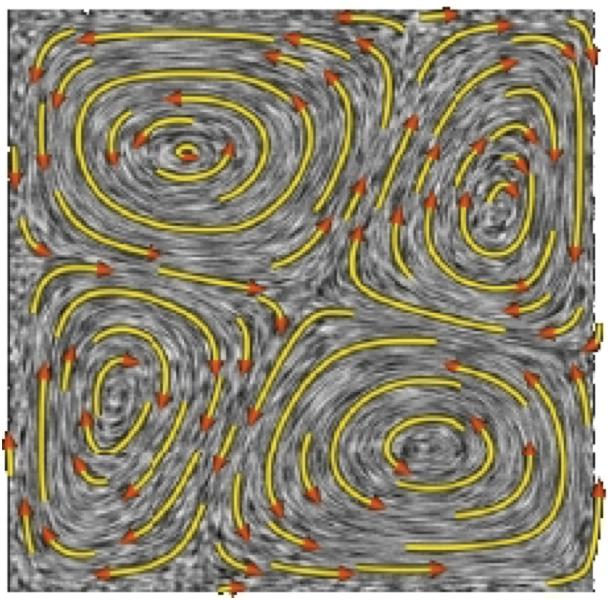}%
\label{fig:fgp}}
\caption{%
Challenging cases of multiple coding ambiguity: we did not code \surfacevolumeabbr and only chose the primary code, which differentiates these visualization techniques (images from 
(a) Weiskopf et al.~\cite{weiskopf2005overview} and 
(b, c) Garcke et al.~\cite{garcke2000continuous}).}
\label{fig:flowcases}
\end{figure}

\begin{figure}[!t]
\centering
\subfloat[%
Node-link+Surface.]{\includegraphics[clip, trim={680 50 0 0},height=0.3\columnwidth]{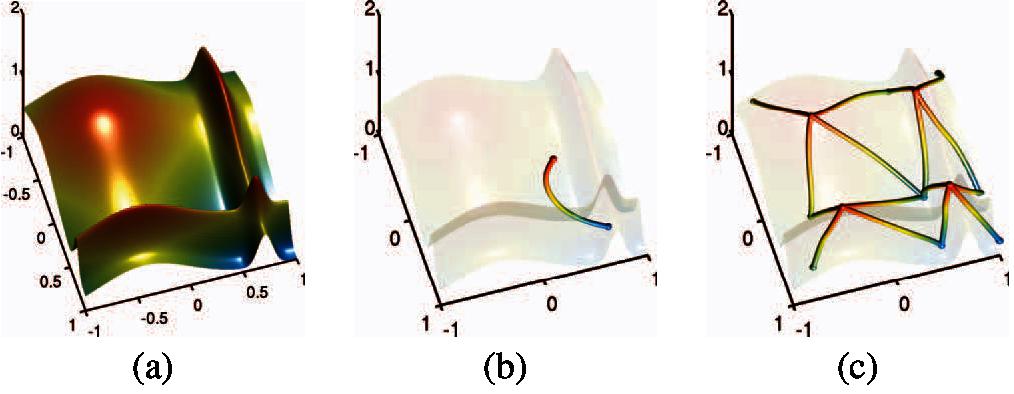}~~~%
\label{fig:nodelinkSurface}}\hfill%
\subfloat[%
Line, Point, and Surface.]{\includegraphics[clip, trim={0 0 0 0},height=0.3\columnwidth]{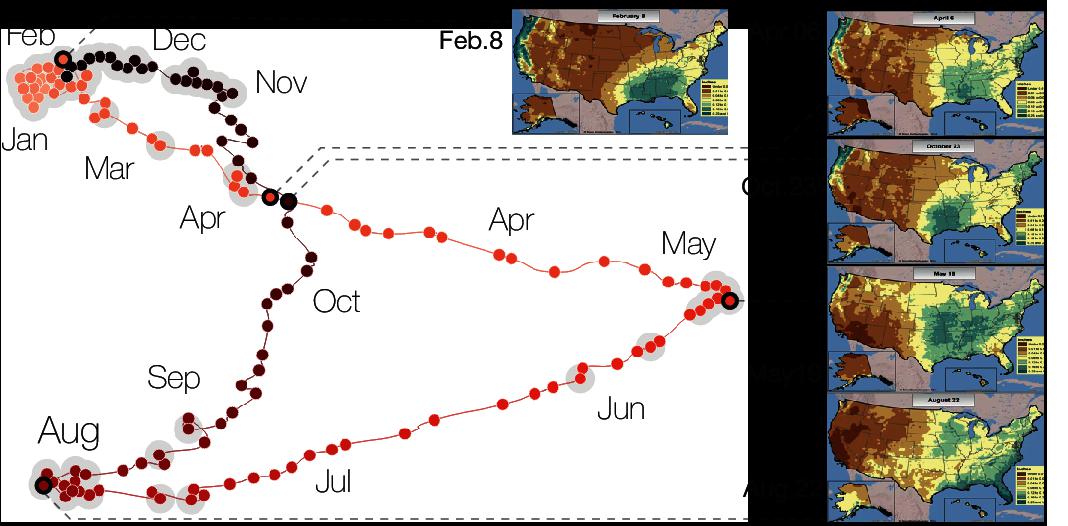}
\label{fig:linepointsurface}}
\caption{Challenging cases of multiple coding ambiguity: two codes signify different aspects of the data that can be separated to stand independently. As a result, multiple codes apply
(images from 
(a) Suits et al.~\cite{suits2000simplification}
and
(b) Bach et al.~\cite{bach2015time}).
}
\label{fig:nodelinkMeshes}
\end{figure}

\noindent Many images showed multiple visual encodings (\eg,  \autoref{fig:flowcases} and \autoref{fig:nodelinkMeshes}), which is one of the primary reasons for the coding inconsistencies we encountered. \tochange{We agreed to code multiple visualization types if these types were distinctive, could be perceived clearly, and were essential.} If multiple visual designs, layered or nested, could be distinguished from one another, we tagged more than one encoding, \eg, Cont.color + Glyph for~\autoref{fig:gc} and node-link + surface for \autoref{fig:nodelinkSurface}.

\begin{figure}[!t]
\centering
\subfloat[\barabbr.]{\includegraphics[clip, trim={0 0 0 0},height=0.11\columnwidth]{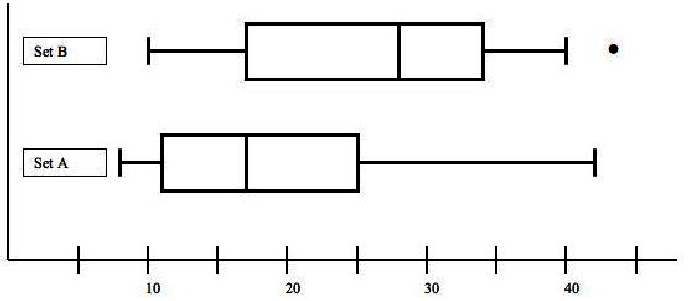}~~~%
\label{fig:b}}\hfill%
\subfloat[%
\barabbr + \pointabbr.]{\includegraphics[clip, trim={0 0 0 0},height=0.11\columnwidth]{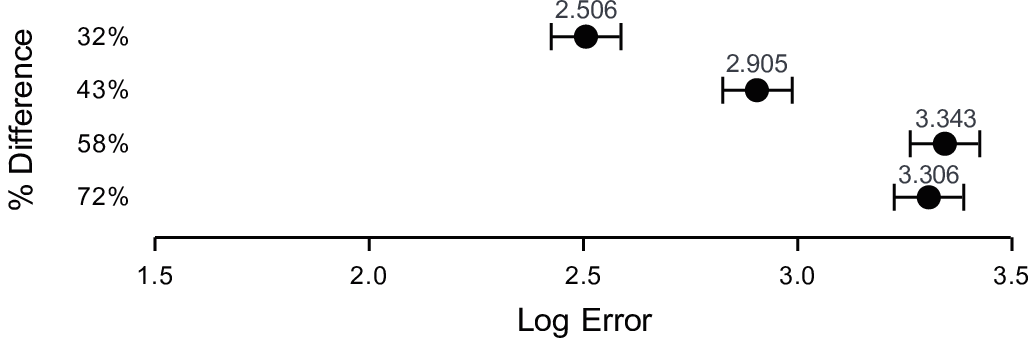}
\label{fig:bp}}\hfill
\subfloat[%
\lineabbr.]{\includegraphics[clip, trim={0 0 0 0},height=0.18\columnwidth]{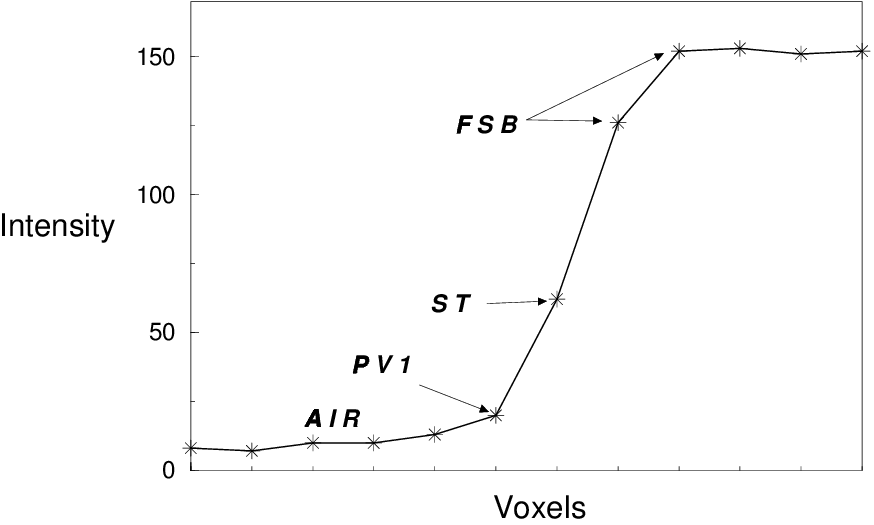}
\label{fig:bl}}
\caption{\jctg{Challenging cases of coding statistical charts: We coded \barabbr only in the  Box-whisker plot in (a) because it shows a group of quartiles. In (b), we coded \pointabbr when points show the mean values. In (c), we excluded \pointabbr, because these points function as an annotation to the \lineabbr and do not represent the underlying data (images from
(a) Chlan and Rheingans~\cite{chlan2005multivariate}, 
(b) Ziemkiewicz and Kosara~\cite{ziemkiewicz2010laws}, and
(c) Lakare et al.~\cite{lakare20003d}).
}}
\label{fig:hardStatChart}
\end{figure}

\jctg{We also decided when (not) to check multiple labels (\eg,~\autoref{fig:hardStatChart}). For example, points appear in both confidence interval and line charts. 
For confidence intervals, we chose to code both \visbar and \vispoint as primary categories because 
the bar length encodes the interval, and because the confidence interval is not always symmetric~(\autoref{fig:bp}).} In contrast, we considered those points that sometimes are added to line charts (\autoref{fig:bl}) as an annotation and did not code them. These decisions were difficult, sometimes inconsistent, and error-prone (\eg, the dot for the average could be an annotation when the drawing of the error bars is symmetric). We identified conflicting understandings through many hours of conversations on individual and difficult images, which often required some coders to compromise their opinion.

\section{\tochange{Validation Experiment with Expert Users}}
\label{sec:validationstudy}

\noindent\tochange{Having finished coding the images, we next designed a controlled experiment to explore whether or not viewers with visualization experience can apply our \typename coding schema to categorize images
(pre-registered at \osflink\ and with IRB approval from The Ohio State University, ID: 2022B0363).}
\tochange{We had 
two high-level working hypotheses:} 
\begin{itemize}[nosep,left=0pt .. \parindent]
\item First, we expected that participants could use our \typename category descriptions and identify images with better-than-chance agreement. 
\item Second, we expected certain image categories to be more difficult for participants and result in less agreement (matrix/grid, glyphs, and continuous color \& greyscale-based representations) based on our own experience.
\end{itemize}

\subsection{Methods}
\label{sec:methods}

\begin{figure*}[tb]
	\centering
\includegraphics[width=\textwidth]{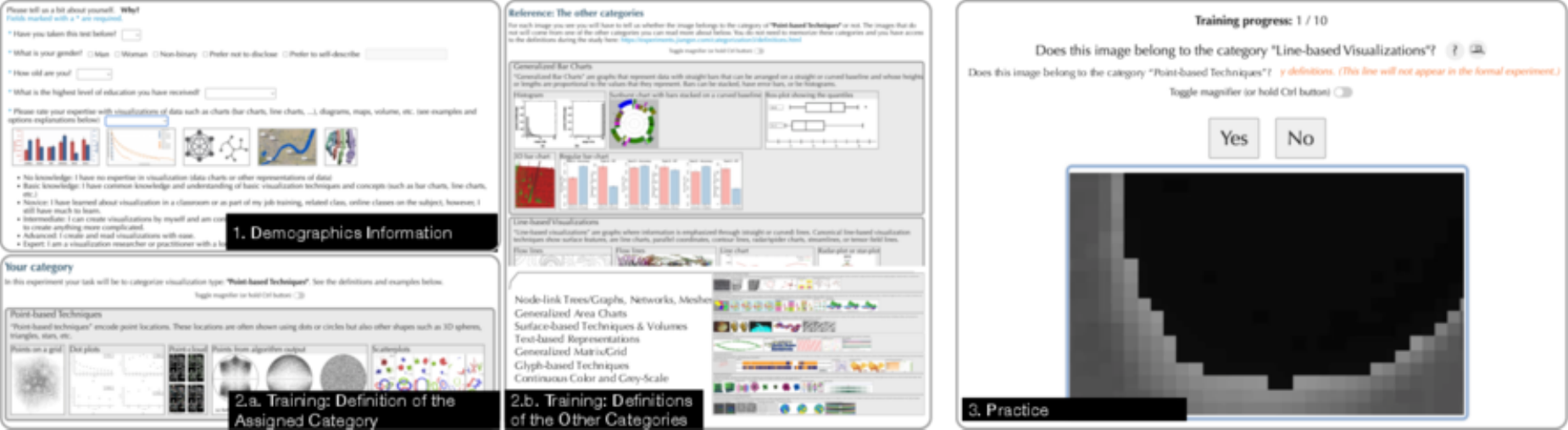}
	\caption{\tochange{\textbf{Empirical evaluation workflow}: We invited 110 participants (\expertExternal) to participate in the study. They first saw a consent form and completed a demographic questionnaire. Then we gave them the definition (corresponding to the text in italics in \autoref{sec:visTypesResults}) and example images for their assigned category. Next, they were invited to familiarize themselves with the other categories, their definitions, and example images. The following screen introduced the study interface. 
 Each participant then completed ten training trials before moving on to the 36 study trials. Readers can try the study at 
\href{http://go.osu.edu/vistype}{\texttt{go.osu.edu/vistype}}.
 }
	}
	\label{fig:expertEvalExp}
\end{figure*}

\noindent%
\textbf{Tasks.}
We assigned each participant a single visualization type category, $c$. 
In each trial, we showed the participant one image from the assigned or another category \tochange{(\autoref{fig:expertEvalExp} shows the interface we used during the experiment)}, and let them answer the 
yes/no question:
\textit{``Does this image belong to category $c$?''}

\textbf{Image Selection.} 
\tochange{%
We pre-selected 
18 images for each visualization type, for a total of 180 images for the 10 types.} 
Each set contained images with the same difficulty levels based on the two coders who had initially labeled them: 2\texttimes\,easy (6 samples, hardness level L$=$1), 
easy-neutral (6 samples, L$=$2), 
2\texttimes\,neutral (2 samples, L$=$3), and easy/neutral-hard or 
2\texttimes\, hard (4 samples, L$=$4).
\tochange{All images had been previously identified as having a single visualization type only.}

\textbf{Study Design and Order of Trials.} 
\tochange{We used a between-subject design with the visualization type (10 categories) as the only independent variable. Dependent variables were \textit{agreement} with our assigned labels for each image and task completion \textit{time}. 
Each participant completed 36 trials, including 18 images from the assigned category (target) and 18 from the other (non-target) categories. For example, when their assigned category was ``generalized bar representation,'' they saw 18 
Bar images and 18 images from 9 other categories, with 2 randomly selected in each of these 9 categories. We ordered these 36 images randomly per participant.}

\textbf{Participants.} 
We recruited 110 participants with some form of visualization education (\eg, having taken or currently taking a visualization course) to experts (\ie, researchers or practitioners with a longer practice in visualization). We call these participants \emph{\expertExternal} to differentiate them from ourselves (\emph{\expertUs}). 
We conducted a power analysis (\osflink)
to address possible biases of the yes\discretionary{/}{}{/}no test \cite{prins2016psychophysics} and to determine the number of participants.
We also recruited participants with diverse visualization backgrounds; \ie, people working in visual analysis, 
encoding methods, evaluation, 
simulation, and algorithmic foundations, experienced with different data types and stages of the 
analysis pipeline.

After the trials, we asked participants to describe their answers for two images they had seen: one for an answer that agreed with our coding and one that did not (if all answers agreed or did not agree, we randomly chose two images). 
We also collected optional free-text comments.

\subsection{Results}

\noindent We collected 3,960 yes/no answers from 110 participants with self-reported expertise levels being `expert' (49\texttimes), `advanced' (23\texttimes), `intermediate' (22\texttimes), and 'novice' (16\texttimes). Among them, 46 participants held a `Ph.D. degree', 41 were enrolled in a `graduate school,'  20 in a `college,' and 3 in a `professional school.'
\tochange{The participants' self-reported gender distribution was 35 women, 73 men, and 2 others 
\jctg{(roughly similar to the gender distribution of IEEE VIS authors now \cite{tovanich:2022:gender})}.}
We also collected 215 image-specific and 32 general comments about the study.

\subsubsection{\textbf{Agreement}}

\begin{figure}[!tb]
\centering
{
\includegraphics[width=\columnwidth,clip, trim={460px 1340px 310 50}]{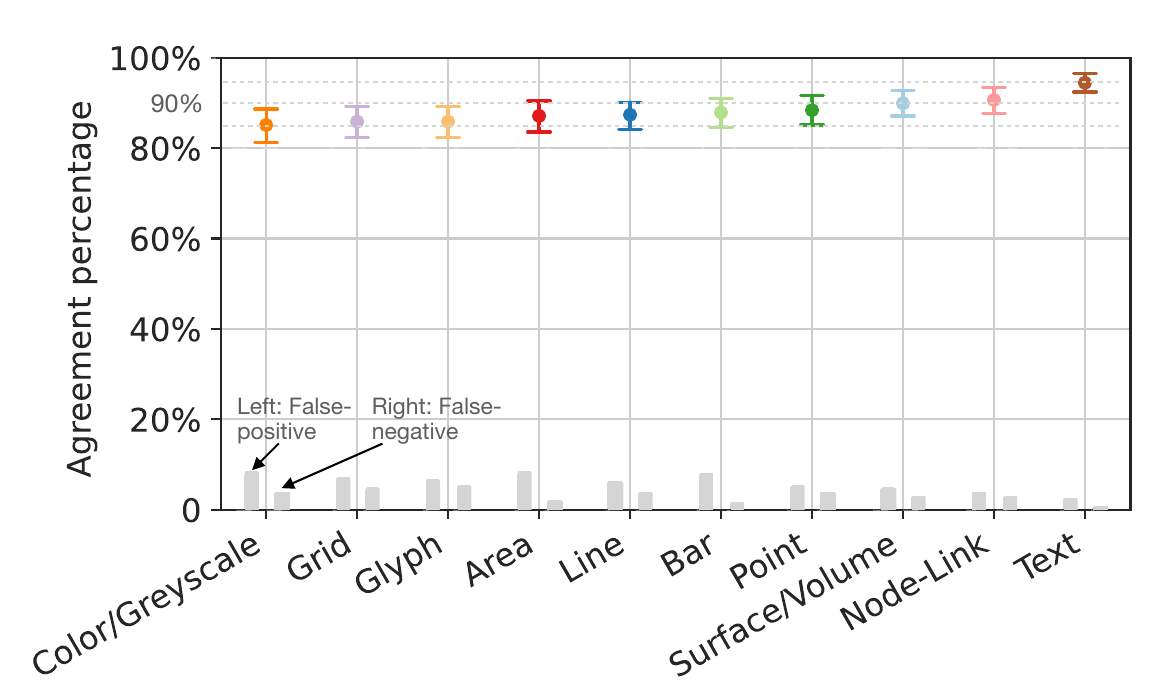}%
}
\caption{\textbf{\jctg{Empirical Results on Agreement Percentage for Each Visualization Type}}: 
the dots indicate the mean accuracy of each category; the error bars show $95\%$ confidence intervals. }
\label{fig:exp:accuracy}
\end{figure}

\jctg{Our first
hypothesis, \ie,
that participants could apply our category definitions, is supported.
Our second 
hypothesis on the difficulty levels is also reflected in the degree of agreement between participants.}
Participants could categorize images with an agreement above 80\% for all 
categories (\autoref{fig:exp:accuracy}).
\textabbr ($94.4\%$),
\nodelinkabbr ($90.7\%$), \surfacevolumeabbr ($89.9\%$), and \pointabbr ($88.3\%$) had the highest percentage of agreement. Matching our expectation, \colorabbr ($85.1\%$), 
\glyphabbr ($85.9\%$), and 
\gridabbr ($85.9\%$) scored slightly lower, and \jctg{\areaabbr ($87.1\%$), \lineabbr ($87.3\%$), and \barabbr ($87.9\%$) were in the middle.} 
\jcfr{The light-grey bars on the bottom of \autoref{fig:exp:accuracy} represent the false-positive rates (Type I error, where a viewer incorrectly affirmed an image as the target type) and the false negative rates (instances where a viewer erroneously denied an image as the target type).  Here, the categories of color\discretionary{/}{}{/}greyscale ($9.6\%$), \areaabbr ($9.6\%$), and \barabbr ($9.1\%$) had the highest false-positive rates, in contrast to grid ($8.3\%$), \glyphabbr ($7.8\%$), \pointabbr ($6.6\%$), \surfacevolumeabbr ($6.1\%$), and \nodelinkabbr ($5.3\%$). \textabbr had the lowest incidences of both false-positives ($3.5\%$) and false-negatives ($2.0\%$). Overall, \glyphabbr ($6.3\%$), \gridabbr ($5.8\%$), and \colorabbr ($5.3\%$) had the highest false negatives, further confirming our hypothesis that these types were more difficult to recognize compared to other types, such as \lineabbr ($5.0\%$), \pointabbr ($5.0\%$), \surfacevolumeabbr ($4.0\%$), \areaabbr ($3.2\%$), \barabbr ($3.0\%$), and \nodelinkabbr ($4.0\%$).}
\autoref{fig:expEvalAgree} and \autoref{fig:exp:hardest} in \appref{sm:resultsExpertExternal} show the image examples that \expertExternal with a Ph.D. degree mostly agree with our labels.

\subsubsection{\tochange{\textbf{Participants' Comments}}}
\tochange{From the 247 comments made by the  participants (\expertExternal),
we found their comments aligned well with our perceived difficulties in coding these images, as discussed in \autoref{sec:discussion}. All comments we quote in this section are from the `advanced experts' holding a `Ph.D. degree,' unless specified otherwise.}

\textit{The Influence of Domain Knowledge.}
In the coding phase, we constantly reminded ourselves to suppress our knowledge of nonvisual aspects of the images to resolve ambiguities.
We can see that a similar difficulty arose for participants 
who left comments that indicated an inclination to consider specific techniques, data, or even preexisting definitions in their coding: 
\textit{``The point-based classification I followed is based on one information visualization book. I am unfamiliar with the scientific visualization literature. If `point-based' term is also used by the scientific visualization community, my choices may be random for such a type of visualization.''}

\tochange{Some of the images would be inherently ambiguous. 
Participants commented on the background knowledge about techniques and disagreed with some example techniques appearing in the training phase. For example, 
we coded violin plots as 
\areaabbr
because a shaded area often represents a kernel density estimate (KDE), where the area under the curve is meaningful in the plot. However, a participant mentioned: 
\textit{``I would like to additionally comment violin-plots [...]
If a probability distribution is shown using a histogram (which is often done to ease implementation), then I would call it a bar chart technique. If proper KDE is computed and shown, it becomes a line-based technique (in combination with the original boxplot, \ie, bar chart technique). That's why I said no for the violin plot example.''} So here, this participant took the graphical elements of lines, which only signify boundaries, instead of the \et of \textit{area} bounded by these \textit{lines}.}

\tochange{We also saw that sometimes participants used a \emph{functional} view of an image rather than our interpretation based on \et. 
Several participants explained, for example, their encoding of a set of stacked circles as a generalized bar chart rather than a point-based representation, as we had done. 
A participant commented: \textit{``I thought that the stacked circles represent the quantity of the x-axis, so I consider it is more close to bar-chart representation.''}}

\textit{The Influence of Data.}
Several participants commented on the fact that some 
images in the study had a resolution too low to read text, indicating that they also wanted more information on some images to help them make correct decisions. 
In our own coding process, we resolved many of these issues through discussions on individual images and iterative revision of our image labels with examples and counter-examples for image coding. 
We also 
sometimes were drawn to 
text in legends and labels 
\jctg{describing the data}
to help us interpret images, especially when categorizations became difficult, even though our goal was not to consider the represented data type.

\textit{Coding Ambiguities.}
Participants also struggled with the type of meshes and surfaces.
\tochange{We had originally put meshes in the 
\surfacevolumeabbr
and later in the category of 
\nodelinkabbr
to emphasize the
connectivity rather than the underlying geometry.
The same challenge was reflected in  participants' comments, \eg:
\textit{``For my category of `surface-based or volume,' I found the differentiation to the group that contains `mesh' difficult for some of the shown examples, as surface extractions are frequently rendered as meshes, and meshes frequently depict a surface of some kind.''}}
\tochange{Another participant stated: \textit{``In CG [computer graphics] context, meshes describe a discretization of a 2D manifold. Displaying such a manifold is just rendering this discretization, since no data given on the manifold is visualized. Think of 1D: if you discretize a straight line into connected line segments, this is not a visualization of a linear graph. There is no underlying data given on that graph, except if you want to visualize the discretization.''} 
This point was reflected in our own conversation. Since the connecting line in a node-link diagram does not always map directly to underlying data, one may consider a mesh to be closer to node-links, even though it is an abstract representation of a surface.}

\textit{Glyph Coding.}
Some participants also mentioned the challenge 
to determine when a primitive becomes a glyph as described in \autoref{sec:ambiguousGlyph}.
There is no agreed-upon threshold of how many data dimensions are encoded in a glyph and when a glyph becomes a chart. 
There seems to be, however, a general consensus that a glyph requires a certain level of complexity to be categorized as such. Yet reaching a consensus on the level of complexity is challenging, as echoed by some participants in our study. Commenting on \autoref{fig:glyphOnly}, one stated that ``\textit{each arrow is the same and it is only their orientation + position that shows a property. 
I would not categorize 2--3 dimensions as
multiple
(the definition given in the study). 
In a similar way, some of the tensor fields provided in the study examples are, I believe, somewhat pushing the definition.
For the vector field, I consider the spiral as part of the glyph.''}

\tochange{One participant mentioned the challenge to label glyphs: \textit{``In doing this study I realized that glyph may have a slightly different definition when thinking about glyphs for abstract data (which typically vary in their shape) as compared with glyphs for continuous spatial data (that might just be little icons, all the same, that vary only in po\-si\-tion\discretionary{/}{}{/}orien\-ta\-tion). 
I used more of the abstract data definition.''}
Another participant wrote
\textit{``I thought that the stacked circles represent the quantity of the x-axis, so I consider it is more close to bar-chart representation.''}}

\textit{Implicit vs. explicit drawings, and continuity.} Our community uses implicit edges to characterize trees and graphs~\cite{nobre2019state, scheibel2020taxonomy}, \eg, treemaps use containment where edges are explicit~\cite{bederson2002ordered}.
An expert called out the ambiguous case between \visnodelink and \vislength, where a figure can be a ``tree'' with edges implicitly shown or a ``bar'' otherwise. The participant (who has a Ph.D. degree) stated:  
\textit{``While this could be understood as a tree structure, this tree just shows the underlying space partitioning of spherical polar coordinates... \eg, one could argue that this is 3D version of a sunburst chart and thus a generalized bar chart.''}
Part of the reason, as the participant explained \textit{``it is unclear whether the data is visualized at all (why are some of the cutoff pyramids visible and not others). So, without knowing what the domain is and what the data are, it is difficult to categorize.''}

\begin{figure}[!t]
\centering
\vspace{-2ex}
\setlength{\figverticaloffset}{11pt}
\setlength{\figureheight}{0.4\columnwidth}
\setlength{\figheightminusverticaloffset}{\figureheight}
\addtolength{\figheightminusverticaloffset}{-\figverticaloffset}
\subfloat[~\hspace{\columnwidth}~%
]
{\raisebox{-\figverticaloffset}[\figheightminusverticaloffset][-\figverticaloffset]{~~~~
\includegraphics[clip, height=0.8\figureheight]{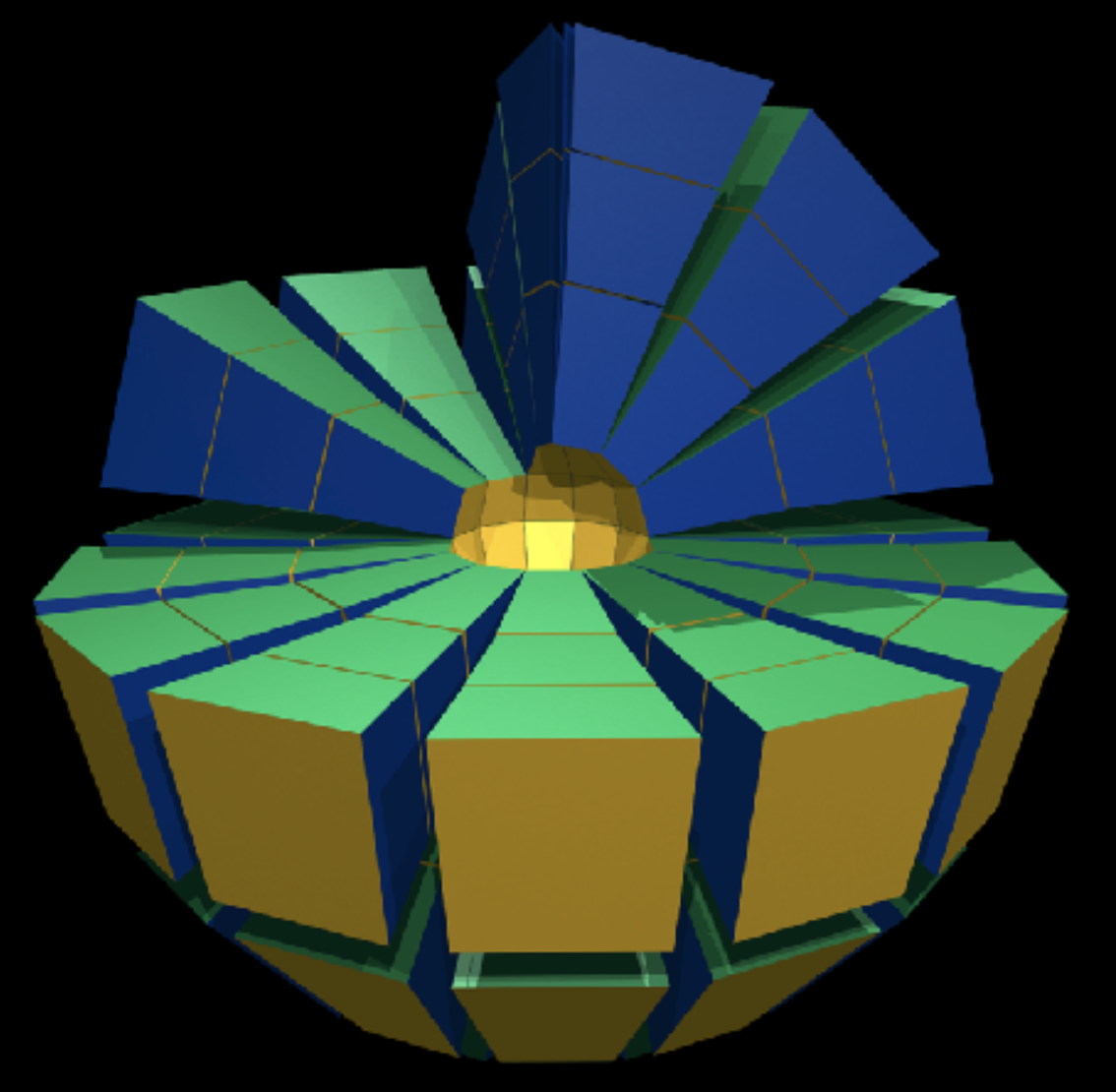}}
\label{fig:leftOnly}}
\hfill%
\subfloat[~\hspace{\columnwidth}~%
]{\raisebox{-\figverticaloffset}[\figheightminusverticaloffset][-\figverticaloffset]{~~~~~\includegraphics[height=0.8\figureheight]{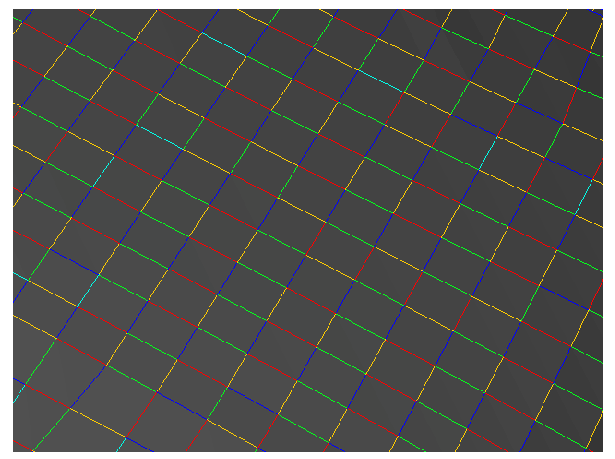}}%
\label{fig:rightOnly}}%
\vspace{-\figverticaloffset}
\caption{\textbf{Challenging cases of coding implicit and explicit stimuli.} (a) could show a tree though the edges are absent or a bar graph.  (b) is too abstract to code without the context.
Images from (a) Weiskopf et al.~\cite{weiskopf2005visualization} and (b) Suits et al.~\cite{suits2000simplification}.}
\label{fig:continuityHard}
\end{figure}

\subsection{\tochange{Summary}}

\noindent{Our findings and participant results reinforce that experts can effectively utilize our categorization methods to distinguish between different categories. 
The presence of ambiguous cases should not be seen as a drawback or a factor that undermines the validity of our typology or the study's overall results.  
Instead, we interpret this ambiguity as evidence of the diverse uses of knowledge and context prevalent in visualization representations in our community.  
In addition, categorization specifically accounts for this ambiguity and the context-specific reinterpretation of category assignments.}

\section{\tochange{Typology Use Case Scenarios}}
\label{sec:cases}

\noindent{Our image-based \typename from the visual appearance of the \et can be useful for various evaluation purposes and downstream tasks. Here, we put our theory into practice and 
conjecture
future research activities
and relevant questions.
}

\subsection{\jctg{Quantification and qualification of visualization: From point evaluation to large-scale evaluation}}
\label{sec:useEval}

\noindent
\jctg{
Measuring the effectiveness of visualization images
is at the core of visualization research. Many controlled lab experiments are by themselves ``point evaluation,'' a comparison of often elegant, yet simplified representations. Here, in order to understand why visualization techniques do or do not work, and to suggest possibilities for new design, it is important to move beyond point evaluation and to introduce generalizable scientific measurement of human visual and behavior principles. In particular, the commonly conclusion that one technique is better than the other may not always be an effective way to think about visualization design because the techniques under investigation may have removed from context uses.  
Our typology offers an opportunity to  reimagine the theory of evaluation. For example, we can broaden empirical data collection by employing a more diverse set of techniques when comparing visualization techniques.  
This broader approach ensures that evaluation recommendations are not constrained by a limited number of data points or techniques.  Moreover, leveraging large data-driven evidence in large-scale empirical studies may also enables the formation of new hypotheses that may not emerge from smaller-scale empirical studies, which often isolate variables but lack the complexity of real-world, context-rich uses. Consequently, the richness of \typename can serve as a valuable benchmark, guiding our interpretations of 
design. 
An example would be to extend Cleveland 
{and McGill}'s 1984 paper~\cite{cleveland1984graphical} to 
study a broader set of \barabbr.
}

\subsection{\jctg{Broadening cross-cutting science beyond human information-processing systems}}
\label{sec:useInfoprocesser}
\noindent
Visualization transcends traditional disciplinary boundaries, drawing significantly 
on advances in cognitive and vision science.
Our visualization images contain rich stimuli representations
that may help gauge and compare information processors in settings beyond natural environments~\cite{rensink2021visualization}.
\jctg{This trend has been reflected, for example, by understanding human-AI differences for chart interpretation in large-language models and supervised learning algorithms~\cite{jiang2022machines, bendeck2024empirical,haehn2018evaluating}. In these cases, AIs can be independent observers or collaborators side-by-side with human observers. Our typology can provide a benchmark image set to assist the tasks of quantifying observers.
}

\subsection{A supporting tool for a standard in visualization}

\noindent
Our work can also facilitate community discussions on standardizing visualization processes. 
For instance,
for the coordinate system we mentioned in \autoref{sec:intro},
\jctg{we could study how bars appear in the \barabbr 
category, observing that a Sunburst chart is just the result of positioning bars through a polar transformation. Thus, we could develop a set of transformation functions to describe the morphological structures in visual layout, 
to inform design, \eg, Wilkson's GoG
~\cite{Wilkinson:2005:GrammarOfGraphics}.}

We can use our data to understand VIS as a scientific discipline. 
\textit{Essential stimuli} enable visualization scholars like us to engage in historical analysis of visualization evolution. 
Our typology can facilitate studying how the field evolved over time and compared to other fields.
For example, we can observe and compare the evolution of rendering styles and methods in a single category, compare design styles across categories, and attempt to reason about possible influences such as individual papers, designers, or rendering hardware.  Finally, when writing an overview article, a textbook, or a lecture series on visualization, categorizations based on visual appearance can highlight the variety of approaches to specific techniques or identify aspects essential to a technique.

\subsection{Limitations and future work}
\label{sec:future}
\noindent
Our work is certainly not without limitations.
The most important one is that our categorization
is influenced by the research team behind it and the dataset we use. It is also influenced by our background and experience in visualization, the prior categorization we have attempted, and the techniques we study daily. 
As such, we release 
our codes for the community to explore the data to develop alternative categorizations. 
Furthermore,
it might be interesting, for instance, to see how D3.js \cite{bostock2011d3} and SVG~\cite{battle2018beagle, hu2019viznet}, 
and their examples
have influenced the types of images published in the community. \jctg{Finally,  it would be also useful to examine how type codes may vary considering data, tasks, and contexts.}

\section{Conclusion}

\noindent
Our image-based typology is a categorization of the visual appearance of \et in our research field
and how viewers can use these stimuli to think and drive concepts. 
The premise is audacious: the message conveyed by visualization rooted in the perception of concepts and the knowledge they contain could be the foundation for all visual sensemaking.
While scholars attempt to evaluate visual representations, bridge distinct information processors, and educate the general public and our students, \typename can bring representations into a space where ideas can be discussed, quantified, and compared.
To this end, our \typename categorize visualizations, enables designers and theorists to cope with visual complexity such that structures and forms of visualizations can be compared, quantified, and studied within and between types. \jcfr{The vast amount of image data within our community has greatly advanced our understanding of how design shapes these types for downstream tasks, enabling a deeper understanding of the message conveyed by types.
}

Our
original motivation was partially rooted in the fact that our community's intellectual contributions were, for a long time, organized in three narrow silos (formerly known as ``InfoVis,'' ``SciVis,'' and ``Visual Analytics''). Yet visualization is a much richer area, with many facets that go far beyond such a simplified consideration \cite{lee2019broadening,isenberg2016visualization}. 
Don Norman says that it is ``things that make us smart'' \cite{norman2014things}. He argues that our ability to reason about complex phenomena is only possible when we externalize our thoughts, data, and concepts. Arguably, data visualization is one of the most compact and sophisticated ways to support `thinking,' and the visualization community creates intriguing and intricate ways for such visual depictions of data. 
By attempting to categorize this rich body of work directly from its source, the images,
and the stimuli in publications, we shed further light on the breadth of our field.
Visualizations are increasingly the currency for understanding intelligent information-processing systems, and we hope that \typename can expand our understanding of visual design rules for all information processors, be they natural or artificial.

\section*{Open Access Statement}
\noindent
\vistypesgooglelink has the \typename dataset.
\osflink has the experimental details, the intermediate and final codes of the \imagesfigSeven images, and  
participants' codes.
\vistypegooglelink~\jctg{in the Suppl.Mat subfolder has 
Google CoLab code to analyze our data online} and 
recoded MASSVIS~\cite{borkin2013makes} types with our typology.
Our VisImageNavigator (\imagelink)
provides a web-based interactive tool to explore and request changes to the tagged \typename image data.
and perform image types with papers and their keywords (\autoref{fig:navigatorTwoHard}).

\vspace{-3mm}

\section*{Image Copyrights}
\noindent We, as authors, state that the images in \autoref{fig:codingprocess}--\ref{fig:dimDistribution}, 
\ref{fig:exp:accuracy}, and the bottom-row plots in \autoref{fig:teaser} (``fre\-quen\-cy\discretionary{/}{}{/}tem\-po\-ral evolution'') are and remain under our own copyright, with the permission to be used here.
They are also available under the \href{https://creativecommons.org/licenses/by/4.0/}{Creative Commons At\-tri\-bu\-tion 4.0 International (\ccLogo\,\ccAttribution\ \mbox{CC BY 4.0})} license and share them at 
\osflink.
All remaining images in the paper are \textcopyright\ IEEE, with permission to be used here.

\vspace{-3mm}
\ifCLASSOPTIONcompsoc
  \section*{Acknowledgments}
\else
  \section*{Acknowledgments}
\fi

\noindent We thank Drew Dimmery at the Hertie School for his help with our power analysis. We also thank Han-Wei Shen for being part of \expertUs in the initial coding phase.
We appreciate the expertise and comments made by
\expertExternal in our validation. We also appreciate the help of about 60 students from the Data Visualization course (CSE5544, Fall 2022/Spring 2023) at The Ohio State University, who piloted our tools. 
The work is supported in part by NSF IIS-1302755, NSF CNS-1531491, and NIST-70NANB13H181.
However, any opinions, findings, conclusions, or recommendations expressed in this material are exclusively our own.

\bibliographystyle{IEEEtranS}
\bibliography{IEEEabrv,ms}

\begin{thebibliography}{100}
\providecommand{\url}[1]{#1}
\csname url@samestyle\endcsname
\providecommand{\newblock}{\relax}
\providecommand{\bibinfo}[2]{#2}
\providecommand{\BIBentrySTDinterwordspacing}{\spaceskip=0pt\relax}
\providecommand{\BIBentryALTinterwordstretchfactor}{4}
\providecommand{\BIBentryALTinterwordspacing}{\spaceskip=\fontdimen2\font plus
\BIBentryALTinterwordstretchfactor\fontdimen3\font minus \fontdimen4\font\relax}
\providecommand{\BIBforeignlanguage}[2]{{%
\expandafter\ifx\csname l@#1\endcsname\relax
\typeout{** WARNING: IEEEtranS.bst: No hyphenation pattern has been}%
\typeout{** loaded for the language `#1'. Using the pattern for}%
\typeout{** the default language instead.}%
\else
\language=\csname l@#1\endcsname
\fi
#2}}
\providecommand{\BIBdecl}{\relax}
\BIBdecl

\bibitem{afzal2012spatial}
S.~Afzal, R.~Maciejewski, Y.~Jang, N.~Elmqvist, and D.~S. Ebert, ``Spatial text visualization using automatic typographic maps,'' \emph{IEEE Trans Vis Comput Graph}, vol.~18, no.~12, pp. 2556--2564, 2012. \href{https://doi.org/10/f4ftpq}
{doi: {{%
10\discretionary{/}{%
}{/}f4ftpq}}}


\bibitem{aigner2011visualization}
W.~Aigner, S.~Miksch, H.~Schumann, and C.~Tominski, \emph{Visualization of Time-oriented Data}.\hskip 1em plus 0.5em minus 0.4em\relax London: Springer, 2011. \href{https://doi.org/10/b834rc}
{doi: {{%
10\discretionary{/}{%
}{/}b834rc}}}


\bibitem{amar2005low}
R.~Amar, J.~Eagan, and J.~Stasko, ``Low-level components of analytic activity in information visualization,'' in \emph{Proc.\ InfoVis}.\hskip 1em plus 0.5em minus 0.4em\relax Los Alamitos: IEEE CS, 2005, pp. 111--117. \href{https://doi.org/10/bwrm27}
{doi: {{%
10\discretionary{/}{%
}{/}bwrm27}}}


\bibitem{athawale2020direct}
T.~M. Athawale, B.~Ma, E.~Sakhaee, C.~R. Johnson, and A.~Entezari, ``Direct volume rendering with nonparametric models of uncertainty,'' \emph{IEEE Trans Vis Comput Graph}, vol.~27, no.~2, pp. 1797--1807, 2021. \href{https://doi.org/10/ghgt7n}
{doi: {{%
10\discretionary{/}{%
}{/}ghgt7n}}}


\bibitem{bach2015time}
B.~Bach, C.~Shi, N.~Heulot, T.~Madhyastha, T.~Grabowski, and P.~Dragicevic, ``Time curves: Folding time to visualize patterns of temporal evolution in data,'' \emph{IEEE Trans Vis Comput Graph}, vol.~22, no.~1, pp. 559--568, 2016. \href{https://doi.org/10/gg2c52}
{doi: {{%
10\discretionary{/}{%
}{/}gg2c52}}}


\bibitem{battle2017position}
L.~Battle, R.~Chang, J.~Heer, and M.~Stonebraker, ``Position statement: The case for a visualization performance benchmark,'' in \emph{Proc.\ DSIA}.\hskip 1em plus 0.5em minus 0.4em\relax Los Alamitos: IEEE CS, 2017, pp. 6:1--6:5. \href{https://doi.org/10/gdm499}
{doi: {{%
10\discretionary{/}{%
}{/}gdm499}}}


\bibitem{battle2018beagle}
L.~Battle, P.~Duan, Z.~Miranda, D.~Mukusheva, R.~Chang, and M.~Stonebraker, ``Beagle: Automated extraction and interpretation of visualizations from the web,'' in \emph{Proc.\ CHI}.\hskip 1em plus 0.5em minus 0.4em\relax New York: ACM, 2018, pp. 594:1--594:8. \href{https://doi.org/10/ggtmxf}
{doi: {{%
10\discretionary{/}{%
}{/}ggtmxf}}}


\bibitem{bederson2002ordered}
B.~B. Bederson, B.~Shneiderman, and M.~Wattenberg, ``Ordered and quantum treemaps: Making effective use of {2D} space to display hierarchies,'' \emph{ACM Trans Graph}, vol.~21, no.~4, pp. 833--854, 2002. \href{https://doi.org/10/fw3shv}
{doi: {{%
10\discretionary{/}{%
}{/}fw3shv}}}


\bibitem{bendeck2024empirical}
A.~Bendeck and J.~Stasko, ``An empirical evaluation of the {GPT-4} multimodal language model on visualization literacy tasks,'' \emph{IEEE Trans Vis Comput Graph}, 2025, to appear. \href{https://doi.org/10/np3q}
{doi: {{%
10\discretionary{/}{%
}{/}np3q}}}


\bibitem{bergman1995rule}
L.~D. Bergman, B.~E. Rogowitz, and L.~A. Treinish, ``A rule-based tool for assisting colormap selection,'' in \emph{Proc.\ Visualization}.\hskip 1em plus 0.5em minus 0.4em\relax Los Alamitos: IEEE CS, 1995, pp. 118--125. \href{https://doi.org/10/cqp28q}
{doi: {{%
10\discretionary{/}{%
}{/}cqp28q}}}


\bibitem{bertin1983semiology}
J.~Bertin, \emph{Semiology of Graphics: Diagrams, Networks, Maps}.\hskip 1em plus 0.5em minus 0.4em\relax Esri, 1983.

\bibitem{bhaniramka2000isosurfacing}
P.~Bhaniramka, R.~Wenger, and R.~Crawfis, ``Isosurfacing in higher dimensions,'' in \emph{Proc.\ Visualization}.\hskip 1em plus 0.5em minus 0.4em\relax Los Alamitos: IEEE CS, 2000, pp. 267--273. \href{https://doi.org/10/bv4sqc}
{doi: {{%
10\discretionary{/}{%
}{/}bv4sqc}}}


\bibitem{bian2020implicit}
R.~Bian, Y.~Xue, L.~Zhou, J.~Zhang, B.~Chen, D.~Weiskopf, and Y.~Wang, ``Implicit multidimensional projection of local subspaces,'' \emph{IEEE Trans Vis Comput Graph}, vol.~27, no.~2, pp. 1558--1568, 2021. \href{https://doi.org/10/ghgt6z}
{doi: {{%
10\discretionary{/}{%
}{/}ghgt6z}}}


\bibitem{borgo2013glyph}
R.~Borgo, J.~Kehrer, D.~H. Chung, E.~Maguire, R.~S. Laramee, H.~Hauser, M.~Ward, and M.~Chen, ``Glyph-based visualization: Foundations, design guidelines, techniques and applications,'' in \emph{EG State of the Art Reports}.\hskip 1em plus 0.5em minus 0.4em\relax Goslar: EG Assoc, 2013, pp. 39--63. \href{https://doi.org/10/f3sttv}
{doi: {{%
10\discretionary{/}{%
}{/}f3sttv}}}


\bibitem{borkin2013makes}
M.~A. Borkin, A.~A. Vo, Z.~Bylinskii, P.~Isola, S.~Sunkavalli, A.~Oliva, and H.~Pfister, ``What makes a visualization memorable?'' \emph{IEEE Trans Vis Comput Graph}, vol.~19, no.~12, pp. 2306--2315, 2013. \href{https://doi.org/10/f5h3pd}
{doi: {{%
10\discretionary{/}{%
}{/}f5h3pd}}}


\bibitem{Borkin2013MASSVIS}
------, ``{MASSVIS} dataset,'' Online dataset: \url{http://massvis.mit.edu/}, visited Sept. 2024.

\bibitem{bostock2011d3}
M.~Bostock, V.~Ogievetsky, and J.~Heer, ``D\textsuperscript{3}: Data-driven documents,'' \emph{IEEE Trans Vis Comput Graph}, vol.~17, no.~12, pp. 2301--2309, 2011. \href{https://doi.org/10/b7bhhf}
{doi: {{%
10\discretionary{/}{%
}{/}b7bhhf}}}


\bibitem{Brinton:1939:GP}
W.~C. Brinton, \emph{Graphic Presentation}.\hskip 1em plus 0.5em minus 0.4em\relax New York: Brinton Assoc., 1939, \href{https://archive.org/search.php?query=external-identifier:"urn:oclc:record:1045601113"}{urn: urn\discretionary{}{:}{:}oclc\discretionary{}{:}{:}record\discretionary{}{:}{:}1045601113}.

\bibitem{Brodlie2012scientific_chap3}
K.~W. Brodlie, ``Visualization techniques,'' in \emph{Scientific Visualization: Techniques and Applications}, K.~W. Brodlie, L.~A. Carpenter, R.~A. Earnshaw, J.~R. Gallop, R.~J. Hubbold, A.~M. Mumford, C.~D. Osland, and P.~Quarendon, Eds.\hskip 1em plus 0.5em minus 0.4em\relax Berlin: Springer, 2012, ch.~3, pp. 37--85. \href{https://doi.org/10/bbrw7g}
{doi: {{%
10\discretionary{/}{%
}{/}bbrw7g}}}


\bibitem{bu2020sinestream}
C.~Bu, Q.~Zhang, Q.~Wang, J.~Zhang, M.~Sedlmair, O.~Deussen, and Y.~Wang, ``Sinestream: Improving the readability of streamgraphs by minimizing sine illusion effects,'' \emph{IEEE Trans Vis Comput Graph}, vol.~27, no.~2, pp. 1634--1643, 2021. \href{https://doi.org/10/ghv58n}
{doi: {{%
10\discretionary{/}{%
}{/}ghv58n}}}


\bibitem{byron2008stacked}
L.~Byron and M.~Wattenberg, ``Stacked graphs--geometry \& aesthetics,'' \emph{IEEE Trans Vis Comput Graph}, vol.~14, no.~6, pp. 1245--1252, 2008. \href{https://doi.org/10/dq8747}
{doi: {{%
10\discretionary{/}{%
}{/}dq8747}}}


\bibitem{cairo2012functional}
A.~Cairo, \emph{The Functional Art: An Introduction to Information Graphics and Visualization}.\hskip 1em plus 0.5em minus 0.4em\relax Berkeley: New Riders, 2012, \href{https://archive.org/search.php?query=external-identifier%3A%22urn%3Aoclc%3Arecord%3A1036960105%22}{urn: urn\discretionary{}{:}{:}oclc\discretionary{}{:}{:}record\discretionary{}{:}{:}1036960105}.

\bibitem{canny1986computational}
J.~Canny, ``A computational approach to edge detection,'' \emph{IEEE Trans Pattern Anal Mach Intell}, vol.~8, no.~6, pp. 679--698, 1986. \href{https://doi.org/10/fn3fdk}
{doi: {{%
10\discretionary{/}{%
}{/}fn3fdk}}}


\bibitem{chen2019looks}
C.~Chen, O.~Li, D.~Tao, A.~Barnett, C.~Rudin, and J.~K. Su, ``\emph{This} looks like \emph{that}: Deep learning for interpretable image recognition,'' in \emph{Proc. NeurIPS}.\hskip 1em plus 0.5em minus 0.4em\relax Red Hook, NY: Curran Associates, Inc., 2019, pp. 8930--8941. \href{https://doi.org/10/kg6p}
{doi: {{%
10\discretionary{/}{%
}{/}kg6p}}}


\bibitem{chen2006visual}
C.~Chen, F.~Ibekwe-SanJuan, E.~SanJuan, and C.~Weaver, ``Visual analysis of conflicting opinions,'' in \emph{Proc.\ VAST}.\hskip 1em plus 0.5em minus 0.4em\relax Los Alamitos: IEEE CS, 2006, pp. 59--66. \href{https://doi.org/10/cf4bwf}
{doi: {{%
10\discretionary{/}{%
}{/}cf4bwf}}}


\bibitem{Chen2020VIS30Kdata}
J.~Chen, M.~Ling, R.~Li, P.~Isenberg, T.~Isenberg, M.~Sedlmair, T.~M{\"o}ller, R.~Laramee, H.-W. Shen, K.~W{\"u}nsche, and Q.~Wang, ``{IEEE} {VIS} figures and tables image dataset,'' Dataset and online search, \href{https://visimagenavigator.github.io/}{\texttt{vi\discretionary{s}{}{s}imag\discretionary{e}{}{e}nav\discretionary{i}{}{i}gator\discretionary{.}{}{.}github\discretionary{.}{}{.}io}}, 2020. \href{https://doi.org/10/kdqd}
{doi: {{%
10\discretionary{/}{%
}{/}kdqd}}}


\bibitem{chen2021vis30k}
J.~Chen, M.~Ling, R.~Li, P.~Isenberg, T.~Isenberg, M.~Sedlmair, T.~M{\"o}ller, R.~S. Laramee, H.-W. Shen, K.~W{\"u}nsche, and Q.~Wang, ``{VIS30K}: A collection of figures and tables from {IEEE} visualization conference publications,'' \emph{IEEE Trans Vis Comput Graph}, vol.~27, no.~9, pp. 3826--3833, 2021. \href{https://doi.org/10/gmsvxd}
{doi: {{%
10\discretionary{/}{%
}{/}gmsvxd}}}


\bibitem{chlan2005multivariate}
E.~B. Chlan and P.~Rheingans, ``Multivariate glyphs for multi-object clusters,'' in \emph{Proc. InfoVis}.\hskip 1em plus 0.5em minus 0.4em\relax Los Alamitos: IEEE CS, 2005, pp. 141--148. \href{https://doi.org/10/fv24bk}
{doi: {{%
10\discretionary{/}{%
}{/}fv24bk}}}


\bibitem{claessen2011flexible}
J.~H. Claessen and J.~J. Van~Wijk, ``Flexible linked axes for multivariate data visualization,'' \emph{IEEE Trans Vis Comput Graph}, vol.~17, no.~12, pp. 2310--2316, 2011. \href{https://doi.org/10/b66s9m}
{doi: {{%
10\discretionary{/}{%
}{/}b66s9m}}}


\bibitem{cleveland1984graphs}
W.~S. Cleveland, ``Graphs in scientific publications,'' \emph{The American Statistician}, vol.~38, no.~4, pp. 261--269, 1984. \href{https://doi.org/10/c3j8hz}
{doi: {{%
10\discretionary{/}{%
}{/}c3j8hz}}}


\bibitem{cleveland1984graphical}
W.~S. Cleveland and R.~McGill, ``Graphical perception: Theory, experimentation, and application to the development of graphical methods,'' \emph{J American Stat Asso}, vol.~79, no. 387, pp. 531--554, 1984. \href{https://doi.org/gdvmwd}
{doi: {{%
gdvmwd}}}


\bibitem{cleveland1985graphical}
------, ``Graphical perception and graphical methods for analyzing scientific data,'' \emph{Science}, vol. 229, no. 4716, pp. 828--833, 1985. \href{https://doi.org/10/fkhq59}
{doi: {{%
10\discretionary{/}{%
}{/}fkhq59}}}


\bibitem{collins2009bubble}
C.~Collins, G.~Penn, and S.~Carpendale, ``Bubble sets: Revealing set relations with isocontours over existing visualizations,'' \emph{IEEE Trans Vis Comput Graph}, vol.~15, no.~6, pp. 1009--1016, 2009. \href{https://doi.org/10/c99shd}
{doi: {{%
10\discretionary{/}{%
}{/}c99shd}}}


\bibitem{dai2016unlocking}
X.~Dai and R.~Prout, ``Unlocking {S}uper {B}owl insights: Weighted word embeddings for {T}witter sentiment classification,'' in \emph{Proc.\ MISNC, SI, DS}.\hskip 1em plus 0.5em minus 0.4em\relax New York: ACM, 2016, pp. 12:1--12:6. \href{https://doi.org/10/mk6h}
{doi: {{%
10\discretionary{/}{%
}{/}mk6h}}}


\bibitem{daniels2010interactive}
J.~Daniels, II, E.~W. Anderson, L.~G. Nonato, and C.~T. Silva, ``Interactive vector field feature identification,'' \emph{IEEE Trans Vis Comput Graph}, vol.~16, no.~6, pp. 1560--1568, 2010. \href{https://doi.org/10/fq6d7n}
{doi: {{%
10\discretionary{/}{%
}{/}fq6d7n}}}


\bibitem{dasu2020sea}
K.~Dasu, K.-L. Ma, J.~Ma, and J.~Frazier, ``{Sea of Genes}: A reflection on visualising metagenomic data for museums,'' \emph{IEEE Trans Vis Comput Graph}, vol.~27, no.~2, pp. 935--945, 2021. \href{https://doi.org/10/ghv58j}
{doi: {{%
10\discretionary{/}{%
}{/}ghv58j}}}


\bibitem{deng2020visimages}
D.~Deng, Y.~Wu, X.~Shu, M.~Xu, J.~Wu, and S.~F.~Y. Wu, ``{VisImages}: A large-scale, high-quality image corpus in visualization publications,'' \emph{IEEE Trans Vis Comput Graph}, vol.~29, no.~7, pp. 3298--3311, 2023. \href{https://doi.org/10/k8vt}
{doi: {{%
10\discretionary{/}{%
}{/}k8vt}}}


\bibitem{dimara2021critical}
E.~Dimara and J.~Stasko, ``A critical reflection on visualization research: Where do decision making tasks hide?'' \emph{IEEE Trans Vis Comput Graph}, vol.~28, no.~1, pp. 1128--1138, 2022. \href{https://doi.org/10/gnb9b2}
{doi: {{%
10\discretionary{/}{%
}{/}gnb9b2}}}


\bibitem{edmunds:surface}
M.~Edmunds, R.~S. Laramee, G.~Chen, N.~Max, E.~Zhang, and C.~Ware, ``Surface-based flow visualization,'' \emph{Comput Graph}, vol.~36, no.~8, pp. 974--990, 2012. \href{https://doi.org/10/f4mhfp}
{doi: {{%
10\discretionary{/}{%
}{/}f4mhfp}}}


\bibitem{fanea2005interactive}
E.~Fanea, S.~Carpendale, and T.~Isenberg, ``An interactive {3D} integration of parallel coordinates and star glyphs,'' in \emph{Proc.\ InfoVis}.\hskip 1em plus 0.5em minus 0.4em\relax Los Alamitos: IEEE CS, 2005, pp. 149--156. \href{https://doi.org/10/fvdg5f}
{doi: {{%
10\discretionary{/}{%
}{/}fvdg5f}}}


\bibitem{felzenszwalb2010object}
P.~F. Felzenszwalb, R.~B. Girshick, D.~McAllester, and D.~Ramanan, ``Object detection with discriminatively trained part-based models,'' \emph{IEEE Trans Pattern Anal Mach Intell}, vol.~32, no.~9, pp. 1627--1645, 2010. \href{https://doi.org/10/fgv7fd}
{doi: {{%
10\discretionary{/}{%
}{/}fgv7fd}}}


\bibitem{feng2020topology}
Z.~Feng, H.~Li, W.~Zeng, S.-H. Yang, and H.~Qu, ``Topology density map for urban data visualization and analysis,'' \emph{IEEE Trans Vis Comput Graph}, vol.~27, no.~2, pp. 828--838, 2021. \href{https://doi.org/10/k8sf}
{doi: {{%
10\discretionary{/}{%
}{/}k8sf}}}


\bibitem{Fisher:2018:MDV}
D.~Fisher and M.~Meyer, \emph{Making Data Visual}.\hskip 1em plus 0.5em minus 0.4em\relax O' Reilly Media, 2018.

\bibitem{franke2020modern}
L.~Franke and D.~Haehn, ``Modern scientific visualizations on the web,'' \emph{Inf}, vol.~7, no.~4, pp. 37:1--37:35, 2020. \href{https://doi.org/ghnnm2}
{doi: {{%
ghnnm2}}}


\bibitem{fuchs2013evaluation}
J.~Fuchs, F.~Fischer, F.~Mansmann, E.~Bertini, and P.~Isenberg, ``Evaluation of alternative glyph designs for time series data in a small multiple setting,'' in \emph{Proc.\ CHI}.\hskip 1em plus 0.5em minus 0.4em\relax New York: ACM, 2013, pp. 3237--3246. \href{https://doi.org/10/gh52rn}
{doi: {{%
10\discretionary{/}{%
}{/}gh52rn}}}


\bibitem{garcke2000continuous}
H.~Garcke, T.~Preu{\ss}er, M.~Rumpf, A.~Telea, U.~Weikard, and J.~van Wijk, ``A continuous clustering method for vector fields,'' in \emph{Proc.\ Visualization}.\hskip 1em plus 0.5em minus 0.4em\relax Los Alamitos: IEEE CS, 2000, pp. 351--358. \href{https://doi.org/10/bft6b9}
{doi: {{%
10\discretionary{/}{%
}{/}bft6b9}}}


\bibitem{givon1986prototypes}
T.~Giv{\'o}n, ``Prototypes: Between {P}lato and {W}ittgenstein,'' in \emph{Noun Classes and Categorization}, C.~Craig, Ed.\hskip 1em plus 0.5em minus 0.4em\relax Amsterdam: John Benjamins Publishing, 1986, pp. 77--102, \href{https://archive.org/search.php?query=external-identifier:"urn:oclc:record:1245623243"}{urn: urn\discretionary{}{:}{:}oclc\discretionary{}{:}{:}record\discretionary{}{:}{:}1245623243}.

\bibitem{gregorski:adaptive}
B.~F. Gregorski, J.~G. Senecal, M.~A. Duchaineau, and K.~I. Joy, ``Adaptive extraction of time-varying isosurfaces,'' \emph{IEEE Trans Vis Comput Graph}, vol.~10, no.~6, pp. 683--694, 2004. \href{https://doi.org/10/cn3st4}
{doi: {{%
10\discretionary{/}{%
}{/}cn3st4}}}


\bibitem{gresh2000weave}
D.~L. Gresh, B.~E. Rogowitz, R.~L. Winslow, D.~F. Scollan, and C.~K. Yung, ``{WEAVE}: A system for visually linking {3-D} and statistical visualizations applied to cardiac simulation and measurement data,'' in \emph{Proc.\ Visualization}.\hskip 1em plus 0.5em minus 0.4em\relax Los Alamitos: IEEE CS, 2000, pp. 489--492. \href{https://doi.org/10/djvdvk}
{doi: {{%
10\discretionary{/}{%
}{/}djvdvk}}}


\bibitem{haehn2018evaluating}
D.~Haehn, J.~Tompkin, and H.~Pfister, ``Evaluating ‘graphical perception’ with {CNN}s,'' \emph{IEEE Trans Vis Comput Graph}, vol.~25, no.~1, pp. 641--650, 2019. \href{https://doi.org/10/gd52dt}
{doi: {{%
10\discretionary{/}{%
}{/}gd52dt}}}


\bibitem{He:2023:BeauVis}
T.~He, P.~Isenberg, R.~Dachselt, and T.~Isenberg, ``{BeauVis}: A validated scale for measuring the aesthetic pleasure of visual representations,'' \emph{IEEE Trans Vis Comput Graph}, vol.~29, no.~1, pp. 363--373, 2023. \href{https://doi.org/10/kt3n}
{doi: {{%
10\discretionary{/}{%
}{/}kt3n}}}


\bibitem{Heer:2010:TVZ}
J.~Heer, M.~Bostock, and V.~Ogievetsky, ``A tour through the visualization zoo: A survey of powerful visualization techniques, from the obvious to the obscure,'' \emph{Queue}, vol.~8, no.~5, pp. 20--30, 2010. \href{https://doi.org/10/k8sw}
{doi: {{%
10\discretionary{/}{%
}{/}k8sw}}}


\bibitem{helman1990surface}
J.~L. Helman and L.~Hesselink, ``Surface representations of two-and three-dimensional fluid flow topology,'' in \emph{Proc. Visualization}.\hskip 1em plus 0.5em minus 0.4em\relax Los Alamitos: IEEE CS, 1990, pp. 6--13. \href{https://doi.org/10/cs4s66}
{doi: {{%
10\discretionary{/}{%
}{/}cs4s66}}}


\bibitem{hlawatsch2011flow}
M.~Hlawatsch, P.~Leube, W.~Nowak, and D.~Weiskopf, ``Flow radar glyphs---{S}tatic visualization of unsteady flow with uncertainty,'' \emph{IEEE Trans Vis Comput Graph}, vol.~17, no.~12, pp. 1949--1958, 2011. \href{https://doi.org/10/cvc442}
{doi: {{%
10\discretionary{/}{%
}{/}cvc442}}}


\bibitem{hlawitschka2005hot}
M.~Hlawitschka and G.~Scheuermann, ``{HOT}-lines: Tracking lines in higher order tensor fields,'' in \emph{Proc.\ VIS}.\hskip 1em plus 0.5em minus 0.4em\relax Los Alamitos: IEEE CS, 2005, pp. 27--34. \href{https://doi.org/10/c77b3m}
{doi: {{%
10\discretionary{/}{%
}{/}c77b3m}}}


\bibitem{hu2019viznet}
K.~Hu, S.~Gaikwad, M.~Hulsebos, M.~A. Bakker, E.~Zgraggen, C.~Hidalgo, T.~Kraska, G.~Li, A.~Satyanarayan, and {\c{C}}.~Demiralp, ``{VizNet}: Towards a large-scale visualization learning and benchmarking repository,'' in \emph{Proc.\ CHI}.\hskip 1em plus 0.5em minus 0.4em\relax New York: ACM, 2019, pp. 662:1--662:12. \href{https://doi.org/10/gf2b85}
{doi: {{%
10\discretionary{/}{%
}{/}gf2b85}}}


\bibitem{hummel2010iris}
M.~Hummel, C.~Garth, B.~Hamann, H.~Hagen, and K.~I. Joy, ``Iris: Illustrative rendering for integral surfaces,'' \emph{IEEE Trans Vis Comput Graph}, vol.~16, no.~6, pp. 1319--1328, 2010. \href{https://doi.org/10/d5xx88}
{doi: {{%
10\discretionary{/}{%
}{/}d5xx88}}}


\bibitem{ingram2010dimstiller}
S.~Ingram, T.~Munzner, V.~Irvine, M.~Tory, S.~Bergner, and T.~M{\"o}ller, ``Dim{S}tiller: Workflows for dimensional analysis and reduction,'' in \emph{Proc.\ VAST}.\hskip 1em plus 0.5em minus 0.4em\relax Los Alamitos: IEEE CS, 2010, pp. 3--10. \href{https://doi.org/10/bhcd6v}
{doi: {{%
10\discretionary{/}{%
}{/}bhcd6v}}}


\bibitem{isenberg2010exploratory}
P.~Isenberg, D.~Fisher, M.~R. Morris, K.~Inkpen, and M.~Czerwinski, ``An exploratory study of co-located collaborative visual analytics around a tabletop display,'' in \emph{Proc.\ VAST}.\hskip 1em plus 0.5em minus 0.4em\relax Los Alamitos: IEEE CS, 2010, pp. 179--186. \href{https://doi.org/10/cpc8vv}
{doi: {{%
10\discretionary{/}{%
}{/}cpc8vv}}}


\bibitem{isenberg2016vispubdata}
P.~Isenberg, F.~Heimerl, S.~Koch, T.~Isenberg, P.~Xu, C.~D. Stolper, M.~Sedlmair, J.~Chen, T.~M{\"o}ller, and J.~Stasko, ``Vispubdata.org: A metadata collection about {IEEE} visualization ({VIS}) publications,'' \emph{IEEE Trans Vis Comput Graph}, vol.~23, no.~9, pp. 2199--2206, 2017. \href{https://doi.org/10/ggwwrv}
{doi: {{%
10\discretionary{/}{%
}{/}ggwwrv}}}


\bibitem{isenberg2016visualization}
P.~Isenberg, T.~Isenberg, M.~Sedlmair, J.~Chen, and T.~M{\"o}ller, ``Visualization as seen through its research paper keywords,'' \emph{IEEE Trans Vis Comput Graph}, vol.~23, no.~1, pp. 771--780, 2017. \href{https://doi.org/10/f92gps}
{doi: {{%
10\discretionary{/}{%
}{/}f92gps}}}


\bibitem{isenberg2013systematic}
T.~Isenberg, P.~Isenberg, J.~Chen, M.~Sedlmair, and T.~M{\"o}ller, ``A systematic review on the practice of evaluating visualization,'' \emph{IEEE Trans Vis Comput Graph}, vol.~19, no.~12, pp. 2818--2827, 2013. \href{https://doi.org/10/f5h29z}
{doi: {{%
10\discretionary{/}{%
}{/}f5h29z}}}


\bibitem{Jacob:2004:Classification}
E.~K. Jacob, ``Classification and categorization: A difference that makes a difference,'' \emph{Lib Trends}, vol.~52, no.~3, pp. 515--540, 2004, \href{http://hdl.handle.net/2142/1686}{hdl: 2142/1686}.

\bibitem{jenny2020cartographic}
B.~Jenny, M.~Heitzler, D.~Singh, M.~Farmakis-Serebryakova, J.~C. Liu, and L.~Hurni, ``Cartographic relief shading with neural networks,'' \emph{IEEE Trans Vis Comput Graph}, vol.~27, no.~2, pp. 1225--1235, 2020. \href{https://doi.org/10/ghv58f}
{doi: {{%
10\discretionary{/}{%
}{/}ghv58f}}}


\bibitem{jeong2010interactive}
W.-K. Jeong, J.~Schneider, S.~Turney, B.~E. Faulkner-Jones, D.~Meyer, R.~Westermann, R.~C. Reid, J.~Lichtman, and H.~Pfister, ``Interactive histology of large-scale biomedical image stacks,'' \emph{IEEE Trans Vis Comput Graph}, vol.~16, no.~6, pp. 1386--1395, 2010. \href{https://doi.org/10/fc7rd9}
{doi: {{%
10\discretionary{/}{%
}{/}fc7rd9}}}


\bibitem{ji2019visual}
X.~Ji, H.-W. Shen, A.~Ritter, R.~Machiraju, and P.-Y. Yen, ``Visual exploration of neural document embedding in information retrieval: Semantics and feature selection,'' \emph{IEEE Trans Vis Comput Graph}, vol.~25, no.~6, pp. 2181--2192, 2019. \href{https://doi.org/10/ggjbsg}
{doi: {{%
10\discretionary{/}{%
}{/}ggjbsg}}}


\bibitem{jiang2022machines}
S.~Jiang, W.-L. Chao, J.~Chen, D.~Haehn, M.~Ling, C.~Shang, and H.~Pfister, ``Are machines more effective than humans for graphical perception tasks?'' \emph{J Vision}, vol.~22, no.~14, pp. 3784--3784, 2022. \href{https://doi.org/10/nkgk}
{doi: {{%
10\discretionary{/}{%
}{/}nkgk}}}


\bibitem{Kehrer2013a}
J.~Kehrer and H.~Hauser, ``Visualization and visual analysis of multifaceted scientific data: A survey,'' \emph{IEEE Trans Vis Comput Graph}, vol.~19, no.~3, pp. 495--513, 2013. \href{https://doi.org/10/f4kwr6}
{doi: {{%
10\discretionary{/}{%
}{/}f4kwr6}}}


\bibitem{keim2000designing}
D.~A. Keim, ``Designing pixel-oriented visualization techniques: Theory and applications,'' \emph{IEEE Trans Vis Comput Graph}, vol.~6, no.~1, pp. 59--78, 2000. \href{https://doi.org/10/fnqrw9}
{doi: {{%
10\discretionary{/}{%
}{/}fnqrw9}}}


\bibitem{kim2010evaluating}
S.~Kim, I.~Woo, R.~Maciejewski, D.~S. Ebert, T.~D. Ropp, and K.~Thomas, ``Evaluating the effectiveness of visualization techniques for schematic diagrams in maintenance tasks,'' in \emph{Proc.\ APGV}.\hskip 1em plus 0.5em minus 0.4em\relax New York: ACM, 2010, pp. 33--40. \href{https://doi.org/10/bs23ww}
{doi: {{%
10\discretionary{/}{%
}{/}bs23ww}}}


\bibitem{Kim:2021:Githru}
Y.~Kim, J.~Kim, H.~Jeon, Y.-H. Kim, H.~Song, B.~Kim, and J.~Seo, ``Githru: Visual analytics for understanding software development history through {G}it metadata analysis,'' \emph{IEEE Trans Vis Comput Graph}, vol.~27, no.~2, pp. 656--666, 2021. \href{https://doi.org/10/gmghvf}
{doi: {{%
10\discretionary{/}{%
}{/}gmghvf}}}


\bibitem{prins2016psychophysics}
F.~A.~A. Kingdom and N.~Prins, \emph{Psychophysics: A Practical Introduction}, 2nd~ed.\hskip 1em plus 0.5em minus 0.4em\relax Amsterdam: Academic Press, 2016. \href{https://doi.org/10/k8r7}
{doi: {{%
10\discretionary{/}{%
}{/}k8r7}}}


\bibitem{koesten2023subjective}
L.~Koesten, D.~Dimmery, M.~Gleicher, and T.~M{\"o}ller, ``Subjective visualization experiences: Impact of visual design and experimental design,'' U. Wien, arXiv preprint 2310.13713, 2023. \href{https://doi.org/10/k8sc}
{doi: {{%
10\discretionary{/}{%
}{/}k8sc}}}


\bibitem{kruschke2020alcove}
J.~K. Kruschke, ``{ALCOVE}: An exemplar-based connectionist model of category learning,'' \emph{Psychol Rev}, vol.~99, no.~1, pp. 22--44, 1992. \href{https://doi.org/10/ftqx2m}
{doi: {{%
10\discretionary{/}{%
}{/}ftqx2m}}}


\bibitem{lakare20003d}
S.~Lakare, M.~Wan, M.~Sato, and A.~Kaufman, ``3d digital cleansing using segmentation rays,'' in \emph{Proc.\ VIS}.\hskip 1em plus 0.5em minus 0.4em\relax Los Alamitos: IEEE CS, 2000, pp. 37--44. \href{https://doi.org/10/d7w9n3}
{doi: {{%
10\discretionary{/}{%
}{/}d7w9n3}}}


\bibitem{lam2011empirical}
H.~Lam, E.~Bertini, P.~Isenberg, C.~Plaisant, and S.~Carpendale, ``Empirical studies in information visualization: Seven scenarios,'' \emph{IEEE Trans Vis Comput Graph}, vol.~18, no.~9, pp. 1520--1536, 2012. \href{https://doi.org/10/drrh6j}
{doi: {{%
10\discretionary{/}{%
}{/}drrh6j}}}


\bibitem{latour2012visualisation}
B.~Latour, ``Wizualizacja i poznanie: zrysowywanie rzeczy razem ({V}isualisation and cognition: Drawing things together),'' \emph{Avant Trends Interdiscip Stud}, vol.~3, no.~T, pp. 207--257, 2012, online: \href{https://philpapers.org/rec/LATVAC-3}{philpapers\discretionary{}{.}{.}org\discretionary{/}{}{/}rec\discretionary{/}{}{/}LATVAC\discretionary{}{-}{-}3}.

\bibitem{lee2020data}
B.~Lee, D.~Brown, B.~Lee, C.~Hurter, S.~Drucker, and T.~Dwyer, ``Data visceralization: Enabling deeper understanding of data using virtual reality,'' \emph{IEEE Trans Vis Comput Graph}, vol.~27, no.~2, pp. 1095--1105, 2021. \href{https://doi.org/10/ghv58h}
{doi: {{%
10\discretionary{/}{%
}{/}ghv58h}}}


\bibitem{lee2019broadening}
B.~Lee, K.~Isaacs, D.~A. Szafir, G.~E. Marai, C.~Turkay, M.~Tory, S.~Carpendale, and A.~Endert, ``Broadening intellectual diversity in visualization research papers,'' \emph{IEEE Comput Graph Appl}, vol.~39, no.~4, pp. 78--85, 2019. \href{https://doi.org/10/gg2j2m}
{doi: {{%
10\discretionary{/}{%
}{/}gg2j2m}}}


\bibitem{lee2010sparkclouds}
B.~Lee, N.~H. Riche, A.~K. Karlson, and S.~Carpendale, ``Sparkclouds: Visualizing trends in tag clouds,'' \emph{IEEE Trans Vis Comput Graph}, vol.~16, no.~6, pp. 1182--1189, 2010. \href{https://doi.org/10/fv5z96}
{doi: {{%
10\discretionary{/}{%
}{/}fv5z96}}}


\bibitem{li2018toward}
R.~Li and J.~Chen, ``Toward a deep understanding of what makes a scientific visualization memorable,'' in \emph{Short Papers of IEEE Visualization/SciVis}.\hskip 1em plus 0.5em minus 0.4em\relax Los Alamitos: IEEE CS, 2018, pp. 26--31. \href{https://doi.org/10/gr633t}
{doi: {{%
10\discretionary{/}{%
}{/}gr633t}}}


\bibitem{lohse1994classification}
G.~L. Lohse, K.~Biolsi, N.~Walker, and H.~H. Rueter, ``{A Classification of Visual Representations},'' \emph{Commun ACM}, vol.~37, no.~12, pp. 36--50, 1994. \href{https://doi.org/10/b92kbj}
{doi: {{%
10\discretionary{/}{%
}{/}b92kbj}}}


\bibitem{lu2024VAID}
Y.~Lu, A.~Wu, H.~Li, Z.~Deng, J.~Lan, J.~Wu, Y.~Wang, H.~Qu, D.~Deng, and Y.~Wu, ``{VAID}: Indexing view designs in visual analytics system,'' in \emph{Proc.\ CHI}.\hskip 1em plus 0.5em minus 0.4em\relax New York: ACM, 2024. \href{https://doi.org/doi.org/mj5d}
{doi: {{%
doi\hspace{.1pt}\discretionary{.}{%
}{.}\hspace{.4pt}org\discretionary{/}{%
}{/}mj5d}}}


\bibitem{Maciejewski:2011:DRT}
R.~Maciejewski, \emph{Data Representations, Transformations, and Statistics for Visual Reasoning}.\hskip 1em plus 0.5em minus 0.4em\relax Cham: Springer, 2011. \href{https://doi.org/10/d4565q}
{doi: {{%
10\discretionary{/}{%
}{/}d4565q}}}


\bibitem{mackinlay1986automating}
J.~Mackinlay, ``Automating the design of graphical presentations of relational information,'' \emph{ACM Trans Graph}, vol.~5, no.~2, pp. 110--141, 1986. \href{https://doi.org/10/dxdkkp}
{doi: {{%
10\discretionary{/}{%
}{/}dxdkkp}}}


\bibitem{mackinlay2007show}
J.~Mackinlay, P.~Hanrahan, and C.~Stolte, ``Show me: Automatic presentation for visual analysis,'' \emph{IEEE Trans Vis Comput Graph}, vol.~13, no.~6, pp. 1137--1144, 2007. \href{https://doi.org/10/fgwbh9}
{doi: {{%
10\discretionary{/}{%
}{/}fgwbh9}}}


\bibitem{malik1997computing}
J.~Malik and R.~Rosenholtz, ``Computing local surface orientation and shape from texture for curved surfaces,'' \emph{Int J Comput Vision}, vol.~23, no.~2, pp. 149--168, 1997. \href{https://doi.org/10/cpx8m4}
{doi: {{%
10\discretionary{/}{%
}{/}cpx8m4}}}


\bibitem{malisiewicz2008recognition}
T.~Malisiewicz and A.~A. Efros, ``Recognition by association via learning per-exemplar distances,'' in \emph{Proc.\ CVPR}.\hskip 1em plus 0.5em minus 0.4em\relax Los Alamitos: IEEE CS, 2008, pp. 42:1--42:8. \href{https://doi.org/10/c73x5z}
{doi: {{%
10\discretionary{/}{%
}{/}c73x5z}}}


\bibitem{medin1978context}
D.~L. Medin and M.~M. Schaffer, ``Context theory of classification learning.'' \emph{Psychol Rev}, vol.~85, no.~3, pp. 207--238, 1978. \href{https://doi.org/10/b4wm44}
{doi: {{%
10\discretionary{/}{%
}{/}b4wm44}}}


\bibitem{meulemans2020simple}
W.~Meulemans, M.~Sondag, and B.~Speckmann, ``A simple pipeline for coherent grid maps,'' \emph{IEEE Trans Vis Comput Graph}, vol.~27, no.~2, pp. 1236--1246, 2021. \href{https://doi.org/10/ghgt53}
{doi: {{%
10\discretionary{/}{%
}{/}ghgt53}}}


\bibitem{moritz2018formalizing}
D.~Moritz, C.~Wang, G.~L. Nelson, H.~Lin, A.~M. Smith, B.~Howe, and J.~Heer, ``Formalizing visualization design knowledge as constraints: Actionable and extensible models in draco,'' \emph{IEEE Trans Vis Comput Graph}, vol.~25, no.~1, pp. 438--448, 2019. \href{https://doi.org/10/cs68}
{doi: {{%
10\discretionary{/}{%
}{/}cs68}}}


\bibitem{munzner2014visualization}
T.~Munzner, \emph{Visualization Analysis and Design}.\hskip 1em plus 0.5em minus 0.4em\relax New York: CRC Press, 2014. \href{https://doi.org/10/gd3xgq}
{doi: {{%
10\discretionary{/}{%
}{/}gd3xgq}}}


\bibitem{nguyen2021modeling}
N.~Nguyen, O.~Strnad, T.~Klein, D.~Luo, R.~Alharbi, P.~Wonka, M.~Maritan, P.~Mindek, L.~Autin, D.~S. Goodsell, and I.~Viola, ``Modeling in the time of {COVID}-19: Statistical and rule-based mesoscale models,'' \emph{IEEE Trans Vis Comput Graph}, vol.~27, no.~2, pp. 722--732, 2021. \href{https://doi.org/10/k8sh}
{doi: {{%
10\discretionary{/}{%
}{/}k8sh}}}


\bibitem{Nielson:1991:TAD}
G.~M. Nielson and B.~Hamann, ``The asymptotic decider: Removing the ambiguity in marching cubes,'' in \emph{Proc.\ Visualization}.\hskip 1em plus 0.5em minus 0.4em\relax Los Alamitos: IEEE CS, 1991, pp. 83--91. \href{https://doi.org/10/fg4d6v}
{doi: {{%
10\discretionary{/}{%
}{/}fg4d6v}}}


\bibitem{nobre2019state}
C.~Nobre, M.~Meyer, M.~Streit, and A.~Lex, ``The state of the art in visualizing multivariate networks,'' \emph{Comput Graph Forum}, vol.~38, no.~3, pp. 807--832, 2019. \href{https://doi.org/10/ghpp2g}
{doi: {{%
10\discretionary{/}{%
}{/}ghpp2g}}}


\bibitem{norman2014things}
D.~Norman, \emph{Things That Make Us Smart: Defending Human Attributes in the Age of the Machine}.\hskip 1em plus 0.5em minus 0.4em\relax New York: Diversion Books, 2014, \href{https://archive.org/search.php?query=external-identifier%3A%22urn%3Aoclc%3Arecord%3A1036960105%22}{urn: urn\discretionary{}{:}{:}oclc\discretionary{}{:}{:}record\discretionary{}{:}{:}1036960105}.

\bibitem{nosofsky1986attention}
R.~M. Nosofsky, ``Attention, similarity, and the identification-ca\-te\-go\-ri\-za\-tion relationship,'' \emph{J Exp Psychol Anim Learn Cognit: Gener}, vol. 115, no.~1, pp. 39--57, 1986. \href{https://doi.org/10/bjgj3w}
{doi: {{%
10\discretionary{/}{%
}{/}bjgj3w}}}


\bibitem{overbye2000new}
T.~J. Overbye and J.~D. Weber, ``New methods for the visualization of electric power system information,'' in \emph{Proc.\ InfoVis}, Los Alamitos, 2000, pp. 131--136. \href{https://doi.org/10/fdk73d}
{doi: {{%
10\discretionary{/}{%
}{/}fdk73d}}}


\bibitem{palmer1978fundamental}
S.~Palmer, ``Fundamental aspects of cognitive representation,'' in \emph{Cognition and Categorization}, E.~Rosch and B.~Lloyd, Eds., 1978, ch.~9, pp. 259--303, \href{https://archive.org/search.php?query=external-identifier%3A%22urn%3Aoclc%3Arecord%3A1409364022%22}{urn: urn\discretionary{}{:}{:}oclc\discretionary{}{:}{:}record\discretionary{}{:}{:}1409364022}.

\bibitem{phan2005flow}
D.~Phan, L.~Xiao, R.~Yeh, and P.~Hanrahan, ``Flow map layout,'' in \emph{Proc.\ InfoVis}.\hskip 1em plus 0.5em minus 0.4em\relax Los Alamitos: IEEE CS, 2005, pp. 219--224. \href{https://doi.org/10/fq44hs}
{doi: {{%
10\discretionary{/}{%
}{/}fq44hs}}}


\bibitem{potter2004pictorial}
M.~C. Potter, A.~Staub, and D.~H. O'Connor, ``Pictorial and conceptual representation of glimpsed pictures,'' \emph{J. Exp. Psychol Hum Percept Perform}, vol.~30, no.~3, p. 478, 2004. \href{https://doi.org/10/c26zf9}
{doi: {{%
10\discretionary{/}{%
}{/}c26zf9}}}


\bibitem{Pu:2000:AlgorithmVis}
P.~Pu and D.~Lalanne, ``Interactive problem solving via algorithm visualization,'' in \emph{Proc.\ InfoVis}.\hskip 1em plus 0.5em minus 0.4em\relax Los Alamitos: IEEE CS, 2000, pp. 145--153. \href{https://doi.org/10/bh73bn}
{doi: {{%
10\discretionary{/}{%
}{/}bh73bn}}}


\bibitem{rees:survey}
D.~Rees and R.~Laramee, ``A survey of information visualization books,'' \emph{Comput Graph Forum}, vol.~38, no.~1, pp. 610--646, 2019. \href{https://doi.org/10/gftgv6}
{doi: {{%
10\discretionary{/}{%
}{/}gftgv6}}}


\bibitem{rensink2021visualization}
R.~A. Rensink, ``Visualization as a stimulus domain for vision science,'' \emph{Journal of Vision}, vol.~21, no.~8, pp. 3--3, 2021. \href{https://doi.org/10/mg29}
{doi: {{%
10\discretionary{/}{%
}{/}mg29}}}


\bibitem{rheingans1995interactive}
P.~Rheingans, M.~Marietta, and J.~Nichols, ``Interactive {3D} visualization of actual anatomy and simulated chemical time-course data for fish,'' in \emph{Proc.\ Visualization}.\hskip 1em plus 0.5em minus 0.4em\relax Los Alamitos: IEEE CS, 1995, pp. 393--396. \href{https://doi.org/10/bst9vq}
{doi: {{%
10\discretionary{/}{%
}{/}bst9vq}}}


\bibitem{rosch1973natural}
E.~H. Rosch, ``Natural categories,'' \emph{Cognit Psychol}, vol.~4, no.~3, pp. 328--350, 1973. \href{https://doi.org/10/cj6qvb}
{doi: {{%
10\discretionary{/}{%
}{/}cj6qvb}}}


\bibitem{sacha2016visual}
D.~Sacha, L.~Zhang, M.~Sedlmair, J.~A. Lee, J.~Peltonen, D.~Weiskopf, S.~C. North, and D.~A. Keim, ``Visual interaction with dimensionality reduction: A structured literature analysis,'' \emph{IEEE Trans Vis Comput Graph}, vol.~23, no.~1, pp. 241--250, 2017. \href{https://doi.org/10/f9zvx5}
{doi: {{%
10\discretionary{/}{%
}{/}f9zvx5}}}


\bibitem{satyanarayan2016vega}
A.~Satyanarayan, D.~Moritz, K.~Wongsuphasawat, and J.~Heer, ``{V}ega-{L}ite: A grammar of interactive graphics,'' \emph{IEEE Trans Vis Comput Graph}, vol.~23, no.~1, pp. 341--350, 2017. \href{https://doi.org/10/f92f32}
{doi: {{%
10\discretionary{/}{%
}{/}f92f32}}}


\bibitem{savva2011revision}
M.~Savva, N.~Kong, A.~Chhajta, L.~Fei-Fei, M.~Agrawala, and J.~Heer, ``{ReVision}: Automated classification, analysis and redesign of chart images,'' in \emph{Proc. UIST}.\hskip 1em plus 0.5em minus 0.4em\relax New York: ACM, 2011, pp. 393--402. \href{https://doi.org/10/bcbnfw}
{doi: {{%
10\discretionary{/}{%
}{/}bcbnfw}}}


\bibitem{scheibel2020taxonomy}
W.~Scheibel, M.~Trapp, D.~Limberger, and J.~D{\"o}llner, ``A taxonomy of treemap visualization techniques.'' in \emph{VISIGRAPP (3: IVAPP)}, 2020, pp. 273--280. \href{https://doi.org/10/gp46vp}
{doi: {{%
10\discretionary{/}{%
}{/}gp46vp}}}


\bibitem{Schulz:2022:TreeVis}
H.-J. Schulz, ``Treevis.net: A tree visualization reference,'' \emph{IEEE Comput Graph Appl}, vol.~31, no.~6, pp. 11--15, 2011. \href{https://doi.org/10/dmgxfs}
{doi: {{%
10\discretionary{/}{%
}{/}dmgxfs}}}


\bibitem{Sedlmair:2012:DSM}
M.~Sedlmair, M.~Meyer, and T.~Munzner, ``Design study methodology: Reflections from the trenches and the stacks,'' \emph{IEEE Trans Vis Comput Graph}, vol.~18, no.~12, pp. 2431--2440, 2012. \href{https://doi.org/10/f4fv7x}
{doi: {{%
10\discretionary{/}{%
}{/}f4fv7x}}}


\bibitem{shneiderman2001ordered}
B.~Shneiderman and M.~Wattenberg, ``Ordered treemap layouts,'' in \emph{Proc. InfoVis}.\hskip 1em plus 0.5em minus 0.4em\relax Los Alamitos: IEEE CS, 2001, pp. 73--78. \href{https://doi.org/10/dtq8rj}
{doi: {{%
10\discretionary{/}{%
}{/}dtq8rj}}}


\bibitem{skau2016arcs}
D.~Skau and R.~Kosara, ``Arcs, angles, or areas: Individual data encodings in pie and donut charts,'' \emph{Comput Graph Forum}, vol.~35, no.~3, pp. 121--130, 2016. \href{https://doi.org/10/f8wcg8}
{doi: {{%
10\discretionary{/}{%
}{/}f8wcg8}}}


\bibitem{song2006atmospheric}
Y.~Song, J.~Ye, N.~Svakhine, S.~Lasher-Trapp, M.~Baldwin, and D.~Ebert, ``An atmospheric visual analysis and exploration system,'' \emph{IEEE Trans Vis Comput Graph}, vol.~12, no.~5, pp. 1157--1164, 2006. \href{https://doi.org/10/d7t3cb}
{doi: {{%
10\discretionary{/}{%
}{/}d7t3cb}}}


\bibitem{stevens2024bioclip}
S.~Stevens, J.~Wu, M.~J. Thompson, E.~G. Campolongo, C.~H. Song, D.~E. Carlyn, L.~Dong, W.~M. Dahdul, C.~Stewart, T.~Berger-Wolf, W.-L. Chao, and Y.~Su, ``{BioCLIP}: A vision foundation model for the tree of life,'' in \emph{Proc.\ CVPR}, 2024, pp. 19\,412--19\,424. \href{https://doi.org/10/mg3n}
{doi: {{%
10\discretionary{/}{%
}{/}mg3n}}}


\bibitem{stokes2022striking}
C.~Stokes, V.~Setlur, B.~Cogley, A.~Satyanarayan, and M.~A. Hearst, ``Striking a balance: Reader takeaways and preferences when integrating text and charts,'' \emph{IEEE Trans Vis Comput Graph}, vol.~29, no.~1, pp. 1233--1243, 2023. \href{https://doi.org/10/mcsk}
{doi: {{%
10\discretionary{/}{%
}{/}mcsk}}}


\bibitem{stoll2005visualization}
C.~Stoll, S.~Gumhold, and H.-P. Seidel, ``Visualization with stylized line primitives,'' in \emph{Proc.\ Visualization}.\hskip 1em plus 0.5em minus 0.4em\relax Los Alamitos: IEEE CS, 2005, pp. 695--702. \href{https://doi.org/10/bh4jp9}
{doi: {{%
10\discretionary{/}{%
}{/}bh4jp9}}}


\bibitem{sucholutsky2023getting}
I.~Sucholutsky, L.~Muttenthaler, A.~Weller, A.~Peng, A.~Bobu, B.~Kim, B.~C. Love, E.~Grant, I.~Groen, J.~Achterberg, J.~B. Tenenbaum, K.~M. Collins, K.~L. Hermann, K.~Oktar, K.~Greff, M.~N. Hebart, N.~Jacoby, Q.~Zhang, R.~Marjieh, R.~Geirhos, S.~Chen, S.~Kornblith, S.~Rane, T.~Konkle, T.~P. O'Connell, T.~Unterthiner, A.~K. Lampinen, K.-R. Müller, M.~Toneva, and T.~L. Griffiths, ``Getting aligned on representational alignment,'' arXiv preprint 2310.13018, 2023. \href{https://doi.org/doi:10/mh8m}
{doi: {{%
doi\discretionary{:}{%
}{:}10\discretionary{/}{%
}{/}mh8m}}}


\bibitem{suits2000simplification}
F.~Suits, J.~T. Klosowski, W.~P. Horn, and G.~Lecina, ``Simplification of surface annotations,'' in \emph{Proc.\ Visualization}.\hskip 1em plus 0.5em minus 0.4em\relax Los Alamitos: IEEE CS, 2000, pp. 235--242. \href{https://doi.org/10/dw25dk}
{doi: {{%
10\discretionary{/}{%
}{/}dw25dk}}}


\bibitem{tanaka1991object}
J.~W. Tanaka and M.~Taylor, ``Object categories and expertise: Is the basic level in the eye of the beholder?'' \emph{Cognit Psychol}, vol.~23, no.~3, pp. 457--482, 1991. \href{https://doi.org/ft55zq}
{doi: {{%
ft55zq}}}


\bibitem{tikhonova2010visualization}
A.~Tikhonova, C.~D. Correa, and K.-L. Ma, ``Visualization by proxy: A novel framework for deferred interaction with volume data,'' \emph{IEEE Trans Vis Comput Graph}, vol.~16, no.~6, pp. 1551--1559, 2010. \href{https://doi.org/10/dvb76h}
{doi: {{%
10\discretionary{/}{%
}{/}dvb76h}}}


\bibitem{tovanich:2022:gender}
N.~Tovanich, P.~Dragicevic, and P.~Isenberg, ``Gender in 30 years of {IEEE} visualization,'' \emph{IEEE Trans Vis Comput Graph}, vol.~28, no.~1, pp. 497--507, 2022. \href{https://doi.org/10/nkkz}
{doi: {{%
10\discretionary{/}{%
}{/}nkkz}}}


\bibitem{treisman1996binding}
A.~Treisman, ``The binding problem,'' \emph{Curr Opin Neurobiol}, vol.~6, no.~2, pp. 171--178, 1996. \href{https://doi.org/10/b6jzkc}
{doi: {{%
10\discretionary{/}{%
}{/}b6jzkc}}}


\bibitem{tufte1998visual}
E.~R. Tufte, \emph{Visual Explanations: Images and Quantities, Evidence and Narrative}.\hskip 1em plus 0.5em minus 0.4em\relax Cheshire: Graphics Press, 1998, \href{https://archive.org/search.php?query=external-identifier%3A%22urn%3Aoclc%3Arecord%3A1280715234%22} {urn: urn\discretionary{}{:}{:}oclc\discretionary{}{:}{:}record\discretionary{}{:}{:}1280715234}.

\bibitem{Tufte1990Envisioning}
------, \emph{Envisioning Information}.\hskip 1em plus 0.5em minus 0.4em\relax Cheshire: Graphics Press, 2001, \href{https://archive.org/search.php?query=external-identifier%3A%22urn%3Aoclc%3Arecord%3A1034670525%22} {urn: urn\discretionary{}{:}{:}oclc\discretionary{}{:}{:}record\discretionary{}{:}{:}1034670525}.

\bibitem{tversky2019mind}
B.~Tversky, \emph{Mind in Motion: How Action Shapes Thought}.\hskip 1em plus 0.5em minus 0.4em\relax UK: Hachette, 2019.

\bibitem{ushizima2012augmented}
D.~Ushizima, D.~Morozov, G.~H. Weber, A.~G.~C. Bianchi, J.~A. Sethian, and E.~W. Bethel, ``Augmented topological descriptors of pore networks for material science,'' \emph{IEEE Trans Vis Comput Graph}, vol.~18, no.~12, pp. 2041--2050, 2012. \href{https://doi.org/10/f4fqtv}
{doi: {{%
10\discretionary{/}{%
}{/}f4fqtv}}}


\bibitem{van2003image}
J.~J. van Wijk, ``Image based flow visualization for curved surfaces,'' in \emph{Proc.\ VIS}.\hskip 1em plus 0.5em minus 0.4em\relax Los Alamitos: IEEE CS, 2003, pp. 123--130. \href{https://doi.org/10/fws2sq}
{doi: {{%
10\discretionary{/}{%
}{/}fws2sq}}}


\bibitem{verma2000flow}
V.~Verma, D.~Kao, and A.~Pang, ``A flow-guided streamline seeding strategy,'' in \emph{Proc.\ Visualization}.\hskip 1em plus 0.5em minus 0.4em\relax Los Alamitos: IEEE CS, 2000, pp. 163--170. \href{https://doi.org/10/drkhm9}
{doi: {{%
10\discretionary{/}{%
}{/}drkhm9}}}


\bibitem{Engelhardt:2002:LanguageOfGraphics}
J.~von Engelhardt, ``The language of graphics,'' Ph.D. dissertation, University of Amsterdam, the Netherlands, 2002, \href{https://hdl.handle.net/11245/1.208097}{hdl: 11245/1.208097}.

\bibitem{wang2020toward}
X.~Wang, Z.~Bylinskii, A.~Hertzmann, and R.~Pepperell, ``Toward quantifying ambiguities in artistic images,'' \emph{ACM Trans Appl Percept}, vol.~17, no.~4, pp. 13:1--13:10, 2020. \href{https://doi.org/10/gm8gm8}
{doi: {{%
10\discretionary{/}{%
}{/}gm8gm8}}}


\bibitem{wang2017edwordle}
Y.~Wang, X.~Chu, C.~Bao, L.~Zhu, O.~Deussen, B.~Chen, and M.~Sedlmair, ``{EdWordle}: Consistency-preserving word cloud editing,'' \emph{IEEE Trans Vis Comput Graph}, vol.~24, no.~1, pp. 647--656, 2018. \href{https://doi.org/10/gcp8bd}
{doi: {{%
10\discretionary{/}{%
}{/}gcp8bd}}}


\bibitem{ward2008multivariate}
M.~O. Ward, ``Multivariate data glyphs: Principles and practice,'' in \emph{Handbook of Data Visualization}.\hskip 1em plus 0.5em minus 0.4em\relax Berlin: Springer, 2008, pp. 179--198. \href{https://doi.org/10/ddmw82}
{doi: {{%
10\discretionary{/}{%
}{/}ddmw82}}}


\bibitem{Ward:2015:IDV}
M.~O. Ward, G.~Grinstein, and D.~Keim, \emph{Interactive Data Visualization: Foundations, Techniques, and Applications}, 2nd~ed.\hskip 1em plus 0.5em minus 0.4em\relax New York: A K Peters/CRC Press, 2015. \href{https://doi.org/10/k8sk}
{doi: {{%
10\discretionary{/}{%
}{/}k8sk}}}


\bibitem{ware2012information}
C.~Ware, \emph{Information Visualization: Perception for Design}, 2nd~ed.\hskip 1em plus 0.5em minus 0.4em\relax San Francisco: Morgan Kaufmann, 2004, \href{https://archive.org/search.php?query=external-identifier%3A%22urn%3Aoclc%3Arecord%3A56344454%22} {urn: urn\discretionary{}{:}{:}oclc\discretionary{}{:}{:}record\discretionary{}{:}{:}56344454}.

\bibitem{ware2020information}
------, \emph{Information Visualization: Perception for Design}, 4th~ed.\hskip 1em plus 0.5em minus 0.4em\relax San Francisco: Morgan Kaufmann, 2020. \href{https://doi.org/10/mj55}
{doi: {{%
10\discretionary{/}{%
}{/}mj55}}}


\bibitem{weiskopf2005visualization}
D.~Weiskopf, M.~Borchers, T.~Ertl, M.~Falk, O.~Fechtig, R.~Frank, F.~Grave, A.~King, U.~Kraus, T.~M{\"u}ller, H.-P. Nollert, I.~R. Mendez, H.~Ruder, T.~Schafhitzel, S.~Sch{\"a}r, C.~Zahn, and M.~Zatloukal, ``Visualization in the {E}instein year 2005: A case study on explanatory and illustrative visualization of relativity and astrophysics,'' in \emph{Proc.\ Visualization}.\hskip 1em plus 0.5em minus 0.4em\relax Los Alamitos: IEEE CS, 2005, pp. 583--590. \href{https://doi.org/10/bbsrhm}
{doi: {{%
10\discretionary{/}{%
}{/}bbsrhm}}}


\bibitem{weiskopf2005overview}
D.~Weiskopf and G.~Erlebacher, ``Overview of flow visualization,'' in \emph{Visualization Handbook}, C.~D. Hansen and C.~R. Johnson, Eds.\hskip 1em plus 0.5em minus 0.4em\relax Oxford: Elsevier, 2005, ch.~12, pp. 261--278. \href{https://doi.org/10/d4hzdf}
{doi: {{%
10\discretionary{/}{%
}{/}d4hzdf}}}


\bibitem{Weiss:2021:Revisited}
M.~Wei{\ss}, K.~Angerbauer, A.~Voit, M.~Schwarzl, M.~Sedlmair, and S.~Mayer, ``Revisited: Comparison of empirical methods to evaluate visualizations supporting crafting and assembly purposes,'' \emph{IEEE Trans Vis Comput Graph}, vol.~27, no.~2, pp. 1204--1213, 2021. \href{https://doi.org/10/ghgt5z}
{doi: {{%
10\discretionary{/}{%
}{/}ghgt5z}}}


\bibitem{Wilkinson:2005:GrammarOfGraphics}
L.~Wilkinson, \emph{The Grammar of Graphics}, 2nd~ed.\hskip 1em plus 0.5em minus 0.4em\relax New York: Springer, 2005. \href{https://doi.org/10/bhsbp7}
{doi: {{%
10\discretionary{/}{%
}{/}bhsbp7}}}


\bibitem{wittgenstein2010philosophical}
L.~Wittgenstein, \emph{Philosophical Investigations}.\hskip 1em plus 0.5em minus 0.4em\relax New York: MacMillan, 1953, \href{https://archive.org/search.php?query=external-identifier:"urn:oclc:record:1245626409"}{urn: urn\discretionary{:}{}{:}oclc\discretionary{:}{}{:}record\discretionary{:}{}{:}1245626409}.

\bibitem{wolfe2021guided}
J.~M. Wolfe, ``{G}uided {S}earch 6.0: An updated model of visual search,'' \emph{Psychon Bull Rev}, vol.~28, no.~4, pp. 1060--1092, 2021. \href{https://doi.org/10/gh2s45}
{doi: {{%
10\discretionary{/}{%
}{/}gh2s45}}}


\bibitem{xia2020smap}
J.~Xia, T.~Chen, L.~Zhang, W.~Chen, Y.~Chen, X.~Zhang, C.~Xie, and T.~Schreck, ``{SMAP}: A joint dimensionality reduction scheme for secure multi-party visualization,'' in \emph{Proc.\ VAST}.\hskip 1em plus 0.5em minus 0.4em\relax Los Alamitos: IEEE CS, 2020, pp. 107--118. \href{https://doi.org/10/k8st}
{doi: {{%
10\discretionary{/}{%
}{/}k8st}}}


\bibitem{Ye:2021:ShuttleSpace}
S.~Ye, Z.~Chen, X.~Chu, Y.~Wang, S.~Fu, L.~Shen, K.~Zhou, and Y.~Wu, ``{S}huttle{S}pace: Exploring and analyzing movement trajectory in immersive visualization,'' \emph{IEEE Trans Vis Comput Graph}, vol.~27, no.~2, pp. 860--869, 2021. \href{https://doi.org/10/ghgt4x}
{doi: {{%
10\discretionary{/}{%
}{/}ghgt4x}}}


\bibitem{ye2022visatlas}
Y.~Ye, R.~Huang, and W.~Zeng, ``{VISA}tlas: An image-based exploration and query system for large visualization collections via neural image embedding,'' \emph{IEEE Trans Vis Comput Graph}, vol.~30, no.~7, pp. 3224--3240, 2024. \href{https://doi.org/10/grq3m3}
{doi: {{%
10\discretionary{/}{%
}{/}grq3m3}}}


\bibitem{yi2007toward}
J.~S. Yi, Y.~ah~Kang, J.~Stasko, and J.~A. Jacko, ``Toward a deeper understanding of the role of interaction in information visualization,'' \emph{IEEE Trans Vis Comput Graph}, vol.~13, no.~6, pp. 1224--1231, 2007. \href{https://doi.org/10/fmrs6r}
{doi: {{%
10\discretionary{/}{%
}{/}fmrs6r}}}


\bibitem{yoghourdjian2020scalability}
V.~Yoghourdjian, Y.~Yang, T.~Dwyer, L.~Lawrence, M.~Wybrow, and K.~Marriott, ``Scalability of network visualisation from a cognitive load perspective,'' \emph{IEEE Trans Vis Comput Graph}, vol.~27, no.~2, pp. 1677--1687, 2021. \href{https://doi.org/10/ghs96j}
{doi: {{%
10\discretionary{/}{%
}{/}ghs96j}}}


\bibitem{yue2018bitextract}
X.~Yue, X.~Shu, X.~Zhu, X.~Du, Z.~Yu, D.~Papadopoulos, and S.~Liu, ``{BitExTract}: Interactive visualization for extracting {B}itcoin exchange intelligence,'' \emph{IEEE Trans Vis Comput Graph}, vol.~25, no.~1, pp. 162--171, 2019. \href{https://doi.org/10/ggjbt3}
{doi: {{%
10\discretionary{/}{%
}{/}ggjbt3}}}


\bibitem{ziemkiewicz2010laws}
C.~Ziemkiewicz and R.~Kosara, ``Laws of attraction: From perceptual forces to conceptual similarity,'' \emph{IEEE Trans Vis Comput Graph}, vol.~16, no.~6, pp. 1009--1016, 2010. \href{https://doi.org/10/cz74g2}
{doi: {{%
10\discretionary{/}{%
}{/}cz74g2}}}


\end{thebibliography}

\vspace{-33pt}
\begin{IEEEbiography}[{\includegraphics[width=1in,height=1.25in,clip,keepaspectratio]{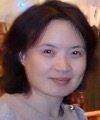}}]{Jian Chen}
 is an associate professor in the Department of 
 Computer Science and Engineering at The Ohio State University.
She did her postdoctoral work in Computer Science and BioMed at Brown University to study evaluation and biological motion. She
received her PhD degree in Computer Science from Virginia Polytechnic Institute and State University 
on 3D interaction. 
Her current research interests include visualization theory, evaluation, and interaction.
\end{IEEEbiography}

\vspace{-38pt}
\vfill
\begin{IEEEbiography}[{\includegraphics[width=1in,height=1.25in,clip,keepaspectratio]{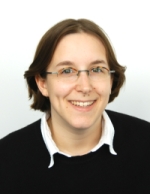}}]
{Petra Isenberg} is a senior research scientist at Inria, Saclay, France in the Aviz team. She received her Diploma degree from the University of Magdeburg, Germany and her PhD degree from the University of Calgary, Canada in 2010. She is particularly interested in exploring how people can most effectively work together when analyzing large and complex data sets on novel display technology such as small touchscreens, wall displays, or tabletops. 
\end{IEEEbiography}

\vspace{-40pt}
\vfill

\begin{IEEEbiography}[{\includegraphics[width=1in,height=1.25in,clip,keepaspectratio]{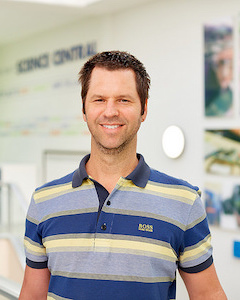}}]
{Robert S. Laramee} is a professor at the University of Nottingham, UK in the School of Computer Science. He received the PhD degree from the Vienna University of Technology, Austria at the Institute of Computer Graphics and Algorithms. Between 2001 and 2006 he was a researcher at the VRVis Research Center 
and a software engineer at AVL 
in the department of Advanced Simulation Technologies. He was a Lecturer, then Associate Professor at Swansea University, Wales in the Department of Computer Science. His research interests are in the areas of scientific visualization, information visualization, and visual analytics. 
\end{IEEEbiography}

\vspace{-40pt}
\vfill
\begin{IEEEbiography}[{\vspace{-3.5ex}\includegraphics[width=1in,height=1.25in,clip,keepaspectratio]{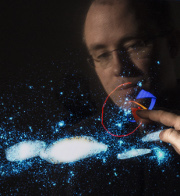}}]
{Tobias Isenberg} is a senior research scientist at Inria, France. He received his PhD degree from the University of Magdeburg, Germany in 2004. Previously he held positions as postdoctoral fellow at the University of Calgary, Canada, and as assistant professor at the University of Groningen, the Netherlands. He is
particularly interested in interactive visualization environments for 3D spatial data as well as illustrative visualization and abstraction techniques.
\end{IEEEbiography}

\vspace{-45pt}
\vfill
\begin{IEEEbiography}[{\includegraphics[width=1in,height=1.25in,clip,keepaspectratio]{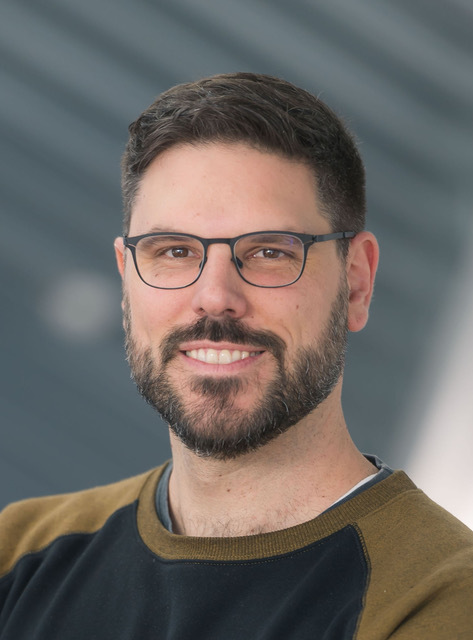}}]
{Michael Sedlmair} is a professor at the University of Stuttgart
and is leading the research group for Visualization and Virtual/Augmented 
Reality (VR/AR) there. He received the PhD degree in Computer Science from University of Munich, Germany in 2010. His specific research interests focus on interactive 
machine learning, interactive visualization in VR/AR, as well as the 
methodological and theoretical frameworks underlying them.
\end{IEEEbiography}

\vspace{-40pt}
\vfill
\begin{IEEEbiography}[{\includegraphics[width=1in,height=1.25in,clip,keepaspectratio]{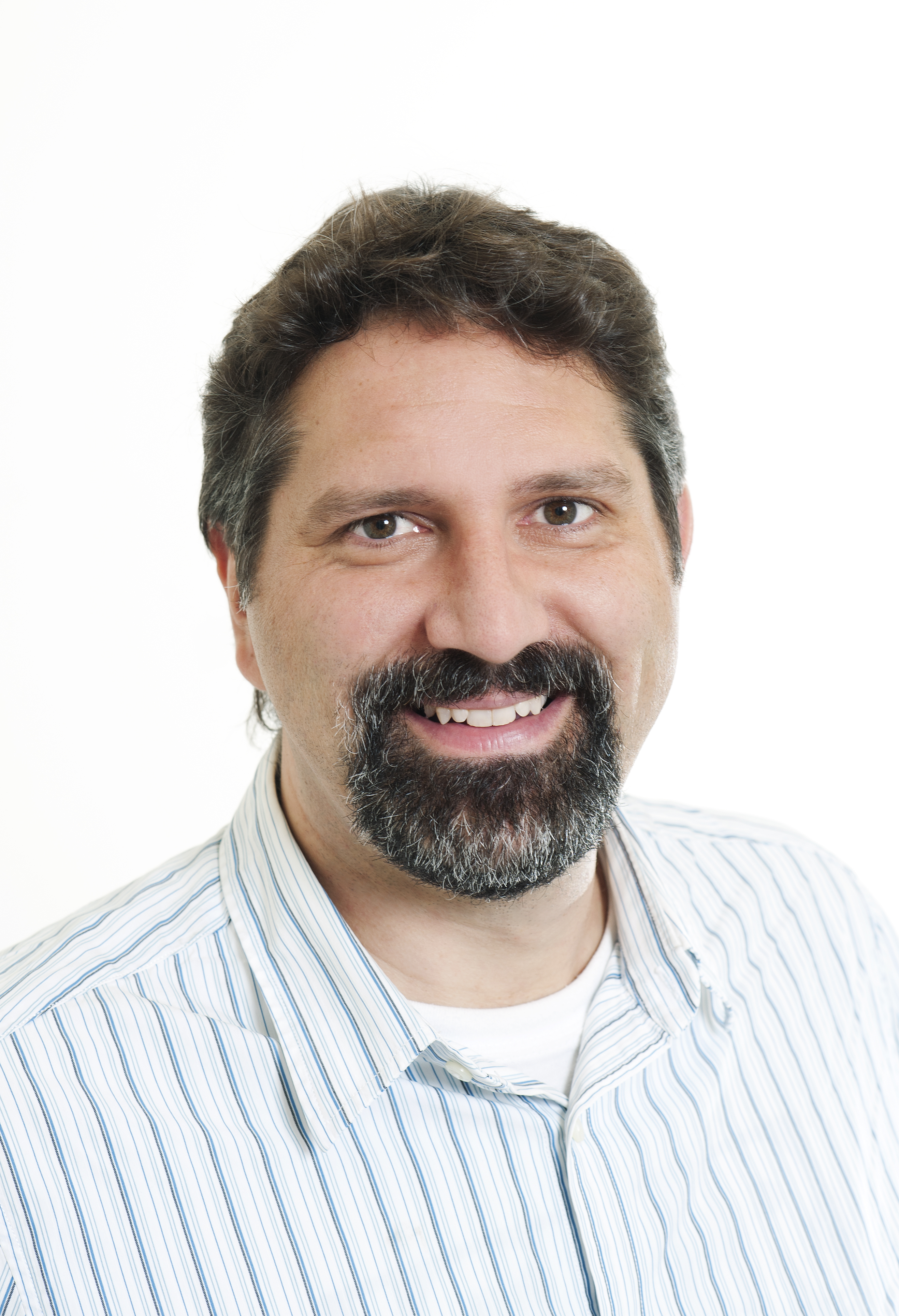}}]
{Torsten M\"oller} is 
a professor at the University of Vienna, Austria. 
Between 1999 and 2012, he was a Computing Science faculty member at Simon Fraser University, Canada. He received his PhD in Computer and Information Science from Ohio State University in 1999 and a Vordiplom (BSc) in mathematical computer science from Humboldt University of Berlin, Germany. He is a senior member of IEEE and ACM, and a member of Eurographics. His research interests include algorithms and tools for analyzing and displaying data with principles rooted in computer graphics, human-computer interaction, image processing, machine learning, and visualization.
\end{IEEEbiography}

\vspace{-40pt}
\vfill
\begin{IEEEbiography}[{\includegraphics[width=1in,height=1.25in,clip,keepaspectratio]{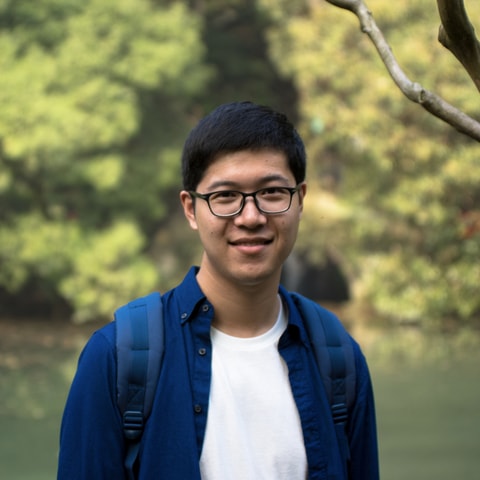}}]{Rui Li} 
is a Ph.D. student in Computer Science and Engineering at Ohio State University. He received the B.E. degree in Computer Science and Engineering from Central South University, China. His research interests include evaluation of data visualization, visual analytics, and data mining.
\end{IEEEbiography}


\clearpage
\appendices

\section{Source Image Data and Our \typename Data} 
\label{sm:imageData}

\noindent\tochange{%
The
first six of the seven years of image data come from the collection of figure and table data from IEEE VIS publications from 1990--2019 in Chen et al.'s VIS30K~\cite{Chen2020VIS30Kdata}. 
We added the 2020 dataset to the VIS30K data collection. 
In doing this, we reused the authors' (our) approach and the meta-data to remove all tables (unless part of a figure) and their open-source model and tools to extract and clean the new figure data of 2020 after scraping all papers from the IEEE Xplore.  The first author collected and curated the image data for the full year. 
Other co-authors checked the final image set. The overall image dataset is accessible at \vistypeImagegooglelink.
}
\jcfr{
This web-based tool (\imagelink) also allows the viewers 
to 
send an email request to update our image codes (\autoref{fig:navigatorTwoHard}).
}

\section{\jctg{What this paper is not?}}
\label{sec:smNotAbout}

\noindent
\jctg{We compare our typology to others 
{for the purpose of} categorization. Our work neither judges the quality of visualizations nor studies the morphological relationships between visual 
{designs}. 
Cleveland's critique of visual constructions~\cite{cleveland1984graphs} and Wilkinson's morphology in 
GoG~\cite{Wilkinson:2005:GrammarOfGraphics} did so. We also do not study how designs relate to meaning in images; Cairo's Functional Art~\cite{cairo2012functional} did this. 
Our work provides a level of categorization of main visual foci or essential stimuli --- so we can clearly describe the main visual focus of attention in these images.
In addition, 
our codes do not cover non-essential elements, \eg, axis and labels, which are indispensable, but do not map to data directly.}

\jctg{Approaches for representing data visualizations fall into two broad categories: those that focus on designers' drawings and those that prioritize viewers' understanding.}
Ours belongs to the latter approach. \jctg{Wilkinson's GoG~\cite{Wilkinson:2005:GrammarOfGraphics} is similar to ours in that it does not use techniques to describe graphics but contains a `generalizable' list of categories.
His categories are perhaps most \textit{typical}---called prototypes~\cite{tversky2019mind}, useful for communication.
Wilkinson cautiously noted that his categorization taxonomy was incomplete, did not represent the visualization domain, and was designed solely for implementation purposes.
In contrast, our categorization is meant to be complete covering a much broader set
and formally coded by experts through iterations. 
Compared to Wilkinson, Heer~\cite{Heer:2010:TVZ} focused on atypical drawings of a set of what he called ``unusual'' and ``exotic'' techniques: \textit{``you don't go to the zoo to see Chihuahuas and raccoons; you go to admire the majestic polar bear, the graceful zebra, and the terrifying Sumatran tiger.''} Also, his focus was specific advances of specific techniques rather than categorizing representations. 
In our typology, we assigned the atypical images by comparing their similarity to the typical ones. 
Finally, in contrast to
Cleveland's vision
to understand visualizations from images---\textit{``a graphical method is successful only if the decoding is effective''}~\cite{cleveland1985graphical}, he enables `\textit{elementary} graphical perception' work through `\textit{elementary} level visual marks in statistical charts'. In this regard, our work broadens the purview of representations at the level of 
essential stimuli.
}


\section{\tochange{Additional Codes: Function, dimensionality, and 
coding transitions}}
\label{sec:smKeywords}

\noindent{\tochange{We further describe how difficult it can be to understand images that were taken out of the context of the text and captions. Thus, additional decisions were needed in the coding process.}
We have also coded two \textit{function codes} and two \textit{dimensionality codes}, as well as experiences concerning the \textit{difficulties} of categorizing visualization images (see definitions in \autoref{tab:funcDim}.)
}

\subsection{Pilot Study: Failed VisKeywords-Based Method}
\label{sm:keywords}

\noindent
The top-21 authors' keywords for specific techniques are:  \textit{bar chart}, 
\textit{cartographic map},
\textit{circular node-link tree}, 
\textit{flow chart}, 
\textit{flowline}, 
\textit{glyphs}, 
\textit{heatmap}, 
\textit{isosurface rendering}, 
\textit{line chart}, 
\textit{matrix}, 
\textit{node-link diagram}, 
\textit{parallel coordinate plot}, 
\textit{pie chart}, 
\textit{point cloud}, 
\textit{scatter plot}, 
\textit{tag cloud}, 
\textit{timeline}, 
\textit{treemap}, 
\textit{volume rendering}, 
\textit{Voronoi diagram}, and 
\textit{wireframe rendering}.

\subsection{\tochange{Pilot study: Failed Bertin's Marks and Channels-based Method}}
\label{sm:Bertin}
\noindent\tochange{One may consider Bertin's marks and channels to describe how we see visualization images. This perspective decodes the complex perceptual process by describing low-level \textit{features}, such as color, shape, size, orientation, curvature, and lines, in the images first, and then treat visual processing as a means for the visual system  to ``bond'' together these features to become a ``single representation of an object''~\cite{treisman1996binding}.} 
\jcfr{This system uses five criteria of categorization: printable on white paper, visible at a glance, reading distance of book or atlas, normal and constant lighting, and readily available graphics means, which unfit our categorization of coding essential stimuli to represent what viewers should see from images. 
}

We demonstrate that relying on low-level features alone is not enough to understand the images
For example, we may observe in \autoref{fig:sm:barSurface} that edges (lines) in (b) (computed using the Canny Edge detector) can describe the design feature of bar edges in (a). However, the bar height in (a) should be the essential stimulus, and the line descriptor does not show the bar structures well. 
In the second example in \autoref{fig:sm:barSurface}c, we also show features defined by the Histograms of Oriented Gradients (HOGs) (d) for the surface in (c). HOGs let us see structures. However, it does not show the `\textit{surface}' in (c).

\subsection{Pilot Study: Consolidation}
\label{sm:consolidation}

\begin{figure*}[!th]
\centering
\subfloat[]{\includegraphics[height=0.4\columnwidth]{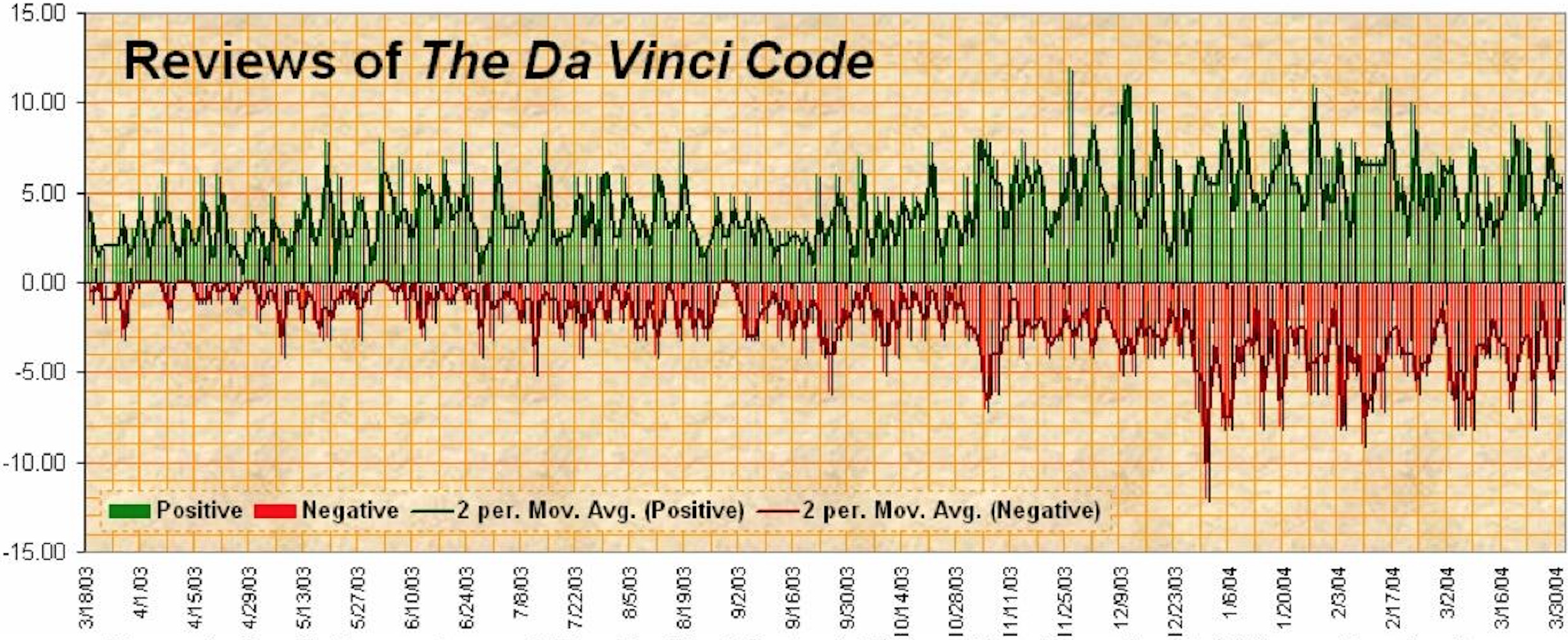}%
}
\hfil
\subfloat[]{\includegraphics[clip, trim={0 30 0 0},height=0.4\columnwidth]{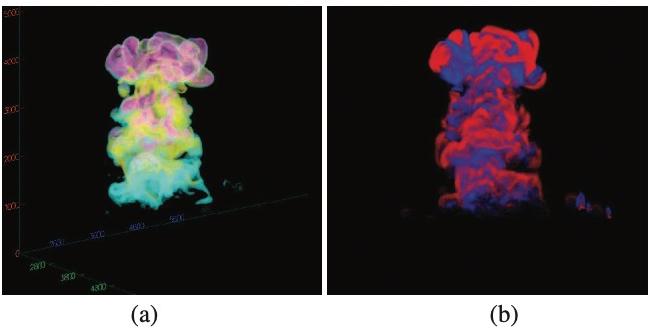}%
}
\caption{\tochange{\textbf{Rationale to Remove Timeline from \typename.} We ask the question: Is there a timeline in the figures? Here shows two examples. (a) yes because of the dates on the x-axis and (b) yes but not so obvious and the judgment counts on the coders' knowledge.  Images from (a) Chen et al.~\cite{chen2006visual} and (b) Song et al.~\cite{song2006atmospheric}.}}
\label{fig:sm:timeline}
\end{figure*}

\noindent\tochange{In \textbf{Phase 3 Consolidation} (\autoref{sec:classificationProcess}) we made several decisions to group stimuli, (1) if two representations share specific stimuli, (\eg, donut charts use length and thus fall into the category of ``\textit{Generalized Bar Representations}'', see \autoref{fig:typecomparisons}) and (2) if a stimulus is closer to one category than to others, \eg, the geometric wireframe mesh is closer to ``\textit{node-link}''. We further devise category-specific characteristics, \eg, the bar examples share common length feature regardless of the coordinate system; ``\textit{surface-based Representations and Volumes}'' resembles the inside and outside boundaries, and ``\textit{Grid}'' shows the family of layout appearances. Some additional examples are below. 
\autoref{fig:sm:timeline} show our rationale for removing timeline from our typology---the judgment depends on the viewer's knowledge because drawing the temporal cue may not always be necessary.}

\tochange{
\jcfr{\textbf{We merged techniques if the visual representations of the essential stimuli are alike.}
For example, flowlines, parallel coordinate plots, and line charts all contain line features. 
\codingdecisions{We merged them into the \lineabbr category.}
}
}

The flow chart category was prevalent and represented the largest number of figures in the other category of our 2006 coding, thus adding and later further expanding to \visschematic. The original flow chart has only captured a limited number of wireframe diagrams. Many charts and diagrams in visualization papers are visually rich and contain techniques to illustrate concepts. 
\codingdecisions{We expanded flowchart to ``General Schematic Representation, schematic images, schematic concept illustration''.}

\tochange{\textbf{We ignored the drawing media used to visualize the data.}
Sometimes, it is challenging to determine whether or not views are hand-drawn or computer-generated. For example, schematic ones can be drawn by algorithms~\cite{kim2010evaluating}. 
``Illustrative visualization'' develops algorithms that would render images 
that appear to be hand-drawn.
The illustrations in the classical book of Bertin's semiotics were also drawn by hand. 
Hence, we chose to ignore the media in the subsequent coding phase.
Instead, we emphasize the elements in the figure rather than the drawing media. For example, a photograph of an environment (\eg, a virtual reality 
) would be coded as ``3D''.}

\begin{figure*}[t]
    \centering
    \includegraphics[width=0.5\columnwidth]{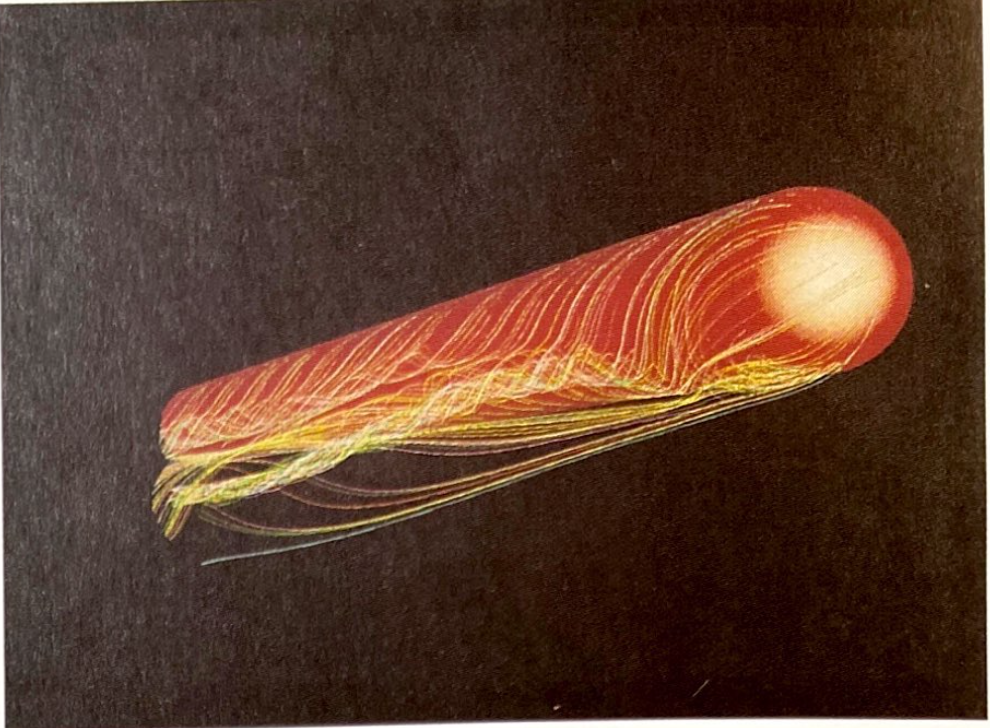}   
    \caption{\tochange{\textbf{Coding context.} In addition to \visline, we incorporated \vissurface  because the geometry of the model itself determines the shape of the lines. Image taken from~\cite{helman1990surface}.}}
    \label{fig:linesurface}
\end{figure*}

\tochange{\textbf{We managed annotation, legend, and context.}
One of the challenges was treating context information, such as annotation marks or color legends. Here, we focus on the primary visual encoding and agree not to code such context information separately. The color legend is in the ``other'' category. 
Context, unless relevant to the data, is not coded. 
\jcfr{Some contextual data, such as geometric models or boundaries, play a crucial role in comprehending visualization techniques and, as such, were coded (\eg, \vissurface is coded besides \visline, see \autoref{fig:linesurface}).}
}

\tochange{\textbf{We avoided including data types in the type names.} For example, scalar, vector, and tensor field visualization techniques can be defined using our typology without mentioning flow fields or tensors.}

\section{Web Interfaces for Annotating Figures}
\label{sm:web-interfaces}

\begin{figure*}[!ht]
	\centering
	\includegraphics[height=0.85\textwidth]{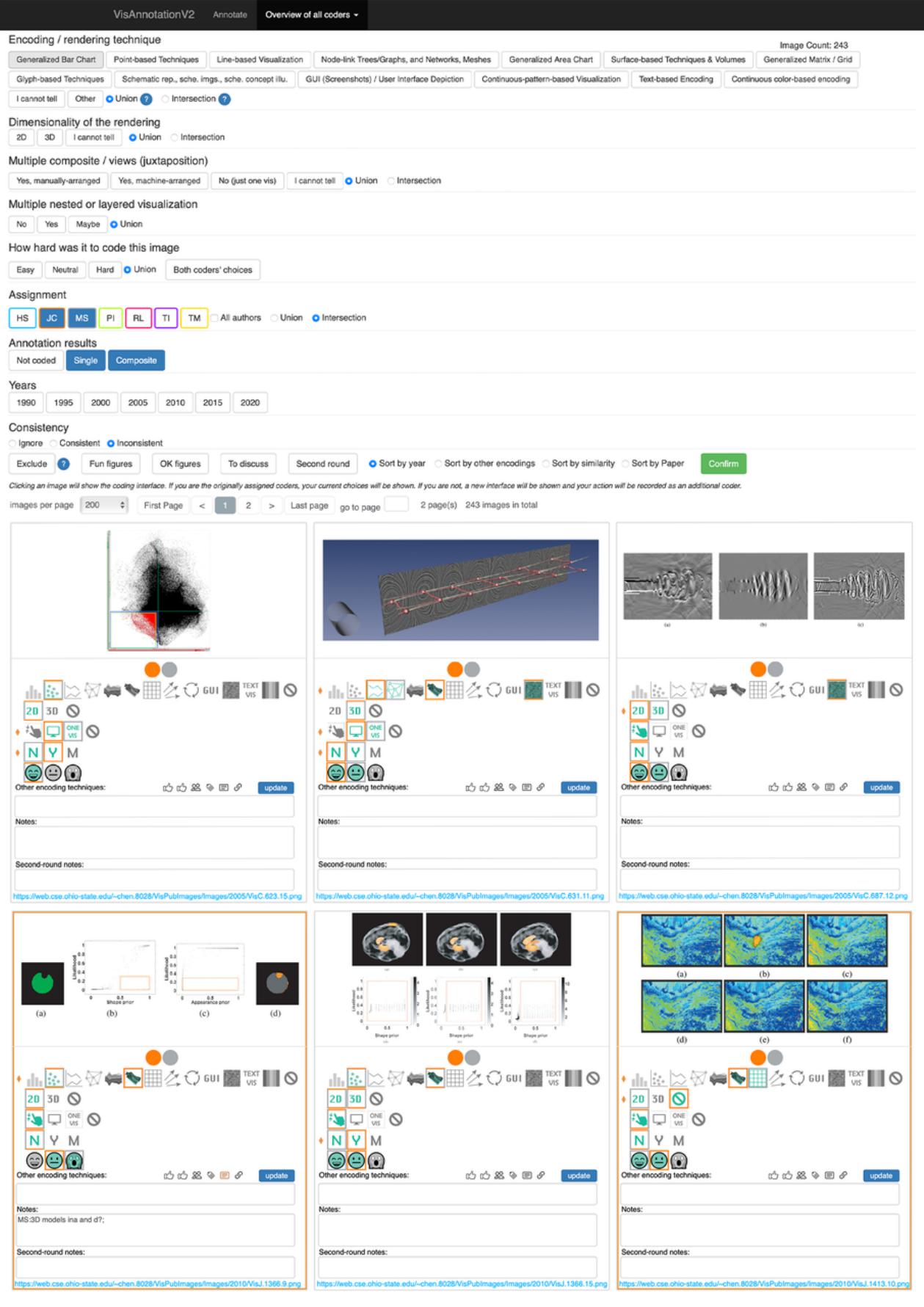}
	\caption{\tochange{\textbf{A screenshot of the web interface for Phase 6.} In Phase 6 we resolved coders' inconsistency through pair-wise comparisons. Two coders results are shown together. For complex images, coders can also look up the images in the same paper coded by other coders through paper-based or image similarity-based search.} 
	}
	\label{fig:web-interface-consistency}
\end{figure*}

\noindent\tochange{We annotated all images and discussed and compared our codings via our own web-based interfaces to support our collaborative work (\autoref{fig:web-interfaces}(a)--(d)).
Our web-based tool automatically loads images and authorizes the users.
The users can tag the given image according to the keyword- or type-based terms.
On the backend of our coding tool, we recorded every button click from each participant during the coding process for post-hoc analyses.
To resolve code consistency, we also implemented a comparative interface (\autoref{fig:web-interface-consistency}) to resolve all coding conflicts. Again, all coders' choices were recorded.}

\section{Additional Results (\expertUs)}
\label{sm:resultsExpertUs}

\begin{figure*}[!th]
\centering
\subfloat[]{
\includegraphics[height=0.3\textwidth]{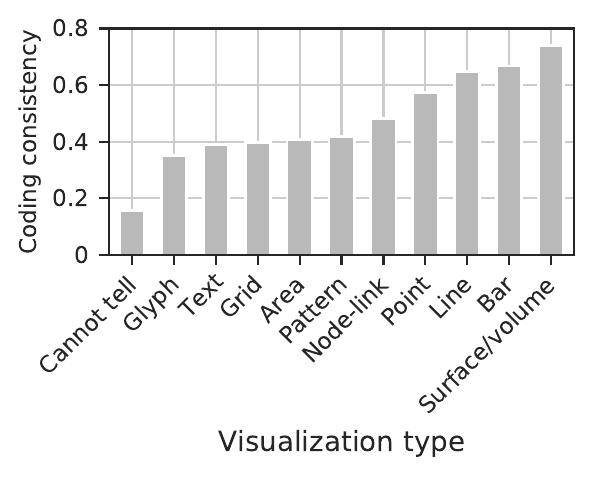}
}
\subfloat[]{
\includegraphics[height=0.3\textwidth]{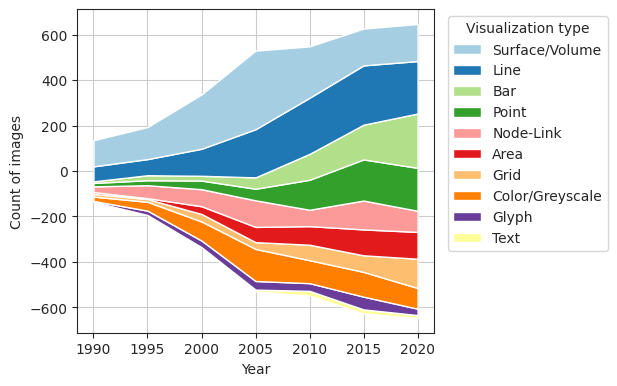}
}
\caption{\textbf{Additional User Study Result on Overall Code Distributions.} (a) The initial consistency of visualization type codes applied to images. We can see that the expert coders (\expertUs) had least consistency related to  \visglyph. (b) The total number of image types by year.}
\label{fig:consistency}
\end{figure*}

\begin{figure*}[!th]
\centering
\subfloat[]{
\includegraphics[height=.4\columnwidth]{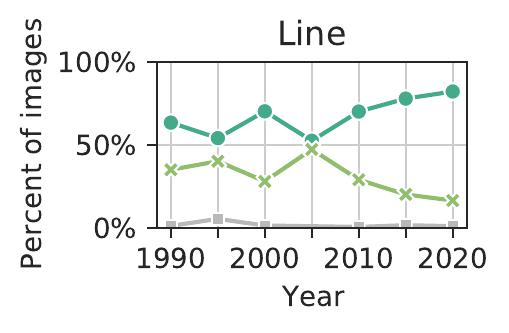}
}
\subfloat[]{
\includegraphics[height=0.4\columnwidth]{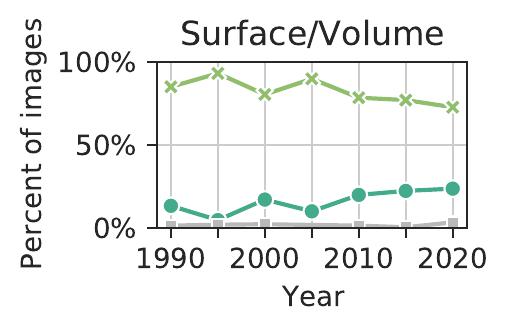}%
}
\caption{\textbf{Additional User Study Result on Dimensionality (color-coded 2D in teal, 3D in lime and `I cannot tell' in gray).}
Temporal overview of the proportions of 2D and 3D images for \vissurface and \visline. We observed a decrease in 3D use over time.}
\label{fig:surfaceline3D}
\end{figure*}

\begin{figure*}[!tp]
  \centering  
  \includegraphics[width=\linewidth]{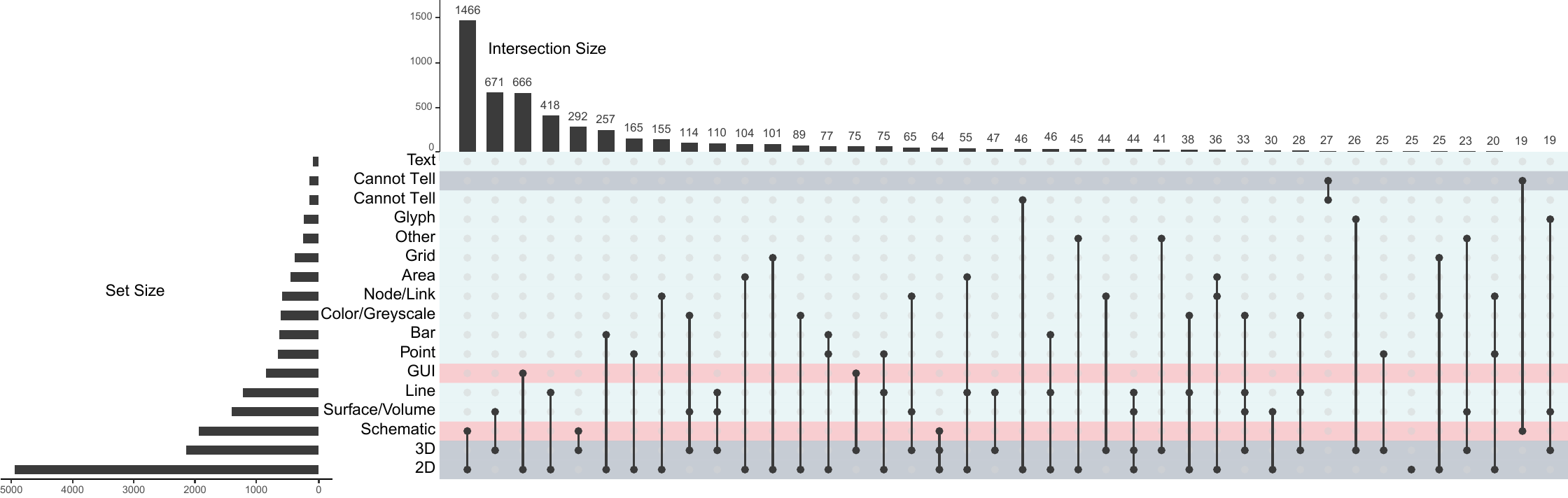}
 \caption{\tochange{\textbf{\typename Co-occurance Results.} Coding results from categorizing IEEE VIS paper images according to visualization types (baby-lue), their dimensionality (gray), and additional image functions (red). There are many 2D Schematics and 2D GUIs. Surface/Volume occurred the most with 3D representation. 2D bars are also frequently used. %
} }
\label{fig:coocc}
\end{figure*}

\noindent\tochange{\autoref{fig:consistency}(a) shows the consistency of the image types
captured in our empirical study. 
Echoing our own experiences and the nature of any categorization, we also found in our discussion that the inconsistent types could be related to personal differences in education, knowledge, and preferences. \autoref{fig:consistency}(b) shows the total number of image types by year. We can see a steady increase in almost all types. 
\autoref{fig:surfaceline3D}
shows \vissurface and \visline uses over time.  
\autoref{fig:coocc} illustrates the co-occurance of these types.}

\section{Additional Results (\expertExternal)}
\label{sm:resultsExpertExternal}

\begin{figure*}[!tb]
\centering
\includegraphics[width=\textwidth]{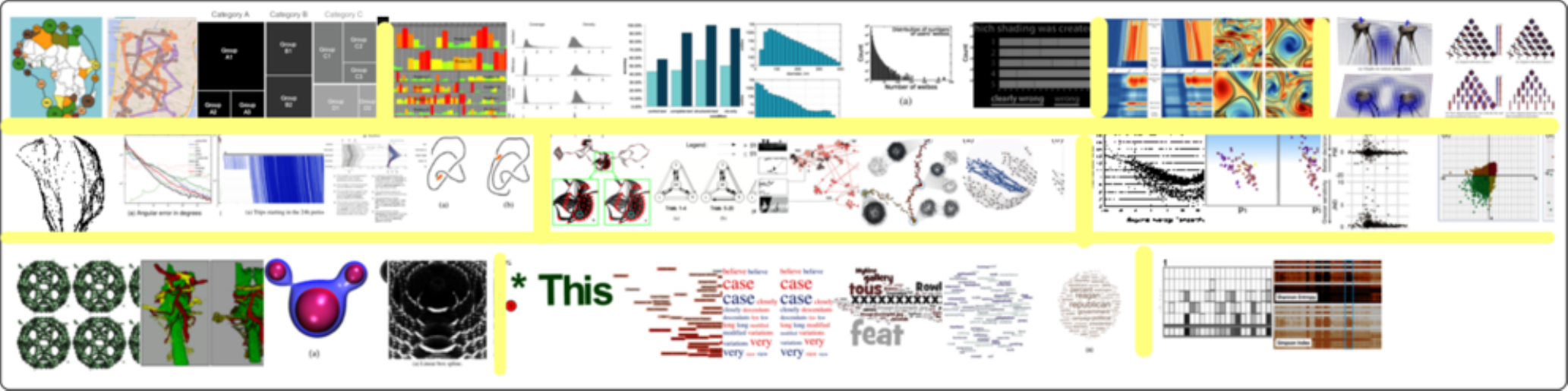}%
\caption{\tochange{\textbf{Example images for which the participants' coding completely matched our own coding results.}
Among these, we see, from top left to bottom right, 3 area, 6 bar, 2 color/grey-scale, 2 glyph,  5 line-based, 5 node-link, 4 point, 4 surface/volume,  6 text representations, and 2 grid. 
All these images had difficulty ratings of easy to neural. 
These examples show that, despite the rich visual features in each of our types, participants can use our category descriptions to identify image types.}
}
\label{fig:expEvalAgree}
\end{figure*}

\begin{figure*}[!tb]
\centering
\includegraphics[width=\textwidth]{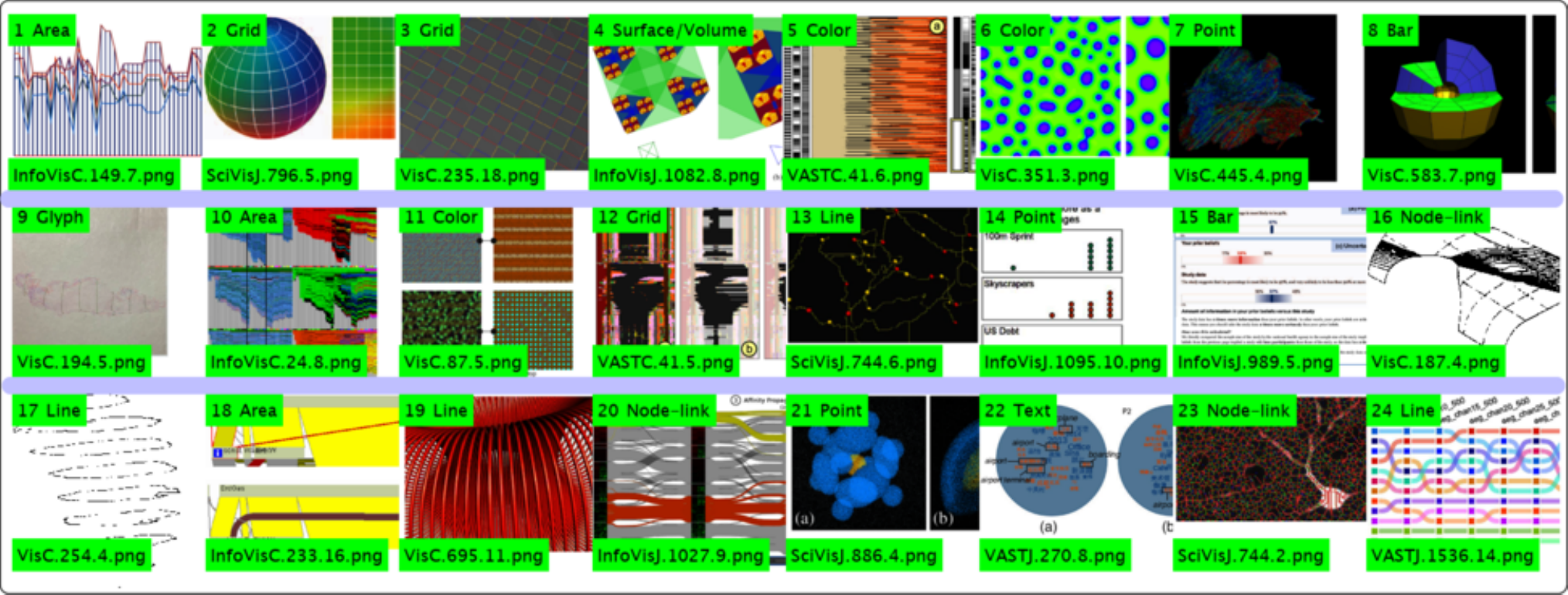}%
\caption{\tochange{\textbf{The Top 24 Most Challenging Images.} Among the 180 target images chosen for evaluation, 24 images are ordered from least to most codes coded by  \expertExternal with a Ph.D. degree, from top left to bottom right. The \expertUs-coded type was tagged on each image. 
\textbf{Intuition.} 
People tend to count on the presence or absence of the `design element', rather than the `essential stimuli'. e.g., people mostly coded 1, \visline; Participants mostly responded no to our code 2. \visgrid and 18. \visarea;  
In our codes, we do not code color when \textit{color} does not encode data. 
Thus, image 22 has only one code \vistext. Participants also coded \viscolor for 22. 
Image 17 is a difficult case. Lines could reveal surfaces rather than the geometry form alone. We coded \visline because we were uncertain about the representation and thus took the visual appearance alone.  
Readers can find the participants' answers using the 
\colorbox{green}{green name tag} 
online at \href{https://go.osu.edu/vistypeexpresult}
{go.osu.edu/visTypeExpResult}.
}
}
\label{fig:exp:hardest}
\end{figure*}

\noindent\tochange{\autoref{fig:expEvalAgree} shows a list of images where all participants chose the exact same labels as we did in our own coding. These examples show that, despite the rich visual features in each of our types, participants can use our category descriptions to identify image types. 
\autoref{fig:exp:hardest} shows some example images \expertExternal participants had mostly disagreed with our answers. 
The cases that \expertExternal found challenging matched those we \expertUs encountered during the coding process. Labels were not made available to indicate the data and the image captions. Participants read the images alone. 
\tochange{
For example, Image 1 was coded by \expertExternal as \textit{Generalized line-based representation}.
Here, it is very difficult to determine the types without knowing what the data represents. 
}
}

\section{Additional Image Examples about Visual Types}
\label{sec:moreExamples}

\noindent
While \autoref{sec:visTypesResults} shows the canonical examples, we present more image results to show the visual appearances of the images in each of the 10 types in~\autoref{fig:sm:typeImages}.

\section{\jctg{Reuse of \typename for non-scholarly images}}

\noindent
\jctg{We demonstrated the reuse of our Typology to code Borkin et al.'s MASSVIS image set~\cite{Borkin2013MASSVIS}, in massive media online images. MASSVIS contained 2,070 images of online media, infographics, and charts. The code comparison is online at \massvislink. 
MASSVIS contained many \barabbr, followed by \areaabbr and \lineabbr. Not surprisingly, it has a limited number of images in the
\surfacevolumeabbr category.  
We also observed that the visualizations in mass medium tended to support fact retrieval since many figures were annotated.
}

\section*{Image Copyright}

\begin{figure*}[t]
    \centering
    \includegraphics[width=\textwidth]{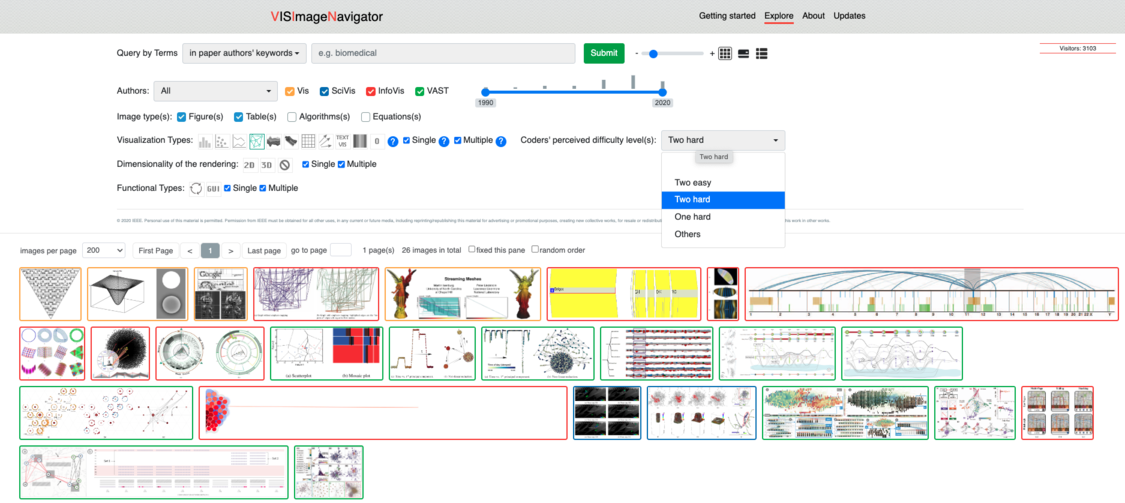}   
    \caption{\textbf{\typename Exploration Tool.} An example query showing the \textit{Generalized node-link, mesh representation}, when both coders labelled \textit{hard}. Readers can try the interface online at \href{https://visimagenavigator.github.io/}{visimagenavigator.github.io}.}
    \label{fig:navigatorTwoHard}
\end{figure*}

\begin{figure*}[!thp]
    \centering
    \begin{subfigure}[b]{0.45\columnwidth}
\includegraphics[height=2\pictureheight]{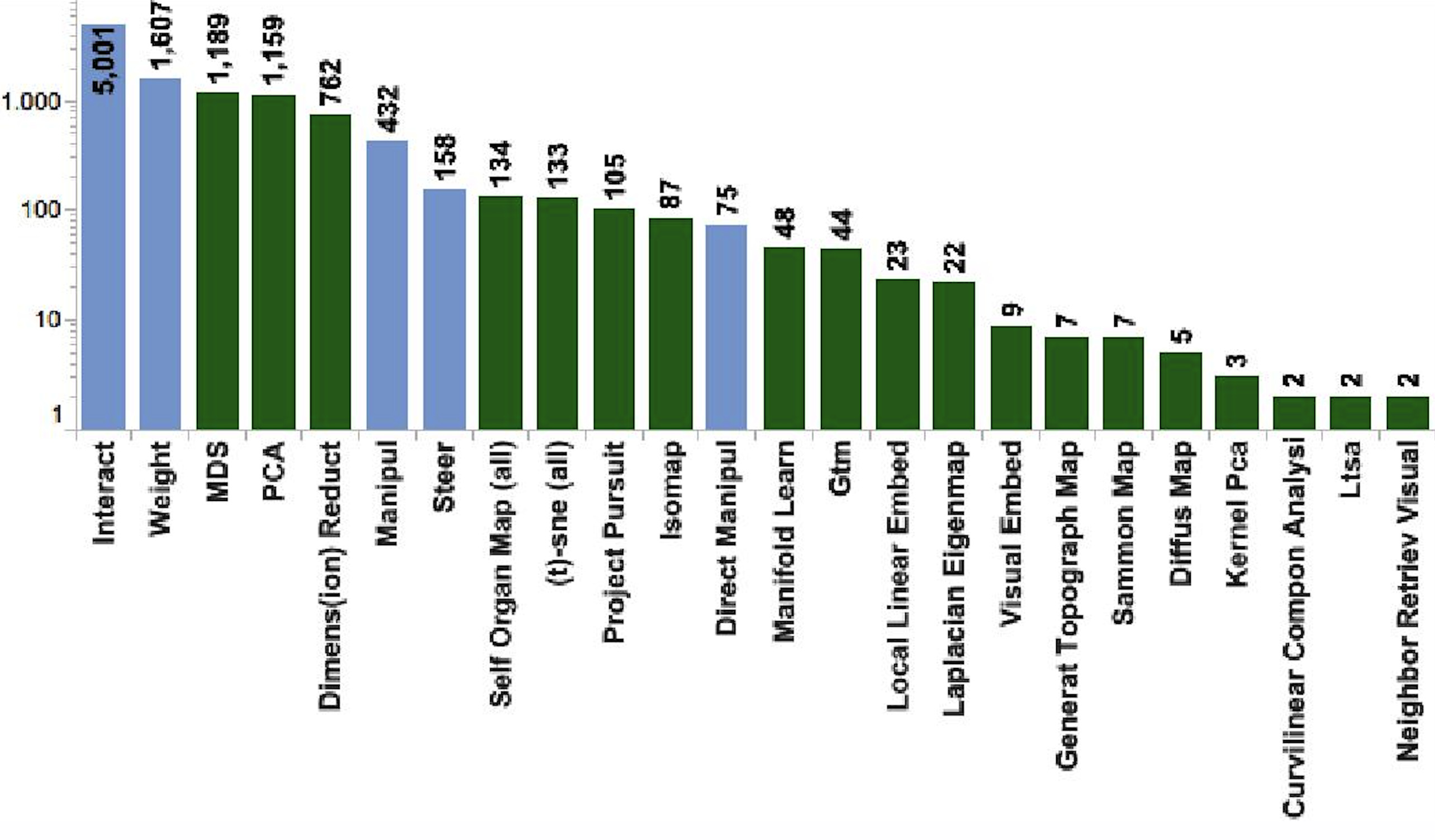}
    \caption{}
    \label{fig:sm:bar}
    \end{subfigure}
     \begin{subfigure}[b]{0.55\columnwidth}
    \includegraphics[height=2\pictureheight]{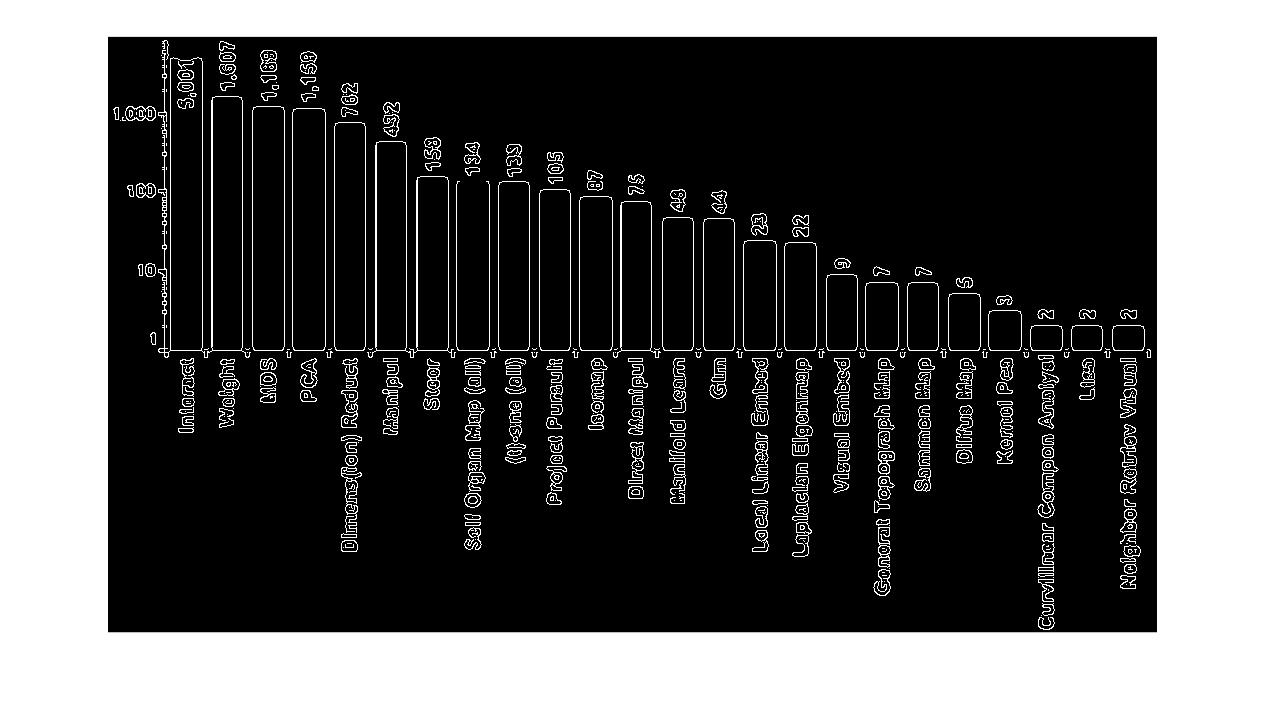}
    \caption{}
    \label{fig:sm:baredge}
    \end{subfigure}
\hspace{0.5cm}
    \begin{subfigure}[b]{0.2\columnwidth}
\includegraphics[height=2\pictureheight]{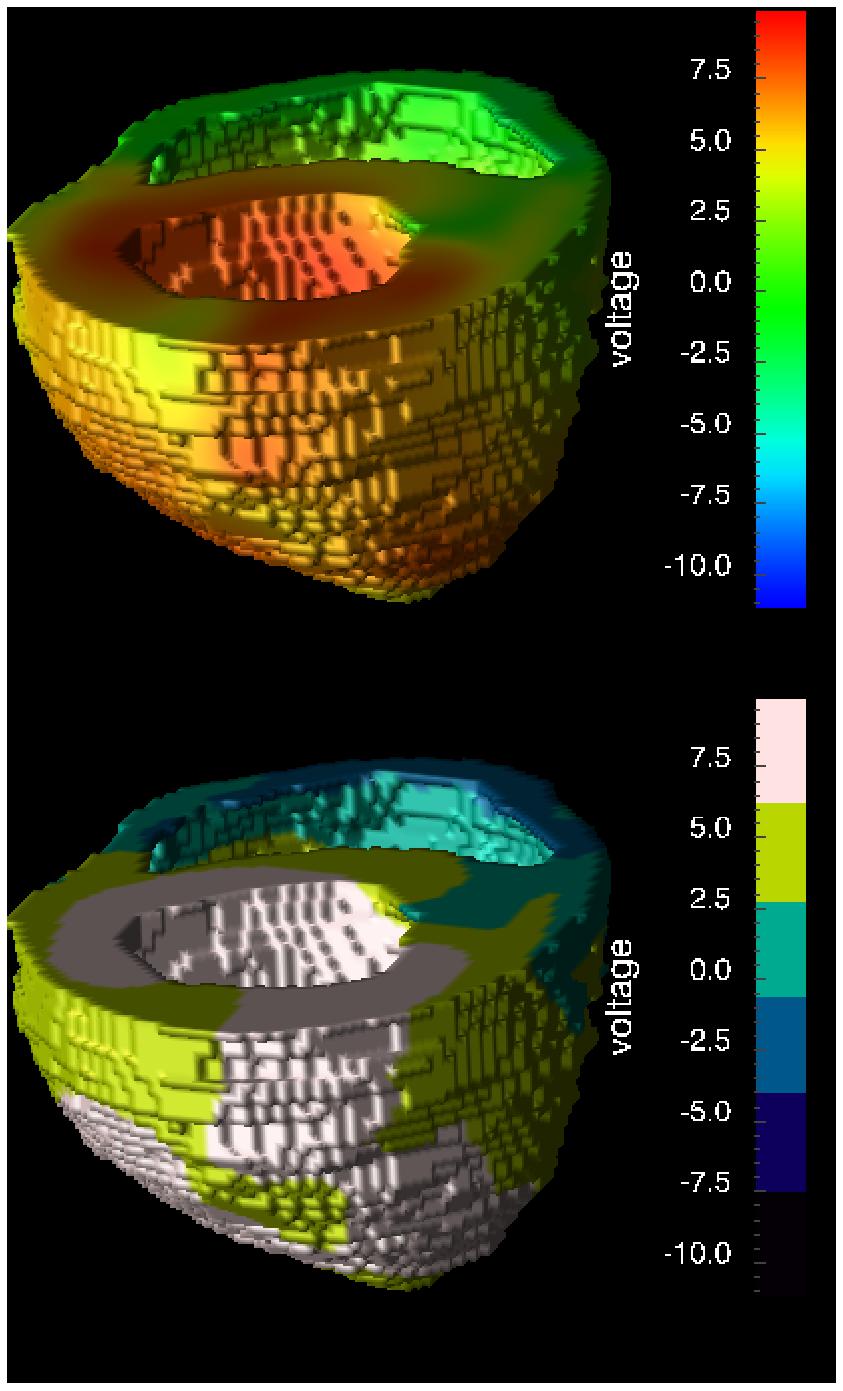}
    \caption{}
    \label{fig:sm:surface}
    \end{subfigure}
 \hspace{0.3cm}
    \begin{subfigure}[b]{0.3\columnwidth}
    \includegraphics[height=2\pictureheight]{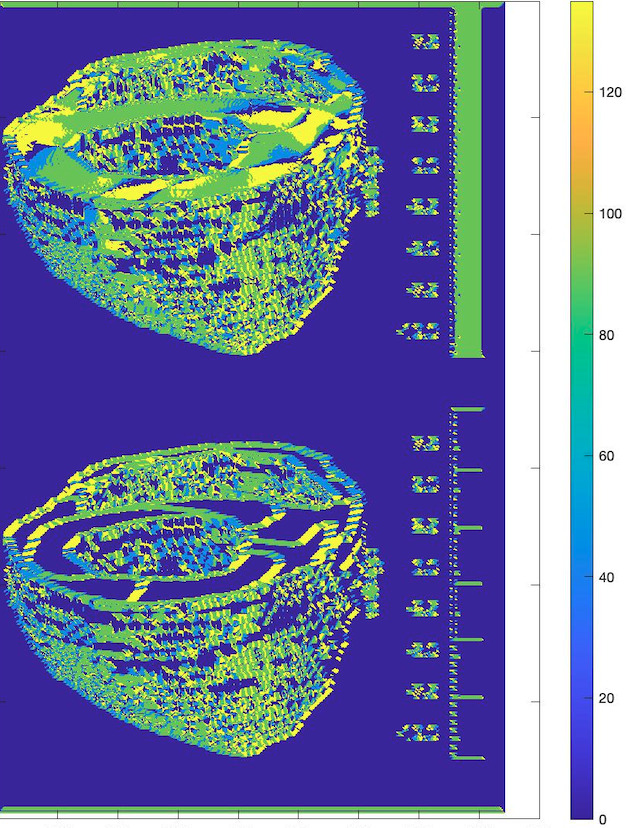}
    \caption{}
    \label{fig:sm:surfacehog}
    \end{subfigure}
    \caption{\jcfr{\textbf{Low-level features (edges and histograms of oriented gradients (HOG))  extracted from images do not necessarily represent how humans see.} (a) The original image in Sacha et al.~\cite{sacha2016visual} and (b) its edges;  (c) The original paper image in Gresh et al.~\cite{gresh2000weave} and (d) its HOGs.}
 }
    \label{fig:sm:barSurface}
\end{figure*}

\begin{figure*}[!ht]
	\subfloat[]{\includegraphics[clip, trim={50 0 0 0}, height=0.3\textwidth]{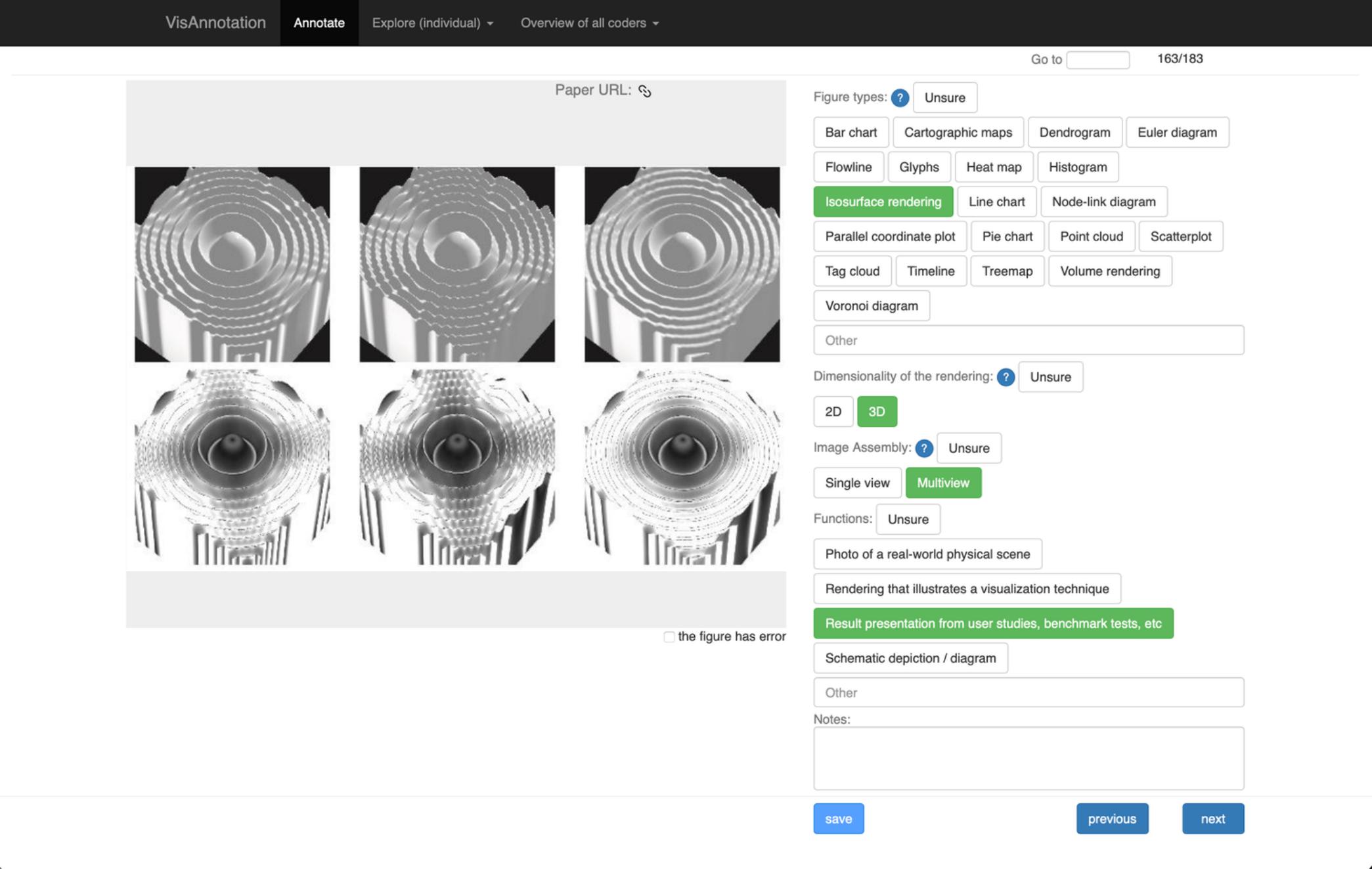}
	\label{fig:web-interface-phase1}}
	\subfloat[]{\includegraphics[clip, trim={50 0 0 0}, height=0.3\textwidth]{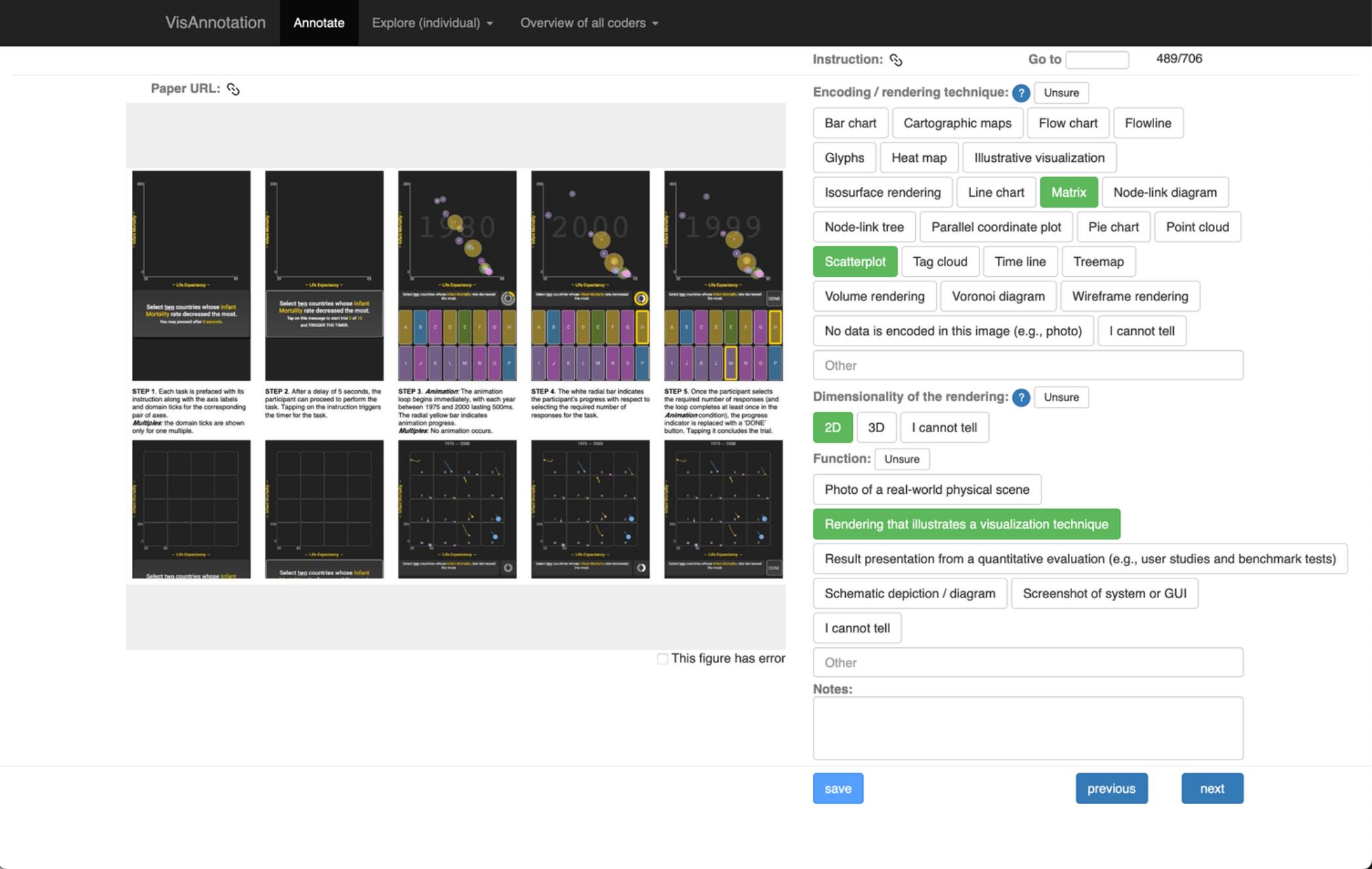}
	\label{fig:web-interface-phase2}}\\[2ex]
	\subfloat[]{\includegraphics[clip, trim={50 0 0 0}, height=0.3\textwidth]{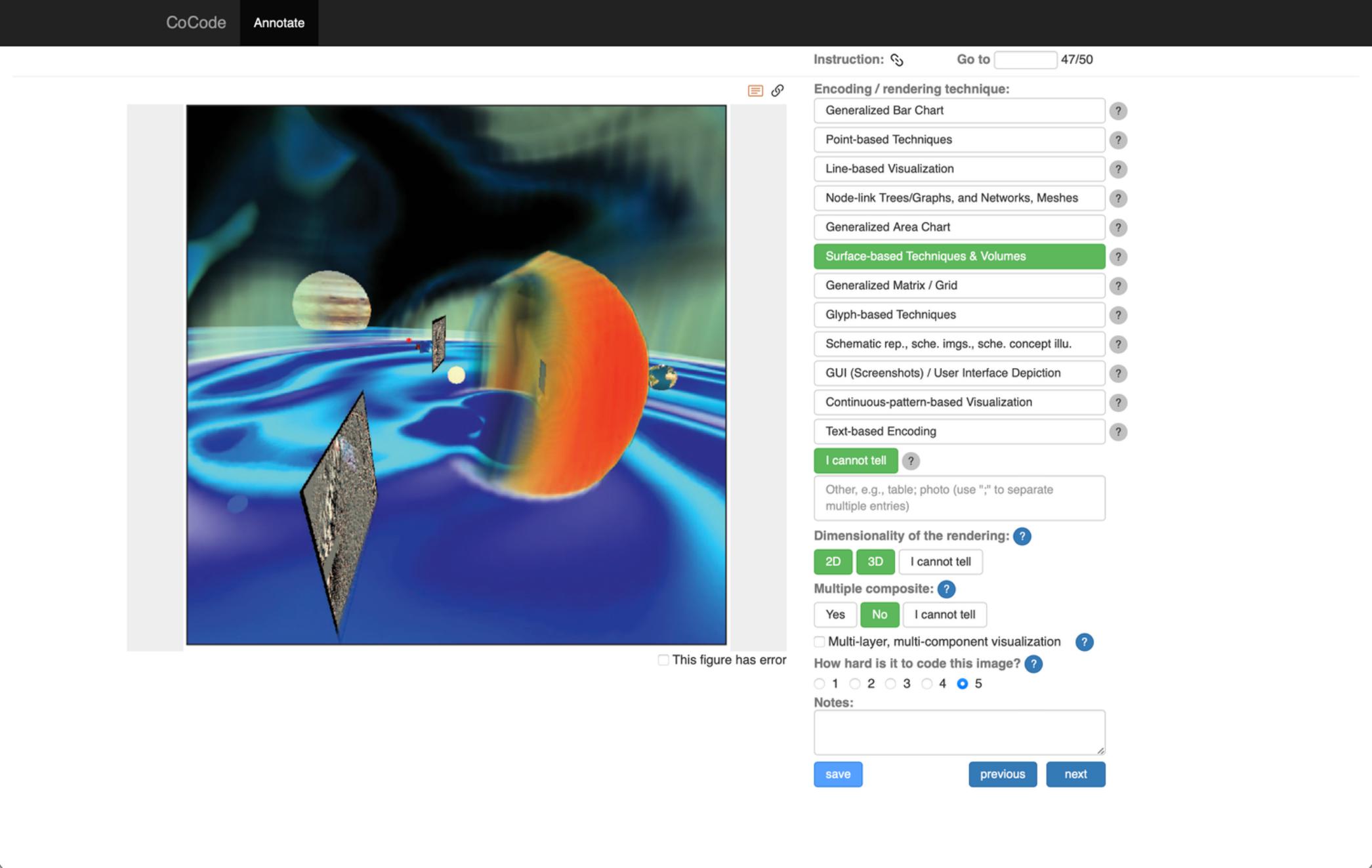}
	\label{fig:web-interface-phase3}}
	\subfloat[]{\includegraphics[clip, trim={50 0 0 0},height=0.3\textwidth]{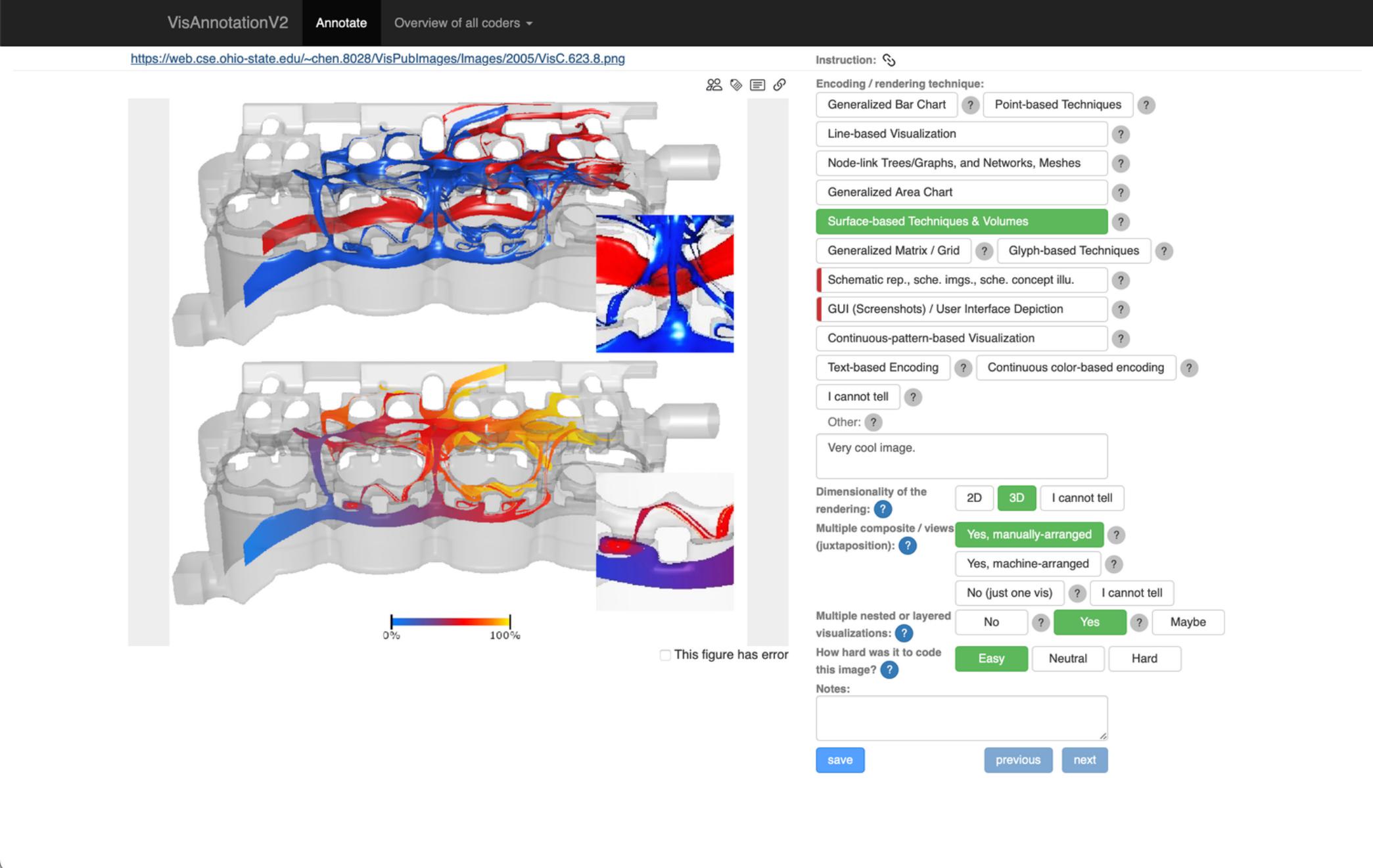}
	\label{fig:web-interface-phase4-5}}
	\caption{\tochange{\textbf{Screenshots of the web interfaces used in Phases 1--3 and Phases 4--5.} Phase 1 focused on techniques derived from authors' keywords:  The coders use the interfaces to code each image. The coding labels and categories were updated reflecting the results of weekly discussions. These label interfaces show that code revolution over time: from
	\textbf{Keyword-based} to Bertin-based to
	\textbf{Type-based} updates.
	}}
	\label{fig:web-interfaces}
\end{figure*}

\begin{figure*}
\includegraphics[width=\textwidth]{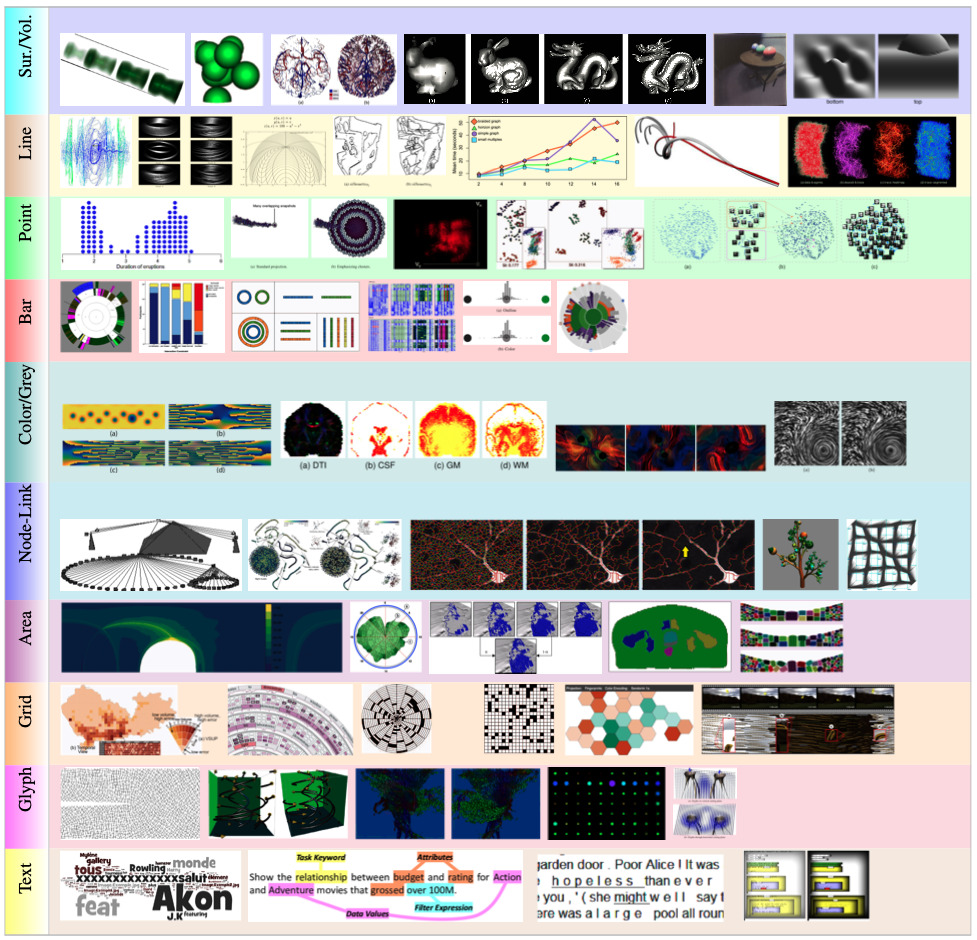}
\caption{\tochange{\textbf{\typename Image Examples. }More single-code image examples in our \typename data.}}
\label{fig:sm:typeImages}
\end{figure*}

\noindent We as authors state that the images in \autoref{fig:navigatorTwoHard}, 
\autoref{fig:sm:baredge}, \autoref{fig:sm:surfacehog}, \autoref{fig:web-interfaces}---
\jctg{\autoref{fig:sm:typeImages}}
in this appendix
are and remain under our own copyright, with permission to be used here.
We have also made them available under the \href{https://creativecommons.org/licenses/by/4.0/}{Creative Commons At\-tri\-bu\-tion 4.0 International (\ccLogo\,\ccAttribution\ \mbox{CC BY 4.0})} license and share them at 
\osflink.
All remaining images are \textcopyright\ IEEE, with the permission to be used here.

\begin{table*}[!ht]
  \caption{\textbf{The visualization \textit{function} and \textit{dimensionality} codes.} We coded images based on their function and dimensionality.  Additional codes not listed here were ``I cannot tell'' to label images that had unclear functions or dimensionalities. }
  \label{tab:funcDim}
  \scriptsize%
	\centering%
  \begin{tabular}{p{0.004\textwidth}p{0.175\textwidth}p{.45\textwidth}p{.295\textwidth}}
  \toprule
 \midrule
 & \textbf{Function Codes} &  \textit{Description} & \textit{Examples} \\
 \midrule
 A. & GUI Screenshots or GUI Photos & 
  Images that show a system or user interface. 
  & a photograph of a person sitting in front of a given system, 
  a figure containing GUI features such as windows, icons, cursor, and pointers (WIMP), or non-WIMP VR/AR interfaces.\\
 B. & Schematic Representations, Concept Illustrations & 
 Often simplified representations showing the appearance, structure, or logic of a process or concept. 
  & workflow diagrams, algorithm diagrams, sketches.\\
 \midrule
 &\textbf{Dimensionality Codes} &  \textit{Description} & \textit{Examples} \\
 \midrule
 & 2D & Flat representations, no specific depth codes added to renditions. & Most statistical charts, most maps, \dots
 \\
 & 3D & Representation with specific depth cues that achieve the perception of 3D (shading, perspective, lighting, …). & Most volume renderings, \dots 

 \\
  \bottomrule
  \end{tabular}\vspace{-1ex}%
\end{table*}

\definecolor{lightgray}{gray}{0.75}

\newcommand{\resheading}[1]{{\small \colorbox{green}{\begin{minipage}{0.0975\textwidth}{\textbf{#1 \vphantom{p\^{E}}}}\end{minipage}}}}

\newlength{\FrameHeight}
\setlength{\FrameHeight}{0.078\textwidth}

\newcommand{\moreSurVol}{
\includegraphics[keepaspectratio, height=\FrameHeight]{moreSurVol/VisC.679.15.png}
\includegraphics[keepaspectratio, height=\FrameHeight]{moreSurVol/VisJ.1198.11.png}
\includegraphics[keepaspectratio, height=\FrameHeight]{moreSurVol/SciVisJ.1301.6.png}
\includegraphics[keepaspectratio, height=\FrameHeight]{moreSurVol/VisC.103.8.png}
\includegraphics[keepaspectratio, height=\FrameHeight]{moreSurVol/VisC.19.10.png}
\includegraphics[keepaspectratio, height=\FrameHeight]{moreSurVol/VisC.87.4.png}
}

\newcommand{\moreLine}{
\includegraphics[keepaspectratio, height=\FrameHeight]{moreLine/VisC.131.6.png}
\includegraphics[keepaspectratio, height=\FrameHeight]{moreLine/VisC.501.3.png}
\includegraphics[keepaspectratio, height=\FrameHeight]{moreLine/VisC.311.10.png}
\includegraphics[keepaspectratio, height=\FrameHeight]{moreLine/VisC.663.7.png}
\includegraphics[keepaspectratio, height=\FrameHeight]{moreLine/InfoVisJ.927.15.png}
\includegraphics[keepaspectratio, height=\FrameHeight]{moreLine/VisJ.1225.14.png}
\includegraphics[keepaspectratio, height=\FrameHeight]{moreLine/SciVisJ.806.1.png}
}

\newcommand{\morePoint}{
\includegraphics[keepaspectratio, height=\FrameHeight]{morePoint/InfoVisJ.1044.2.png}
\includegraphics[keepaspectratio, height=\FrameHeight]{morePoint/VASTJ.1.9.png}
\includegraphics[keepaspectratio, height=\FrameHeight]{morePoint/VisC.469.11.png}
\includegraphics[keepaspectratio, height=\FrameHeight]{morePoint/VisJ.1281.15.png}
\includegraphics[keepaspectratio, height=\FrameHeight]{morePoint/InfoVisJ.1558.1.png}
}

\newcommand{\moreBar}{
\includegraphics[keepaspectratio, height=\FrameHeight]{moreBar/InfoVisC.57.1.png}
\includegraphics[keepaspectratio, height=\FrameHeight]{moreBar/VASTC.195.6.png}
\includegraphics[keepaspectratio, height=\FrameHeight]{moreBar/InfoVisJ.918.10.png}
\includegraphics[keepaspectratio, height=\FrameHeight]{moreBar/VASTC.19.12.png}
\includegraphics[keepaspectratio, height=\FrameHeight]{moreBar/InfoVisJ.1009.6.png}
\includegraphics[keepaspectratio, height=\FrameHeight]{moreBar/VASTJ.101.1.png}
}

\newcommand{\moreText}{
\includegraphics[keepaspectratio, height=\FrameHeight]{moreText/VASTC.147.9.png}
\includegraphics[keepaspectratio, height=\FrameHeight]{moreText/InfoVisJ.369.4.png}
\includegraphics[keepaspectratio, height=\FrameHeight]{moreText/InfoVisJ.489.9.png}
\includegraphics[keepaspectratio, height=\FrameHeight]{moreText/InfoVisC.24.3.png}
}

\newcommand{\moreNodeLink}{
\includegraphics[keepaspectratio, height=\FrameHeight]{moreNodeLink/InfoVisC.74.4.png}
\includegraphics[keepaspectratio, height=\FrameHeight]{moreNodeLink/VASTJ.1.14.png}
\includegraphics[keepaspectratio, height=\FrameHeight]{moreNodeLink/SciVisJ.744.2.png}
\includegraphics[keepaspectratio, height=\FrameHeight]{moreNodeLink/VisC.79.5.png}
\includegraphics[keepaspectratio, height=\FrameHeight]{moreNodeLink/InfoVisJ.1100.15.png}
}

\newcommand{\moreGrid}{
\includegraphics[keepaspectratio, height=\FrameHeight]{moreGrid/VASTJ.839.5.png}
\includegraphics[keepaspectratio, height=\FrameHeight]{moreGrid/VASTJ.1160.7.png}
\includegraphics[keepaspectratio, height=\FrameHeight]{moreGrid/InfoVisJ.935.9.png}
\includegraphics[keepaspectratio, height=\FrameHeight]{moreGrid/SciVisJ.891.2.png}
\includegraphics[keepaspectratio, height=\FrameHeight]{moreGrid/SciVisJ.1005.1.png}
}

\newcommand{\moreGlyph}{
\includegraphics[keepaspectratio, height=\FrameHeight]{moreGlyph/VisC.11.7.png}
\includegraphics[keepaspectratio, height=\FrameHeight]{moreGlyph/VisC.351.11.png}
\includegraphics[keepaspectratio, height=\FrameHeight]{moreGlyph/VisC.73.8.png}
\includegraphics[keepaspectratio, height=\FrameHeight]{moreGlyph/VisC.77.13.png}
\includegraphics[keepaspectratio, height=\FrameHeight]{moreGlyph/VisJ.1595.9.png}
}

\newcommand{\moreArea}{
\includegraphics[keepaspectratio, height=\FrameHeight]{moreArea/SciVisJ.796.18.png}
\includegraphics[keepaspectratio, height=\FrameHeight]{moreArea/SciVisJ.935.4.png}
\includegraphics[keepaspectratio, height=\FrameHeight]{moreArea/VisC.227.3.png}
\includegraphics[keepaspectratio, height=\FrameHeight]{moreArea/VisC.283.10.png}
\includegraphics[keepaspectratio, height=\FrameHeight]{moreArea/VisJ.1261.4.png}
}

\newcommand{\moreColor}{
\includegraphics[keepaspectratio, height=\FrameHeight]{moreColor/SciVisJ.827.5.png}
\includegraphics[keepaspectratio, height=\FrameHeight]{moreColor/SciVisJ.975.7.png}
\includegraphics[keepaspectratio, height=0.7\FrameHeight]{moreColor/VisC.175.1.png}
\includegraphics[keepaspectratio, height=\FrameHeight]{moreColor/VisC.639.3.png}
}

\end{document}